\RequirePackage[l2tabu,orthodox]{nag}
\RequirePackage{fix-cm}
\documentclass[12pt,a4paper]{article}

\usepackage{fixltx2e}

\usepackage[font=small]{caption}
\usepackage[text={450pt,650pt},headheight={14.5pt},centering]{geometry}

\usepackage{lmodern}

\usepackage{graphicx}
\usepackage{booktabs}
\usepackage{subfig}
\usepackage{tikz}
\usetikzlibrary{plotmarks}
\usetikzlibrary{calc}

\tikzstyle arrow=[thick,-latex]
\tikzstyle box=[draw,rounded corners,thick,outer sep=4pt]
\tikzstyle B node=[outer sep=0pt]
\tikzstyle Q node=[inner sep=1pt,outer sep=0pt]

\tikzstyle S-mat=[circle,thick,draw=black,fill=black!20!white,inner sep=2pt]

\usepackage[numbers,merge]{natbib}
\usepackage{collnat}

\usepackage{amsmath,amssymb}
\usepackage{mathtools}
\usepackage{bbm}
\usepackage{slashed}
\usepackage{braket}

\usepackage{dsfont}

\usepackage{hyperref}

\usepackage{xspace}

\numberwithin{equation}{section}

\makeatletter
 \let\old@startsection=\@startsection
 \let\oldl@section=\l@section
 \renewcommand{\@startsection}[6]{\old@startsection{#1}{#2}{#3}{#4}{#5}{#6\mathversion{bold}}}
 \renewcommand{\l@section}[2]{\oldl@section{\mathversion{bold}#1}{#2}}
\makeatother

\DeclareMathOperator{\diag}{diag}

\def\XXint#1#2#3{{\setbox0=\hbox{$#1{#2#3}{\int}$}
    \vcenter{\hbox{$#2#3$}}\kern-.5\wd0}}

\newcommand{\AdS}{\textup{AdS}}
\newcommand{\CFT}{\textup{CFT}}
\newcommand{\Sphere}{\textup{S}}
\newcommand{\CP}{\textup{CP}}
\newcommand{\Torus}{\textup{T}}

\newcommand{\Smat}{\mathcal{S}}
\newcommand{\Tmat}{\mathcal{T}}
\newcommand{\Rmat}{\mathcal{R}}

\newcommand{\mat}[1]{\mathbb{#1}}
\newcommand{\matId}{\mathds{1}}

\newcommand{\Ael}{\mathsf{A}}
\newcommand{\Bel}{\mathsf{B}}
\newcommand{\Cel}{\mathsf{C}}
\newcommand{\Del}{\mathsf{D}}
\newcommand{\Eel}{\mathsf{E}}
\newcommand{\Fel}{\mathsf{F}}

\newcommand{\comm}[2]{[#1,#2]}
\newcommand{\acomm}[2]{\{#1,#2\}}

\newcommand{\abs}[1]{{\left| #1 \right|}}

\newcommand{\alg}[1]{\mathfrak{#1}}
\newcommand{\grp}[1]{\mathrm{#1}}

\newcommand{\algD}[1]{\alg{d}(2,1;#1)}
\newcommand{\grpD}[1]{\grp{D}(2,1;#1)}

\newcommand{\algSL}{\alg{sl}}

\newcommand{\algSU}{\alg{su}}
\newcommand{\grpSU}{\grp{SU}}

\newcommand{\algU}{\alg{u}}

\newcommand{\algPSU}{\alg{psu}}

\newcommand{\gen}[1]{\mathfrak{#1}}

\newcommand{\genQ}{\gen{Q}}
\newcommand{\genS}{\gen{S}}
\newcommand{\genH}{\gen{H}}
\newcommand{\genB}{\gen{B}}
\newcommand{\genP}{\gen{P}}
\newcommand{\genK}{\gen{P}^\dag}
\newcommand{\genM}{\gen{M}}

\newcommand{\rep}[1]{\mathbf{#1}}

\newcommand{\Integers}{\mathbbm{Z}}

\newcommand{\superN}{\mathcal{N}}

\newcommand{\ie}{\textit{i.e.}\xspace}
\newcommand{\eg}{\textit{e.g.}\xspace}

\newcommand{\fixedspaceL}[2]{\mathrlap{#2}\phantom{#1}}
\newcommand{\fixedspaceR}[2]{\phantom{#1}\mathllap{#2}}

\newcommand{\smallL}{\mbox{\itshape\tiny L}}
\newcommand{\smallR}{\mbox{\itshape\tiny R}}

\newcommand{\smallLL}{\mbox{\itshape\tiny LL}}
\newcommand{\smallLR}{\mbox{\itshape\tiny LR}}
\newcommand{\smallRL}{\mbox{\itshape\tiny RL}}
\newcommand{\smallRR}{\mbox{\itshape\tiny RR}}

\newcommand{\smallLLp}{\mbox{\itshape\tiny LL${}'$}}
\newcommand{\smallLpL}{\mbox{\itshape\tiny L${}'$L}}

\newcommand{\smallLRp}{\mbox{\itshape\tiny LR${}'$}}

\newcommand{\smallRLp}{\mbox{\itshape\tiny RL${}'$}}
\newcommand{\smallRpL}{\mbox{\itshape\tiny R${}'$L}}

\newcommand{\smallRRp}{\mbox{\itshape\tiny RR${}'$}}
\newcommand{\smallRpR}{\mbox{\itshape\tiny R${}'$R}}

\newcommand{\ZpIIext}{Z^+_-}
\newcommand{\ZmIIext}{Z^-_+}

\begin{document}

\thispagestyle{empty}

\begin{flushright}\footnotesize\ttfamily
ITP-UU-12/46\\
SPIN-12/43
\end{flushright}
\vspace{5em}

\begin{center}
\textbf{\Large\mathversion{bold} A dynamic $\algSU(1|1)^2$ S-matrix for $\AdS_3/\CFT_2$}

\vspace{2em}

\textrm{\large Riccardo Borsato, Olof Ohlsson Sax and Alessandro Sfondrini} 

\vspace{2em}

\textit{Institute for Theoretical Physics and Spinoza Institute,\\ Utrecht University, 3508 TD Utrecht, The Netherlands}

\vspace{1em}

\texttt{R.Borsato@uu.nl, O.E.OlssonSax@uu.nl, A.Sfondrini@uu.nl}

%%%%%%%%

\end{center}

\vspace{6em}

\begin{abstract}\noindent
We derive the S-matrix for the $\algD{\alpha}^2$ symmetric spin-chain of $\AdS_3/\CFT_2$ by considering the centrally extended $\algSU(1|1)^2$ algebra acting on the spin-chain excitations. The S-matrix is determined uniquely up to four scalar factors, which are further constrained by a set of crossing relations. The resulting scattering includes non-trivial processes between magnons of different masses that were previously overlooked.
\end{abstract}

%%%%%%%%%%%%%%%%%%%%%%%%%%%%%%%%%%%%%%%%%%%%%%%%%%%%%%%%%%%%%%%%%%%%%%%%%%%
\newpage

\section{Introduction}
\label{sec:introduction}
The AdS/CFT correspondence \cite{Maldacena:1997re,Witten:1998qj,Gubser:1998bc} conjectures the equivalence of two seemingly different theories: a string theory on some $\AdS_{d+1}\times X$, where $X$ is a compact space, and a conformal field theory (CFT) on the $d$-dimensional boundary of AdS. A significant success in the investigation of the correspondence was obtained in the case of type IIB strings on $\AdS_5\times \Sphere^5$ and $\mathcal{N}=4$ Super-Yang-Mills (SYM) in the 't Hooft limit, where it was found that integrability techniques could be applied to the computation of the string energy spectrum, or equivalently to the spectrum of the dilatation operator in SYM, see \cite{Arutyunov:2009ga,Beisert:2010jr} for a review; later, similar integrable structures were found also for different instances of the AdS/CFT correspondence.

A particularly interesting case is the one of the $\AdS_3/\CFT_2$ correspondence. Historically, this was one of the earliest examples of holography \cite{Brown:1986nw}. In particular, we are interested in the case of type IIB strings on $\AdS_3\times \Sphere^3\times \Sphere^3\times \Sphere^1$, a background that preserves 16 supercharges. Due to the special properties of 2-dimensional CFTs, and to the fact that the background can be supported by a pure NSNS flux, the resulting string theory and dual conformal field theory are relatively well understood by means of representation theory of chiral algebras \cite{Giveon:1998ns,Elitzur:1998mm,Maldacena:2000hw,Maldacena:2000kv}. The case of a RR background appears less manageable in terms of the NSR formalism \cite{Berkovits:1999im}, but is strikingly similar to the $\AdS_5\times \Sphere^5$ background. Importantly, the Green-Schwarz action on $\AdS_3\times \Sphere^3 \times \Sphere^3\times \Sphere^1$ can essentially be rewritten as a non-linear sigma model on the $\Integers_4$-graded supersymmetric coset
\begin{equation*}
  \frac{\grpD{\alpha}\times \grpD{\alpha}}{\grpSU(1,1)\times \grpSU(2)\times \grpSU(2)} ,
\end{equation*}
which guarantees classical integrability \cite{Babichenko:2009dk,Zarembo:2010yz}. However, the action for the coset sigma-model does not contain two massless modes of the original string theory, which have to be put in by hand \cite{Babichenko:2009dk,Sundin:2012gc}. Classical integrability was later extended to a family of backgrounds interpolating between pure RR and pure NSNS \cite{Cagnazzo:2012se}.

Focusing on a pure RR background and setting aside the issue of how the massless modes should be included in the full quantum theory, it was possible to use classical integrability to write down finite gap equations for the massive modes, out of which a plausible form of the all-loop Bethe ansatz was proposed, yielding an alternating $\algD{\alpha}^2$ spin-chain \cite{OhlssonSax:2011ms}. This was then compared to near plane wave calculations in the string theory, finding partial agreement~\cite{Rughoonauth:2012qd}. The problem of incorporating the massless modes is still not well understood; a first step in this direction was taken in \cite{Sax:2012jv}, where the decoupling of four additional massless modes in the limit where one of the $\Sphere^3$'s blows up, which gives an $\AdS_3\times \Sphere^3\times \Torus^4$ background upon compactification, was studied in detail.

In this paper we will focus on the quantum integrability of the massive modes of $\AdS_3\times \Sphere^3\times \Sphere^3\times \Sphere^1$ strings, or equivalently of the $\algD{\alpha}^2$ spin-chain. We bootstrap the all-loop two-body S-matrix out of the symmetries of the theory, in a way conceptually similar to what was done for $\AdS_5\times \Sphere^5$ strings and $\mathcal{N}=4$ SYM in \cite{Beisert:2005tm}. To do so, we study the $\algD{\alpha}^2$ symmetry algebra, which is broken to a centrally extended $\algSU(1|1)^2$ algebra by the choice of the vacuum, similarly to what was found from studying the symmetries of giant magnons in~\cite{David:2008yk}. Invariance under this symmetry, together with unitarity requirements and a discrete $\Integers_2$ symmetry between excitations of left and right chirality yields essentially a unique S-matrix, which turns out to satisfy the Yang-Baxter equation, and hence is compatible with integrability. This S-matrix generalizes the result of~\cite{Beisert:2005wm,David:2010yg} by including both left- and right-movers, as well as excitations of different masses. To completely determine the S-matrix one needs to fix four antisymmetric phases that play a role similar to the dressing phase of $\AdS_5\times \Sphere^5$ strings~\cite{Arutyunov:2004vx,Beisert:2006ez,Beisert:2006ib,Volin:2009uv,Vieira:2010kb}. These must obey a set of crossing equations whose solution seemingly cannot be given in terms of simple expressions involving the Beisert-Eden-Staudacher dressing phase \cite{Beisert:2006ez}. Furthermore, some of these phases yield non-trivial scattering processes between modes of different mass that were not accounted for in \cite{Babichenko:2009dk,OhlssonSax:2011ms}. 

The plan of the paper is as follows. In section \ref{sec:d21a-spin-chain}, we review the construction of the alternating $\algD{\alpha}^2$ symmetric spin-chain. In section~\ref{sec:centrally-extended-algebra} we discuss the $\algSU(1|1)^2$ algebra and how to construct its extension. In section~\ref{sec:centrally-extended-representations} we discuss the resulting representations, obtaining the dispersion relations and level matching conditions. Section \ref{sec:S-matrix} is devoted to finding the S-matrix out of the extended $\algSU(1|1)^2$ symmetry; we also discuss how this can be related to the $\algPSU(2|2)$ invariant S-matrix of \cite{Beisert:2005tm}, and derive the crossing equations. Some technical details are given in the appendices.

The derivation of the Bethe ansatz equations resulting from this all-loop S-matrix, together with the comparison with string theory predictions~\cite{Babichenko:2009dk,Rughoonauth:2012qd} will be presented in a separate paper~\cite{Borsato:2012ss}. 

\bigskip\noindent
\textbf{Note added:} At the final stage of the preparation of this article, another work aimed at finding the all-loop $\AdS_3/\CFT_2$ S-matrix appeared~\cite{Ahn:2012hw}. The technique used as well as the results found there differ from the ones presented here; in particular the authors there reproduce the conjectured Bethe ansatz of \cite{Babichenko:2009dk,OhlssonSax:2011ms}, while our S-matrix leads to modification of the Bethe equations~\cite{Borsato:2012ss}. For a further comparison of the results of this paper with those of~\cite{Ahn:2012hw}, see appendix~\ref{sec:beisert-su11-S-matrix}.

\section{The alternating \texorpdfstring{$\algD{\alpha}^2$}{d(2,1;a) x d(2,1;a)} symmetric spin-chain}
\label{sec:d21a-spin-chain}

In~\cite{OhlssonSax:2011ms}, an alternating spin-chain with $\algD{\alpha}^2$ symmetry was constructed. In this section we will review that construction. The two copies of the exceptional superalgebra $\algD{\alpha}$ describe, respectively, left- and right-movers in $\AdS_3 \times \Sphere^3 \times \Sphere^3 \times \Sphere^1$. The parameter $\alpha$ is related to the ratio of the radii $R_+$ and $R_-$ of the two three-spheres in the string background. This background preserves 16 supersymmetries provided $R_+$, $R_-$ and the $\AdS_3$ radius $R_{\AdS}$ satisfy the relation
\begin{equation}
  \frac{1}{R_{\AdS}^2} = \frac{1}{R_+^2} +  \frac{1}{R_-^2} .
\end{equation}
This defines a one-parameter family of backgrounds, parametrized by
\begin{equation}
  \alpha = \frac{R_{\AdS}^2}{R_+^2} = 1 - \frac{R_{\AdS}^2}{R_-^2} ,
\end{equation}
with $0<\alpha<1$.

In the weak coupling limit, the left- and right-movers decouple. The analysis of~\cite{OhlssonSax:2011ms} was therefore focused on the left-moving sector only. When higher order corrections are accounted for, these sectors start to interact with each other. In this paper we will construct an all-loop two-particle S-matrix for the full spin-chain. However, in order to better understand the general setup it is useful to start by discussing the symmetries of the spin-chain at weak coupling.

The left-movers are described by an alternating spin-chain, with odd and even sites in the $\algD{\alpha}$ representations $(-\tfrac{\alpha}{2};\tfrac{1}{2};0)$ and $(-\tfrac{1-\alpha}{2};0;\tfrac{1}{2})$, respectively. In analogy with the gauge theory construction of spin-chains in, \eg, $\superN=4$ super Yang-Mills we will refer to these representations as the field modules, and correspondingly call the states sitting at a specific site ``fields'', even though there is not an obvious correspondence to fundamental fields in the underlying CFT.

The bosonic subalgebra of $\algD{\alpha}$ is given by $\algSL(2) \otimes \algSU(2)_+ \otimes \algSU(2)_-$, where we have added the subscripts $\pm$ on the $\algSU(2)$ factors to be able to distinguish them. We denote the corresponding triplets of generators by\footnote{%
  The $\algSL(2)$ generators of $\algD{\alpha}$ are often denoted $\gen{S}_\mu$. However, $\genS$ will be a supercharge of $\algU(1|1)$ in the next section, so here we denote the $\algSL(2)$ generators by $\gen{J}_\mu$.%
} %
$\gen{J}_0$, $\gen{J}_{\pm}$ for the $\algSL(2)$ algebra; $\gen{L}_5$, $\gen{L}_{\pm}$ for $\algSU(2)_+$; and $\gen{R}_8$, $\gen{R}_{\pm}$ for $\algSU(2)_-$. In addition there are eight supercharges $\gen{Q}_{\pm\pm\pm}$ transforming as a tri-spinor under the bosonic subalgebra. The commutation relations of the $\algD{\alpha}$ algebra are given in appendix~\ref{sec:d21a-algebra}. When we later consider left- and right-movers at the same time, we need to introduce some extra notation in order to distinguish the generators of each sector. For now we will keep the notation simpler, since we only focus on the left-moving sector.

\begin{figure}
  \centering
  \subfloat[\label{fig:dynkin-d21a-orig}]{
    \begin{tikzpicture}
      [
      thick,
      node/.style={shape=circle,draw,thick,inner sep=0pt,minimum size=5mm}
      ]
      \useasboundingbox (-1.1cm,-1.1cm) rectangle (1.6cm,1.1cm);

      \node (v1) at (-0.38cm,  0.65cm) [node] {};
      \node (v2) at ( 0.75cm,  0.00cm) [node] {};
      \node (v3) at (-0.38cm, -0.65cm) [node] {};

      \draw (v2.south west) -- (v2.north east);
      \draw (v2.north west) -- (v2.south east);

      \draw (v1) -- (v2);
      \draw (v2) -- (v3);
    \end{tikzpicture}
  }
  \hspace{2cm}
  \subfloat[\label{fig:dynkin-d21a-dual}]{
    \begin{tikzpicture}
      [
      thick,
      node/.style={shape=circle,draw,thick,inner sep=0pt,minimum size=5mm}
      ]
      \useasboundingbox (-1.1cm,-1.1cm) rectangle (1.6cm,1.1cm);

      \node (v1) at (-0.38cm,  0.65cm) [node] {};
      \node (v2) at ( 0.75cm,  0.00cm) [node] {};
      \node (v3) at (-0.38cm, -0.65cm) [node] {};

      \draw (v1.south west) -- (v1.north east);
      \draw (v1.north west) -- (v1.south east);

      \draw (v2.south west) -- (v2.north east);
      \draw (v2.north west) -- (v2.south east);

      \draw (v3.south west) -- (v3.north east);
      \draw (v3.north west) -- (v3.south east);

      \draw (v1) -- (v2);
      \draw (v2) -- (v3);
      \draw [double,double distance=3pt] (v3) -- (v1);
    \end{tikzpicture}
  }
  
  \caption{Two Dynkin diagrams for $\algD{\alpha}$. Diagram~\protect\subref{fig:dynkin-d21a-orig} corresponds to the choice of positive roots in~\eqref{eq:d21a-pos-roots-I}, while diagram~\protect\subref{fig:dynkin-d21a-dual} corresponds to the choice in~\eqref{eq:d21a-pos-roots-II}.}

  \label{fig:d21a-dynkin-diagrams}
\end{figure}
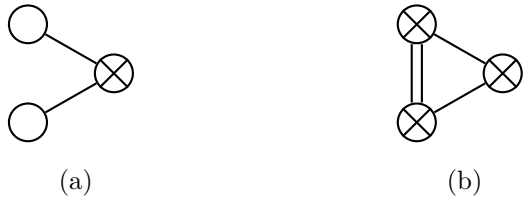

In a superalgebra, there are in general several non-equivalent choices of simple roots, with different choices leading to different Cartan matrices and Dynkin diagrams. Here we will be considering two different such choices. First we have the mixed form, where the simple roots, up to some normalization factors, are given by
\begin{equation}\label{eq:d21a-pos-roots-I}
  \gen{L}_+ , \qquad
  \gen{Q}_{+--} , \qquad
  \gen{R}_+ .
\end{equation}
This corresponds to the Dynkin diagram in figure~\ref{fig:dynkin-d21a-orig}, 
with the Cartan matrix given by
\begin{equation}\label{eq:cartan-matrix-I}
  A =
  \begin{pmatrix}
    \phantom{+}4\alpha & -2\alpha & 0 \\
    -2\alpha & 0 & -2(1-\alpha) \\
    0 & -2(1-\alpha) & \phantom{+}4(1-\alpha)
  \end{pmatrix},
\end{equation}
Alternatively, we can use a fermionic set of simple roots, such as
\begin{equation}\label{eq:d21a-pos-roots-II}
  \gen{Q}_{++-} , \qquad
  \gen{Q}_{-++} , \qquad
  \gen{Q}_{+-+} .
\end{equation}
The corresponding Dynkin diagram is shown in figure~\ref{fig:dynkin-d21a-dual}, and the Cartan matrix now reads
\begin{equation}\label{eq:cartan-matrix-II}
  A =
  \begin{pmatrix}
    0 & 2\alpha & -2 \\
    2\alpha & 0 & 2(1-\alpha) \\
    -2 & 2(1-\alpha) & 0
  \end{pmatrix}.
\end{equation}
In both these bases, the generators 
\begin{equation}\label{eq:d21a-pos-roots-common}
  \gen{J}_+, \qquad \gen{L}_+, \qquad \gen{R}_+, \qquad \gen{Q}_{++-}, \qquad \gen{Q}_{+-+}, \qquad \gen{Q}_{+++} 
\end{equation}
corresponds to positive roots. Hence, the only difference between the two cases is whether $\gen{Q}_{+--}$ or $\gen{Q}_{-++}$ is considered a raising operator.

\paragraph{Spin-chain representations.}

The two representations at the odd and even sites of the spin-chain are short representations of $\algD{\alpha}$. The representation $(-\tfrac{\alpha}{2};\tfrac{1}{2};0)$ at the odd sites consists of bosons $\phi^{(n)}_{\pm}$ transforming as a doublet under $\algSU(2)_+$, and fermions $\psi^{(n)}_{\pm}$ forming a doublet under $\algSU(2)_-$. The indices $n$ indicate the $\algSL(2)$ level of the fields. Similarly, the $(-\tfrac{1-\alpha}{2};0;\tfrac{1}{2})$ representation consists of the bosons $\tilde{\phi}_{\pm}^{(n)}$ and fermions $\tilde{\psi}_{\pm}^{(n)}$, which form doublets under $\algSU(2)_-$ and $\algSU(2)_+$, respectively. 

The action of the $\algD{\alpha}$ generators on the representations can be found in appendix~\ref{sec:d21a-algebra}. Using these expressions, it is straightforward to check that the states $\phi_+^{(0)}$ and $\tilde{\phi}_+^{(0)}$ are annihilated by the charges in~\eqref{eq:d21a-pos-roots-common}, as well as by $\gen{Q}_{+--}$ and $\gen{Q}_{-++}$. Additionally, $\phi_+^{(0)}$ is killed by $\gen{Q}_{-+-}$, while $\tilde{\phi}_+^{(0)}$ is killed by $\gen{Q}_{--+}$. Hence, these states are highest weight in both the bases discussed above. Since both highest weight states are annihilated by two extra supercharges, the two representations are 1/2-BPS.

\paragraph{The spin-chain ground state.}

As a ground state for the alternating spin-chain we use the state
\begin{equation}
  \ket{0}_L = (\phi_+^{(0)}\tilde{\phi}_+^{(0)})^L ,
\end{equation}
which is the highest weight state in the short $(-\frac{L}{2};\frac{L}{2};\frac{L}{2})$ representation. Note that the length $L$ refers to the number of pairs of odd and even sites, so that the number of sites in the full spin-chain actually is $2L$. This state is annihilated by $\gen{Q}_{+--}$ and $\gen{Q}_{-++}$ and is therefore 1/4-BPS. Hence, it satisfies the shortening condition
\begin{equation}
  \acomm{\gen{Q}_{+--}}{\gen{Q}_{-++}} \ket{0}_L = \gen{H} \ket{0}_L = 0 , \qquad
  \gen{H} = - \gen{J}_0 - \alpha \, \gen{L}_5 - (1-\alpha) \, \gen{R}_8 ,
\end{equation}
where we have introduced the left-moving spin-chain Hamiltonian $\gen{H}$.

\paragraph{Spin-chain excitations.}

Excited states are constructed by replacing one or more of the fields in the ground states by descendant states from the same module. If we only want to calculate the classical charges of a state, the exact positions of the excitations are not important, only the field content of the state. We can then denote an excited state by listing all pairs of odd and even sites where at least one field is different from the ground state. For example we write
\begin{equation}
  \begin{aligned}
    \ket{\phi_-^{(0)} \tilde{\phi}_+^{(0)}}_L &= (\phi_+^{(0)}\tilde{\phi}_+^{(0)})^{L-1} (\phi_-^{(0)}\tilde{\phi}_+^{(0)}) , \\
    \ket{\psi_+^{(2)} \tilde{\phi}_-^{(1)} \psi_-^{(0)} \tilde{\phi}_+^{(0)}}_L &= (\phi_+^{(0)}\tilde{\phi}_+^{(0)})^{L-2} (\psi_-^{(2)}\tilde{\phi}_-^{(1)}\psi_-^{(0)}\tilde{\phi}_+^{(0)}) .
  \end{aligned}
\end{equation}
The charges of all states with an excitation at a single site is shown in table~\ref{tab:charges-single-excitation}. These charges are given as the difference between the charge of the excited state and the ground state, which has $(-J_0,L_5,R_8) = (\frac{L}{2},\frac{L}{2},\frac{L}{2})$, where we denote by $J_0$ the eigenvalue of $\alg{J}_0$, and so on.
\begin{table}
  \centering
  \begin{tabular}{cccccl}
    \toprule
    & $-\delta J_0^L$ & $\delta L_5$ & $\delta R_8$ & $H$ \\
    \midrule
    $\ket{\phi_+^{(n)}\tilde{\phi}_+^{(0)}}_L$ & $n$ & $0$ & $0$ & $n$ \\
    $\ket{\phi_-^{(n)}\tilde{\phi}_+^{(0)}}_L$ & $n$ & $-1$ & $0$ & $n+\alpha$ \\
    $\ket{\psi_+^{(n)}\tilde{\phi}_+^{(0)}}_L$ & $n+\frac{1}{2}$ & $-\frac{1}{2}$ & $+\frac{1}{2}$ & $n+\alpha$ \\
    $\ket{\psi_-^{(n)}\tilde{\phi}_+^{(0)}}_L$ & $n+\frac{1}{2}$ & $-\frac{1}{2}$ & $-\frac{1}{2}$ & $n+1$ \\
    \midrule
    $\ket{\phi_+^{(0)}\tilde{\phi}_+^{(n)}}_L$ & $n$ & $0$ & $0$ & $n$ \\
    $\ket{\phi_+^{(0)}\tilde{\phi}_-^{(n)}}_L$ & $n$ & $0$ & $-1$ & $n+1-\alpha$ \\
    $\ket{\phi_+^{(0)}\tilde{\psi}_+^{(n)}}_L$ & $n+\frac{1}{2}$ & $+\frac{1}{2}$ & $-\frac{1}{2}$ & $n+1-\alpha$ \\
    $\ket{\phi_+^{(0)}\tilde{\psi}_-^{(n)}}_L$ & $n+\frac{1}{2}$ & $-\frac{1}{2}$ & $-\frac{1}{2}$ & $n+1$ \\
    \bottomrule
  \end{tabular}
  \caption{The charges of states with a single excitation on either an odd or an even site relative to the charges of the ground state of length $L$.}
  \label{tab:charges-single-excitation}
\end{table}

The lightest excited states in table~\ref{tab:charges-single-excitation} are
\begin{equation*}
  \ket{\phi_-^{(0)}\tilde{\phi}_+^{(0)}}_L , \qquad
  \ket{\psi_+^{(0)}\tilde{\phi}_+^{(0)}}_L , \qquad
  \ket{\phi_+^{(0)}\tilde{\phi}_-^{(0)}}_L  \qquad \text{and} \qquad
  \ket{\phi_+^{(0)}\tilde{\psi}_+^{(0)}}_L ,
\end{equation*}
the first two state having mass $\alpha$ and the second two $1-\alpha$. These play the role of fundamental excitations in the spin-chain. The states containing heavier excitations carry the same charges as a state with several fundamental excitations, and can hence be thought of as composite states. The first two such states are
\begin{equation}\label{eq:fundamental-excitations}
  \ket{\psi_-^{(0)}\tilde{\phi}_+^{(0)}}_L \qquad \text{and} \qquad \ket{\phi_+^{(0)}\tilde{\psi}_-^{(0)}}_L .
\end{equation}
These states are degenerate with each other and with the states
\begin{equation*}
  \ket{\phi_-^{(0)}\tilde{\psi}_+^{(0)}}_L \qquad \text{and} \qquad \ket{\psi_+^{(0)}\tilde{\phi}_-^{(0)}}_L ,
\end{equation*}
and hence correspond to double excitations. Another instance is given by
\begin{equation*}
  \ket{\phi_+^{(1)}\tilde{\phi}_+^{(0)}}_L \qquad \text{and} \qquad \ket{\phi_+^{(0)}\tilde{\phi}_+^{(1)}}_L ,
\end{equation*}
which are again double excitations consisting of
\begin{equation*}
  \ket{\psi_+^{(0)}\tilde{\psi}_+^{(0)}}_L .
\end{equation*}
In the same way any other excitation with a positive $\algSL(2)$ quantum number is related to a state containing a chain of $\psi_+^{(0)}\tilde{\psi}_+^{(0)}$ excitations, \eg,
\begin{equation}
  \ket{\psi_+^{(n)} \tilde{\phi}_+^{(0)}}_L \sim \ket{(\psi_+^{(0)}\tilde{\psi}_+^{(0)})^n \psi_+^{(0)}\tilde{\phi}_+^{(0)}}_L , \qquad
  \ket{\phi_+^{(0)} \tilde{\psi}_+^{(n)}}_L \sim \ket{\phi_+^{(0)}\tilde{\psi}_+^{(0)} (\psi_+^{(0)}\tilde{\psi}_+^{(0)})^n}_L .
\end{equation}
Hence, all the excited states in table~\ref{tab:charges-single-excitation} are related to a state containing only the fundamental excitations in~\eqref{eq:fundamental-excitations}. The first step in the understanding of the symmetries of the spin-chain is therefore to study the properties of these excitations.\footnote{% 
  For a similar discussion of composite excitations of the ABJM spin-chain see~\cite{Klose:2010ki}.%
} %

In~\cite{Sax:2012jv}, a number of closed subsectors of the alternating $\algD{\alpha}$ spin-chain were found. In particular there are two closely related sectors $\algSU(2|1)_+$ and $\algSU(2|1)_-$, consisting of the ground state $\ket{0}_L$, and the excitations $\phi_-^{(0)}$, $\psi_+^{(0)}$ and $\tilde{\phi}_-^{(0)}$, $\tilde{\psi}_+^{(0)}$, respectively. This corresponds to restricting to states containing excitations of the same mass. In this paper we will also consider more general states, containing excitations of two different masses at once.

\subsection{Symmetries of the ground state}
\label{sec:ground-state-symmetries}

By choosing a ground state such as $\ket{0}_L$, we break the symmetry of the spin-chain. The residual symmetry preserving the ground state is generated by the supercharges $\genQ_{+--}$ and $\genQ_{-++}$ and the spin-chain Hamiltonian $\gen{H}$. Together, these generators form an $\algSU(1|1)$ subalgebra of $\algD{\alpha}$. In order to describe the representations of this reduced algebra it is convenient to use a shorter notation. We thus introduce the composite fields
\begin{equation*}
    Z = \phi_+^{(0)} \tilde{\phi}_+^{(0)} , \qquad
    \phi^1 = \phi_-^{(0)} \tilde{\phi}_+^{(0)} , \qquad
    \psi^1 = \psi_+^{(0)} \tilde{\phi}_+^{(0)} , \qquad
    \phi^3 = \phi_+^{(0)} \tilde{\phi}_-^{(0)} , \qquad
    \psi^3 = \phi_+^{(0)} \tilde{\psi}_+^{(0)} ,
\end{equation*}
where we used the indices $1$ and $3$ to distinguish the odd and even sites; these fields are all in the left-moving sector. Using the expressions from appendix~\ref{sec:d21a-algebra}, we find that the $\algSU(1|1)$ generators act on the spin-chain excitations as
\begin{equation}
  \begin{aligned}
    \gen{Q}_{+--} \ket{\phi^1}_L &= 0 , &
    \gen{Q}_{+--} \ket{\psi^1}_L &= \fixedspaceL{\sqrt{1-\alpha} \ket{\phi^3},}{\sqrt{\alpha} \ket{\phi^1}_L,} \\
    \gen{Q}_{-++} \ket{\phi^1}_L &= \fixedspaceL{\sqrt{1-\alpha} \ket{\phi^3},}{\sqrt{\alpha} \ket{\psi^1}_L,} &
    \gen{Q}_{-++} \ket{\psi^1}_L &= 0 , \\
  \end{aligned}
\end{equation}
and
\begin{equation}
  \begin{aligned}
    \gen{Q}_{+--} \ket{\phi^3}_L &= 0 , &
    \gen{Q}_{+--} \ket{\psi^3}_L &= \sqrt{1-\alpha} \ket{\phi^3}_L , \\
    \gen{Q}_{-++} \ket{\phi^3}_L &= \sqrt{1-\alpha} \ket{\psi^3}_L , &
    \gen{Q}_{-++} \ket{\psi^3}_L &= 0 .
  \end{aligned}
\end{equation}
Hence, the excitations on both odd and even sites transform in similar $(1|1)$ multiplets of $\algSU(1|1)$, differing only in the value of the central charge $\gen{H}$.

It will also be useful to consider the additional generator
\begin{equation}
  \gen{B}_{\smallL} = - \tfrac{1}{2} \gen{L}_5 - \tfrac{1}{2} \gen{R}_8,
\end{equation}
which extends the algebra to $\algU(1|1)$. This charge acts on the spin-chain states by
\begin{equation}
  \gen{B}_{\smallL} \ket{0}_L = -\tfrac{L}{2} \ket{0}_{\smallL} , \qquad
  \gen{B}_{\smallL} \ket{\phi}_L = -\left(\tfrac{L}{2}-\tfrac{1}{2}\right) \ket{\phi}_{\smallL} , \qquad
  \gen{B}_{\smallL} \ket{\phi}_L = -\tfrac{L}{2} \ket{\psi}_{\smallL} .
\end{equation}
Hence, $\gen{B}_{\smallL}$ does not preserve the ground state.

\subsection{The full \texorpdfstring{$\algD{\alpha}^2$}{d(2,1;a) x d(2,1;a)} spin-chain}

So far we have only considered the weak coupling limit of the spin-chain, where the left- and right-movers decouple. A general state can then be described as the direct product of two alternating spin-chains of the type described above. In order to distinguish the two sectors we will denote the right-moving fields by a bar. We hence have a composite ground state $\bar{Z}$, and excitations $\bar{\phi}^{\bar{1}}$, $\bar{\psi}^{\bar{1}}$, $\bar{\phi}^{\bar{3}}$ and $\bar{\psi}^{\bar{3}}$. These excitations transform under a second $\algSU(1|1)$ algebra, obtained from the other copy of $\algD{\alpha}$.

In the full spin-chain these sectors are coupled to each other through local interactions. Hence, the spin-chain states are no longer of direct product form. Instead we need to pair up sites of the two $\algD{\alpha}$ spin-chains with each other. The ground state of the full spin-chain is then given by
\begin{equation}
  \ket{0}_L = 
  \Bigg|
    \underbrace{\begin{pmatrix}Z\\\bar{Z}\end{pmatrix} \dotsm \begin{pmatrix}Z\\\bar{Z}\end{pmatrix}}_{L}
  \Bigg\rangle .
\end{equation}
For the excitations we use the notation
\begin{equation}\label{eq:spin-chain-excitations}
    \phi^i = \begin{pmatrix} \phi^i \\ \bar{Z} \end{pmatrix} , \qquad 
    \psi^i = \begin{pmatrix} \psi^i \\ \bar{Z} \end{pmatrix} , \qquad 
    \bar{\phi}^{\bar{\imath}} = \begin{pmatrix} Z \\ \bar{\phi}^{\bar{\imath}} \end{pmatrix} , \qquad 
    \bar{\psi}^{\bar{\imath}} = \begin{pmatrix} Z \\ \bar{\psi}^{\bar{\imath}} \end{pmatrix} , 
\end{equation}
with $i=1,3$ and $\bar{\imath} = \bar{1}, \bar{3}$. With a small abuse of notation we indicate in the same way the excitations of the $\algD{\alpha}$ chain and the ones of the full $\algD{\alpha}^2$ chain; hopefully this will not generate confusion, since from now on we will be focusing only on the latter. In order to construct an S-matrix, we only need to consider asymptotic states, where all the excitations are well separated: there is therefore no need to introduce notation for the case where one or more excitation would sit on the same composite site.

As above, the ground state breaks each of the $\algD{\alpha}$ factors, so that the excitations in~\eqref{eq:spin-chain-excitations} transform non-trivially under an $\algSU(1|1)^2$ algebra.

\section{The centrally extended \texorpdfstring{$\algSU(1|1)^2$}{su(1|1) x su(1|1)} algebra}
\label{sec:centrally-extended-algebra}

In the previous section we saw that the excitations of the $\algD{\alpha}^2$ spin-chain transform under an $\algSU(1|1)^2$ algebra preserving the ground state. One of the charges of this algebra is the spin-chain Hamiltonian $\genH$, which is a function of the coupling contant $h$. In this section we will see how the $\algSU(1|1)^2$ algebra is deformed to accomodate for this coupling dependence. Our final goal is to understand the S-matrix of the $\AdS_3$ spin-chain. However, in this section we find it useful to keep the notation generic. We will therefore consider representations whose parameters we leave unspecified. In section~\ref{sec:d21a-S-matrix}, we will connect back to the notation of section~\ref{sec:d21a-spin-chain}.

\subsection{The \texorpdfstring{$\algU(1|1)$}{su(1|1)} algebra}
\label{sec:u11-algebra}

The algebra $\algU(1|1)$ consists of two supercharges $\genQ$ and $\genS$, a central charge $\genH$, and an outer automorphism $\genB$. The non-trivial commutation relations read
\begin{equation}
    \acomm{\genQ}{\genS} = \genH , \qquad
    \comm{\genB}{\genQ} = - \tfrac{1}{2} \genQ , \qquad
    \comm{\genB}{\genS} = + \tfrac{1}{2} \genS .
\end{equation}
This algebra has a two dimensional representation with the charges acting on a bosonic state $\ket{\phi}$ and a fermionic state $\ket{\psi}$, with the action given by
\begin{equation}
  \begin{aligned}
    \genQ \ket{\fixedspaceL{\psi}{\phi}} &= v \ket{\psi} , &
    \genS \ket{\fixedspaceL{\psi}{\phi}} &= 0 , &
    \genH \ket{\fixedspaceL{\psi}{\phi}} &= H \ket{\phi} , &
    \genB \ket{\fixedspaceL{\psi}{\phi}} &= (B - \tfrac{1}{2}) \ket{\phi} , \\
    \genQ \ket{\psi} &= 0 , &
    \genS \ket{\psi} &= H/v \ket{\phi} , &
    \genH \ket{\psi} &= H \ket{\psi} , &
    \genB \ket{\psi} &= (B - 1) \ket{\psi} .
  \end{aligned}
\end{equation}
This representation is labelled by the two numbers $H$ and $B$. The coefficient $v$ is not physically relevant, but parametrizes the difference in normalization of the states $\ket{\phi}$ and $\ket{\psi}$. Denoting this representation by $(\rep{1}|\rep{1})_{H,B}$, we note that the tensor product between two such states decomposes as
\begin{equation}
  (\rep{1}|\rep{1})_{H,B} \otimes (\rep{1}|\rep{1})_{H',B'} = (\rep{1}|\rep{1})_{H+H',B+B'-1/2} \oplus (\rep{1}|\rep{1})^*_{H+H',B+B'-1} ,
\end{equation}
where the asterisk on the second representation on the right hand side indicates that the statistics of the states have been exchanged, so that the highest weight state is fermionic instead of bosonic.

\subsection{The \texorpdfstring{$\algSU(1|1)^2$}{su(1|1) x su(1|1)} algebra}
\label{sec:su112-alg}

We now consider a direct product of two copies of the $\algU(1|1)$ algebra. Hence we have two copies of each charge. The two copies (anti-)commute with each other, so that the supercharges satisfy
\begin{equation}\label{eq:su112-algebra}
  \begin{aligned}
    \acomm{\genQ_{\smallL}}{\genS_{\smallL}} &= \genH_{\smallL} , \qquad &
    \acomm{\genQ_{\smallL}}{\genQ_{\smallR}} &= 0 , \qquad &
    \acomm{\genQ_{\smallL}}{\genS_{\smallR}} &= 0 , \\
    \acomm{\genQ_{\smallR}}{\genS_{\smallR}} &= \genH_{\smallR} , &
    \acomm{\genS_{\smallL}}{\genS_{\smallR}} &= 0 , &
    \acomm{\genQ_{\smallR}}{\genS_{\smallL}} &= 0 .
  \end{aligned}
\end{equation}
Inspired by the notation in $\AdS_3$ we will refer to the two copies of the algebra as left- and right-moving. We also introduce the combinations
\begin{equation}
  \genH = \genH_{\smallL} + \genH_{\smallR} , \qquad
  \genM = \genH_{\smallL} - \genH_{\smallR} .
\end{equation}
In terms of these generators, the first line above reads
\begin{equation}
  \acomm{\genQ_{\smallL}}{\genS_{\smallL}} = \tfrac{1}{2} \left(\genH + \genM\right) , \qquad
  \acomm{\genQ_{\smallR}}{\genS_{\smallR}} = \tfrac{1}{2} \left(\genH - \genM\right) .
\end{equation}
In the spin-chain $\genH$ plays the role of the energy. This charge will depend on the momentum of the excitations, as well as on the coupling constant. The charge $\genM$ will be treated as an additional central charge labeling the representation. Its eigenvalue appears as a mass in the dispersion relation, and we will assume that it is independent of the momentum.\footnote{%
  When the $\algSU(1|1)^2$ algebra is embedded into $\algPSU(2|2)$, $\genM$ is a particular combination of the Cartan generators of the two bosonic $\algSU(2)$ subalgebra, and is hence quantized. In the $\AdS_3$ case, $\genM$ can in principle receive quantum corrections.%
} %
Below we will freely switch between the notations $\genH_{\smallL}$, $\genH_{\smallR}$ and $\genH$, $\genM$ depending on what is convenient in a particular context.

In addition to the above generators we have two automorphisms $\genB_{\smallL}$ and $\genB_{\smallR}$
\begin{equation}
  \comm{\genB_{\smallL}}{\genQ_{\smallL}} = -\frac{1}{2} \genQ_{\smallL} , \;\;
  \comm{\genB_{\smallL}}{\genS_{\smallL}} = +\frac{1}{2} \genS_{\smallL} , \;\;
  \comm{\genB_{\smallR}}{\genQ_{\smallR}} = -\frac{1}{2} \genQ_{\smallR} , \;\;
  \comm{\genB_{\smallR}}{\genS_{\smallR}} = +\frac{1}{2} \genS_{\smallR} .
\end{equation}
The inclusion of these generators gives the algebra $\algU(1|1)^2$. However, while the charges $\genB_{\smallL}$ and $\genB_{\smallR}$ are conserved, they do not preserve the ground state of the spin-chain. The symmetry acting on the excitations is therefore $\algSU(1|1)^2$.

In the next section we will introduce two additional central extensions to the algebra, which appear in some of the anti-commutators in~\eqref{eq:su112-algebra}. Hence the centrally extended algebra will no longer be of direct product form.

\paragraph{Representations.}
Before turning on any central extensions let us briefly consider representations of the undeformed $\algSU(1|1)^2$ algebra. Since the algebra is a direct product, any irreducible representation can be written as a tensor product of a left-moving and a right-moving part. For later convenience we will consider the two generators $\genQ_{\smallL}$ and $\genS_{\smallR}$ to be lowering operators, while $\genS_{\smallL}$ and $\genQ_{\smallR}$ are raising operators. A highest weight state hence satisfies
\begin{equation}
  \genS_{\smallL} \ket{\text{h.w.}} = 0 , \qquad
  \genQ_{\smallR} \ket{\text{h.w.}} = 0 .
\end{equation}

As usual in a superalgebra we make a distinction between \emph{short} and \emph{long} representation. In the $\algSU(1|1)^2$ algebra we have the two shortening conditions $H_{\smallL} = 0$ and $H_{\smallR} = 0$. A highest weight state with vanishing $H_{\smallR}$ is annihilated by the supercharge $\genS_{\smallR}$. Such a representation will be referred to as a \emph{left-moving} representation. The simplest case is given by the product $(\rep{1}|\rep{1})_{H,B}\otimes\rep{1}$. The two states in this representation are $\ket{\phi}$ and $\ket{\psi}$, with the generators acting as
\begin{equation}
  \begin{aligned}
    \genQ_{\smallL} \ket{\fixedspaceL{\psi}{\phi}} &= v \ket{\psi} , &
    \genS_{\smallL} \ket{\fixedspaceL{\psi}{\phi}} &= 0 , &
    \genH_{\smallL} \ket{\fixedspaceL{\psi}{\phi}} &= H \ket{\phi} , &
    \genB_{\smallL} \ket{\fixedspaceL{\psi}{\phi}} &= (B-\tfrac{1}{2}) \ket{\phi} , \\
    \genQ_{\smallL} \ket{\psi} &= 0 , &
    \genS_{\smallL} \ket{\psi} &= H/v \ket{\phi} , &
    \genH_{\smallL} \ket{\psi} &= H \ket{\psi} , &
    \genB_{\smallL} \ket{\psi} &= (B-1) \ket{\psi} , \\
    \genQ_{\smallR} \ket{\fixedspaceL{\psi}{\phi}} &= 0 , &
    \genS_{\smallR} \ket{\fixedspaceL{\psi}{\phi}} &= 0 , &
    \genH_{\smallR} \ket{\fixedspaceL{\psi}{\phi}} &= 0 , &
    \genB_{\smallR} \ket{\fixedspaceL{\psi}{\phi}} &= 0 , \\
    \genQ_{\smallR} \ket{\psi} &= 0 , &
    \genS_{\smallR} \ket{\psi} &= 0 , &
    \genH_{\smallR} \ket{\psi} &= 0 , &
    \genB_{\smallR} \ket{\psi} &= 0 .
  \end{aligned}
\end{equation}
We also have \emph{right-moving} representations with $H_{\smallL}=0$, whose highest weight states are annihilated by $\genQ_{\smallL}$. Here we will consider the representation $\rep{1}\otimes(\rep{1}|\rep{1})_{H,B}$, with the states $\ket{\bar{\phi}}$ and $\ket{\bar{\psi}}$ satisfying\footnote{Notice that we chose the same representation in the left and right sectors consistently with the embedding in~$\alg{d}(2,1;\alpha)^2$.}
\begin{equation}
  \begin{aligned}
    \genQ_{\smallL} \ket{\fixedspaceL{\psi}{\bar{\phi}}} &= 0 , &
    \genS_{\smallL} \ket{\fixedspaceL{\psi}{\bar{\phi}}} &= 0 , &
    \genH_{\smallL} \ket{\fixedspaceL{\psi}{\bar{\phi}}} &= 0 , &
    \genB_{\smallL} \ket{\fixedspaceL{\psi}{\bar{\phi}}} &= 0 , \\
    \genQ_{\smallL} \ket{\bar{\psi}} &= 0 , &
    \genS_{\smallL} \ket{\bar{\psi}} &= 0 , &
    \genH_{\smallL} \ket{\bar{\psi}} &= 0 , &
    \genB_{\smallL} \ket{\bar{\psi}} &= 0 , \\
    \genQ_{\smallR} \ket{\fixedspaceL{\psi}{\bar{\phi}}} &= v \ket{\bar{\psi}} , &
    \genS_{\smallR} \ket{\fixedspaceL{\psi}{\bar{\phi}}} &= 0 , &
    \genH_{\smallR} \ket{\fixedspaceL{\psi}{\bar{\phi}}} &= H \ket{\bar{\phi}} , &
    \genB_{\smallR} \ket{\fixedspaceL{\psi}{\bar{\phi}}} &= (B-\tfrac{1}{2}) \ket{\bar{\phi}} , \\
    \genQ_{\smallR} \ket{\bar{\psi}} &= 0 , &
    \genS_{\smallR} \ket{\bar{\psi}} &= H/v \ket{\bar{\phi}} , &
    \genH_{\smallR} \ket{\bar{\psi}} &= H \ket{\bar{\psi}} , &
    \genB_{\smallR} \ket{\bar{\psi}} &= (B-1) \ket{\bar{\psi}} .
  \end{aligned}
\end{equation}
In the case of the $\algSU(1|1)^2$ algebra, we can obtain additional short representations by considering any tensor product of simple $\algU(1|1)$ states in the left- or right-moving sector. However, when we add additional central charges the chirality constraint will be deformed and the number of short multiplets will be
reduced.

\subsection{Spin-chain states}

Our goal is to describe a spin-chain with excitations transforming under the centrally extended $\algSU(1|1)^2$ algebra. Before discussing the full algebra and its representations we introduce some notation for describing the states of this spin-chain. The spin-chain ground state is a state where each site of the chain is occupied by an $\algSU(1|1)^2$ singlet which we call $Z$. For a spin-chain of length $L$ we hence have
\begin{equation}
  \ket{0}_L = \ket{Z^L},
\end{equation}
where the exponent indicates that the state contains $L$ consecutive sites occupied by the state $Z$. We will usually suppress the subscript $L$ denoting the length of the chain.

While the ground state preserves the $\algSU(1|1)^2$ algebra, it is charged under the automorphisms $\genB_{\smallL}$ and $\genB_{\smallR}$, which act by
\begin{equation}
  \genB_{\smallL} \ket{0}_L = - \frac{L}{2} \ket{0}_L , \qquad
  \genB_{\smallR} \ket{0}_L = - \frac{L}{2} \ket{0}_L .
\end{equation}
Since the combination $\genB_{\smallL} - \genB_{\smallR}$ does annihilate the ground state, it is useful to introduce two new generators
\begin{equation}
  \tilde{\genB} = \genB_{\smallL} + \genB_{\smallR} , \qquad
  \genB = \genB_{\smallL} - \genB_{\smallR} .
\end{equation}
We can now introduce states with a single excitation at site $n$
\begin{equation}
  \ket{\mathcal{X}_{(n)}}_L = \ket{Z^{n-1} \mathcal{X} Z^{L-n}}_L ,
\end{equation}
where $\mathcal{X}$ is any excitations $\phi$, $\psi$, $\bar{\phi}$ or $\bar{\psi}$. From such states we can construct states of a specific momentum by
\begin{equation}\label{eq:momentum-state}
  \ket{\mathcal{X}_p}_L = \sum_n e^{ipn} \ket{\mathcal{X}_{(n)}}_L.
\end{equation}
It is now straightforward to construct multi-magnon states of the form $\ket{\phi_{p_1} \bar{\psi}_{p_2} \cdots \phi_{p_n}}_L$. We will always consider an asymptotic limit where $L\to\infty$, and where all excitations can be considered to be ordered and well separated. In general we will let each excitation carry different charges under $\genB_{\smallL}$, $\genB_{\smallR}$ and $\genM$. As we will see below, the energy $\genH$ is given as a function of the momenta.

\subsection{Length-changing}

The generators of the centrally extended algebra have the possibility of mixing states of different length. To capture this we introduce some additional notation. The insertion of the symbol $Z^+$ ($Z^-$) in a state indicates the insertion (removal) of a vacuum site at the specified location. For a single excitation, equation~\eqref{eq:momentum-state} shows that adding or removing a vacuum site to the right of the excitation, such as in the state $\ket{\phi_p Z^\pm}$, does nothing except change the total length of the chain. However, if we insert or remove an excitation to the left of the excitation we can move it through the excitation
\begin{equation}
  \ket{Z^+\phi_p} = e^{-ip} \ket{\phi_p Z^+} , \qquad
  \ket{Z^-\phi_p} = e^{+ip} \ket{\phi_p Z^-} .
\end{equation}
In states with multiple excitations we will use the above result to move all occurrences of $Z^\pm$ to the far right, thereby trading them for additional momentum dependent phase factors.

\subsection{The centrally extended \texorpdfstring{$\algSU(1|1)^2$}{su(1|1) x su(1|1)} algebra}
\label{sec:central-extension-I}

The $\algSU(1|1)^2$ can be extended by two additional central charges $\genP$ and $\genK$, which appear in the anti-commutators between supercharges from different sectors. In order to figure out the possible length-changing actions of the supercharges, it is useful to consider the charges under the automorphisms $\genB_{\smallL}$ and $\genB_{\smallR}$ of the single excitation states and the supercharges. These are collected in table~\ref{tab:charges-B1-B2}. 
\begin{table}
  \centering
  \begin{tabular}{cll}
    \toprule
    & $\phantom{+}\mathllap{2}B_{\smallL}$ & $\phantom{+}\mathllap{2}B_{\smallR}$ \\
    \midrule
    $\ket{\fixedspaceL{\psi}{0}}$ & $-L$ & $-L$ \\
    $\ket{\fixedspaceL{\psi}{\phi}}$ & $-L+2B$ & $-L$ \\
    $\ket{\fixedspaceL{\psi}{\psi}}$ & $-L+2B-1$ & $-L$ \\
    $\ket{\fixedspaceL{\psi}{\bar{\phi}}}$ & $-L$ & $-L+2B$ \\
    $\ket{\fixedspaceL{\psi}{\bar{\psi}}}$ & $-L$ & $-L+2B-1$ \\
    \midrule
    $\genQ_{\smallL}$ & $-1$ & $\phantom{+}0$ \\
    $\genS_{\smallL}$ & $+1$ & $\phantom{+}0$ \\
    $\genQ_{\smallR}$ & $\phantom{+}0$ & $-1$ \\
    $\genS_{\smallR}$ & $\phantom{+}0$ & $+1$ \\
    \bottomrule
  \end{tabular}
  \caption{%
    Charges under the automorphisms $\genB_{\smallL}$ and $\genB_{\smallR}$ of the ground
    state, the single excitation states in the left- and right-moving multiplets
    as well as of the supercharges. All the spin-chain states have length
    $L$. To make the table less cluttered, the charges have been rescaled.%
  }
  \label{tab:charges-B1-B2}
\end{table}

Conservation of the charges $\genB_{\smallL}$ and $\genB_{\smallR}$ as well as the fermion number restricts the possible extensions. As an example, we note that a non-trivial state obtained by acting with $\genQ_{\smallR}$ on $\ket{\phi}$ should be fermionic and have the charges
\begin{equation}
  B_{\smallL} = -\frac{L-1}{2} + B - \frac{1}{2}, \qquad
  B_{\smallR} = -\frac{L+1}{2}.
\end{equation}
To construct a state with these charges we would need to change the length in different ways in the left- and right-moving sectors. While such a construction might be possible, it leads to a picture where the charges do not act locally on a single spin-chain. Moreover, we would need to allow for additional 1/4-BPS states with different numbers of left- and right-movers. However, this does not seem to be compatible with the string-theory spectrum, where there is a unique BMN-type ground state~\cite{Babichenko:2009dk,Gomis:2002qi}. For a further discussion of the resulting construction see appendix~\ref{sec:central-extension-II}. 

If we instead try to make the action of $\genS_{\smallR}$ on $\ket{\phi}$ non-trivial we need to find a state with charges
\begin{equation}
  B_{\smallL} = -\frac{L-1}{2} + B - \frac{1}{2}, \qquad
  B_{\smallR} = -\frac{L-1}{2} .
\end{equation}
Such a state is given by $\ket{\psi Z^-}$, \ie, a single excitation $\psi$ in a spin-chain whose length is one less than that of the original state. If we allow the length to change we should therefore allow this action. The $\algSU(1|1)^2$ algebra then requires the action of $\genQ_{\smallR}$ on $\ket{\psi_p}$ to be non-trivial, giving the state $\ket{\phi_p Z^+}$. We also introduce a bosonic generator $\genP = \acomm{\genQ_{\smallL}}{\genQ_{\smallR}}$. Comparing the charges we see that $\genP$ should take $\ket{\phi}$ to $\ket{\phi Z^+}$. In the same way we introduce $\genK = \acomm{\genS_{\smallL}}{\genS_{\smallR}}$. Hence we are lead to consider the algebra
\begin{equation}
  \begin{aligned}
    \acomm{\genQ_{\smallL}}{\genS_{\smallL}} &= \genH_{\smallL} , \qquad &
    \acomm{\genQ_{\smallL}}{\genQ_{\smallR}} &= \genP , \qquad &
    \acomm{\genQ_{\smallL}}{\genS_{\smallR}} &= 0 , \\
    \acomm{\genQ_{\smallR}}{\genS_{\smallR}} &= \genH_{\smallR} , &
    \acomm{\genS_{\smallL}}{\genS_{\smallR}} &= \genK , &
    \acomm{\genS_{\smallL}}{\genQ_{\smallR}} &= 0 .
  \end{aligned}
\end{equation}
The non-trivial commutation relations involving the automorphisms $\genB_{\smallL}$ and $\genB_{\smallR}$ are given by
\begin{equation}
  \begin{aligned}
    \comm{\genB_{\smallL}}{\genQ_{\smallL}} &= -\tfrac{1}{2} \genQ_{\smallL} , \; &
    \comm{\genB_{\smallL}}{\genS_{\smallL}} &= +\tfrac{1}{2} \genS_{\smallL} , \; &
    \comm{\genB_{\smallL}}{\genP} &= -\tfrac{1}{2} \genP , \; &
    \comm{\genB_{\smallL}}{\genK} &= +\tfrac{1}{2} \genK , \\
    \comm{\genB_{\smallR}}{\genQ_{\smallR}} &= -\tfrac{1}{2} \genQ_{\smallR} , &
    \comm{\genB_{\smallR}}{\genS_{\smallR}} &= +\tfrac{1}{2} \genS_{\smallR} , &
    \comm{\genB_{\smallR}}{\genP} &= -\tfrac{1}{2} \genP , &
    \comm{\genB_{\smallR}}{\genK} &= +\tfrac{1}{2} \genK .
  \end{aligned}
\end{equation}

The centrally extended algebra also has a larger set of automorphisms generated by a triplet $\tilde{\genB}$, $\tilde{\genB}_{\pm}$ satisfying
\begin{equation}
  \comm{\tilde{\genB}}{\tilde{\genB}_{\pm}} = \pm \tilde{\genB}_{\pm} , \qquad
  \comm{\tilde{\genB}_+}{\tilde{\genB}_-} = -2 \tilde{\genB} ,
\end{equation}
where $\tilde{\genB} = \genB_{\smallL} + \genB_{\smallR}$ is the same charge we defined in the previous section. Under this $\algSL(2)$ algebra the supercharges form two doublets $(\genS_{\smallL},\genQ_{\smallR})$ and $(\genS_{\smallR},\genQ_{\smallL})$,
\begin{equation}
  \begin{aligned}
    \comm{\tilde{\genB}}{\genQ_{\smallL}} &= -\tfrac{1}{2} \genQ_{\smallL} , \; &
    \comm{\tilde{\genB}}{\genS_{\smallR}} &= +\tfrac{1}{2} \genS_{\smallR} , \; &
    \comm{\tilde{\genB}_+}{\genQ_{\smallL}} &= +\genS_{\smallR} , \; &
    \comm{\tilde{\genB}_-}{\genS_{\smallR}} &= -\genQ_{\smallL} , \\
    \comm{\tilde{\genB}}{\genQ_{\smallR}} &= -\tfrac{1}{2} \genQ_{\smallR} , &
    \comm{\tilde{\genB}}{\genS_{\smallL}} &= +\tfrac{1}{2} \genS_{\smallL} , &
    \comm{\tilde{\genB}_+}{\genQ_{\smallR}} &= +\genS_{\smallL} , &
    \comm{\tilde{\genB}_-}{\genS_{\smallL}} &= -\genQ_{\smallR} , 
  \end{aligned}
\end{equation}
while the central charges split into a singlet ($\genM$) and a triplet ($\genP$, $\genH$, $\genK$). The non-trivial commutation relations take the form
\begin{equation}
  \begin{aligned}
    \comm{\tilde{\genB}}{\genK} &= +\genK , \quad &
    \comm{\tilde{\genB}_-}{\genK} &= -\genH , \quad &
    \comm{\tilde{\genB}_-}{\genH} &= -2\genP , \\
    \comm{\tilde{\genB}}{\fixedspaceL{\genK}{\genP}} &= -\genP , &
    \comm{\tilde{\genB}_+}{\fixedspaceL{\genK}{\genP}} &= +\genH , &
    \comm{\tilde{\genB}_+}{\genH} &= +2\genK .
  \end{aligned}
\end{equation}
Since the charges $\tilde{\genB}$ and $\tilde{\genB}_{\pm}$ act non-trivially on the ground state we will mainly consider the automorphism generated by $\genB = \genB_{\smallL} - \genB_{\smallR}$.

\section{Representations of the extended \texorpdfstring{$\algSU(1|1)^2$}{su(1|1) x su(1|1)} algebra}
\label{sec:centrally-extended-representations}

\subsection{The left-moving representation}
\label{sec:representations}

In this section we will consider the generalization of the left-moving representation considered from section~\ref{sec:su112-alg}. It now takes the form 
\begin{equation}\label{eq:chiral-rep}
  \begin{aligned}
    \genQ_{\smallL} \ket{\phi_p} &= a_p \ket{\psi_p} , \qquad &
    \genQ_{\smallL} \ket{\psi_p} &= 0 , \\
    \genS_{\smallL} \ket{\phi_p} &= 0 , \qquad &
    \genS_{\smallL} \ket{\psi_p} &= b_p \ket{\phi_p} , \\
    \genQ_{\smallR} \ket{\phi_p} &= 0 , \qquad &
    \genQ_{\smallR} \ket{\psi_p} &= c_p \ket{\phi_p\,Z^+} , \\
    \genS_{\smallR} \ket{\phi_p} &= d_p \ket{\psi_p\,Z^-} , \qquad &
    \genS_{\smallR} \ket{\psi_p} &= 0 ,
  \end{aligned}
\end{equation}
where we have parametrized the representation by $a_p$, $b_p$, $c_p$ and $d_p$. The subscript indicates that the values of these coefficients in general depends on the momentum of the excitation. Note that the symbols $Z^\pm$ indicating that a single field $Z$ to be inserted or removed always appear directly after the excitation. We can then calculate the value of the central charges
\begin{equation}
  \begin{aligned}
    \genH_{\smallL} \ket{\phi_p} &= a_p b_p \ket{\phi_p} , \quad &
    \fixedspaceL{\genK}{\genP} \ket{\phi_p} &= a_p c_p \ket{\phi_p Z^+} , \\
    \genH_{\smallR} \ket{\phi_p} &= c_p d_p \ket{\phi_p} , \quad &
    \genK \ket{\phi_p} &= b_p d_p \ket{\phi_p Z^-} . 
  \end{aligned}
\end{equation}
The automorphisms $\genB_{\smallL}$ and $\genB_{\smallR}$ still act by
\begin{equation}
  \begin{aligned}
    \genB_{\smallL} \ket{\fixedspaceL{\psi_p}{\phi_p}} &= \left(-\tfrac{L}{2}+B\right) \ket{\phi_p} , &
    \genB_{\smallR} \ket{\fixedspaceL{\psi_p}{\phi_p}} &= -\tfrac{L}{2} \ket{\fixedspaceL{\psi_p}{\phi_p}} , \\
    \genB_{\smallL} \ket{\psi_p} &= \left(-\tfrac{L}{2}+B-\tfrac{1}{2}\right) \ket{\psi_p} , \qquad &
    \genB_{\smallR} \ket{\psi_p} &= -\tfrac{L}{2} \ket{\psi_p} .
  \end{aligned}
\end{equation}
For the generator $\genB$ this gives
\begin{equation}
  \genB \ket{\phi_p} = B \ket{\phi_p} , \qquad
  \genB \ket{\psi_p} = \left(B -\tfrac{1}{2}\right) \ket{\psi_p} .
\end{equation}

\paragraph{Shortening condition.}
The state $\ket{\phi_p}$ is a highest weight state and therefore satisfies the conditions
\begin{equation}
  \genS_{\smallL} \ket{\phi_p} = \genQ_{\smallR} \ket{\phi_p} = 0.
\end{equation}
Without the central extension, $\ket{\phi_p}$ would satisfy the chirality constraint $\genS_{\smallR} \ket{\phi_p} = 0$. Here this condition takes the form
\begin{equation}
  (\genH_{\smallR} \genQ_{\smallL} - \genP \genS_{\smallR}) \ket{\phi_p} = (a_p c_p d_p - a_p c_p d_p) \ket{\psi_p} = 0.
\end{equation}
Since this particular combination of the lowering operators $\genQ_{\smallL}$ and $\genS_{\smallR}$ vanishes this representation is short.

When we turn off the central extensions $\genP$ and $\genK$, the generator $\genH_{\smallR}$ vanishes. This seem to make the operator $\genH_{\smallR} \genQ_{\smallL} - \genP \genS_{\smallR}$ identically zero. However, a careful treatment of the limit shows that $\genH_{\smallR}$ goes to zero faster than $\genP$. Hence we can recover the chirality constraint of the undeformed representation.

The state $\ket{\phi_p}$ must also be annihilated by the anti-commutator 
\begin{equation}
  \acomm{\genQ_{\smallL}}{\genH_{\smallR} \genQ_{\smallL} - \genP \genS_{\smallR}} \ket{\phi_p} = (\genH_{\smallL} \genH_{\smallR} - \genP\genK) \ket{\phi_p} .
\end{equation}
This shortening condition will play an important role below. It is easy to directly check that
\begin{equation}
  \label{eq:shortening-condition}
  (\genH_{\smallL} \genH_{\smallR} - \genP\genK) \ket{\phi_p} = (a_p b_p c_p d_p - a_p b_p c_p d_p) \ket{\phi_p} = 0.
\end{equation}
Note that the above charge is central and hence also annihilates the state $\ket{\psi_p}$.

\paragraph{Physical states.}
The actual symmetry preserved by the spin-chain ground state is the undeformed $\algSU(1|1)^2$, in which $\genP$ and $\genK$ vanish. Hence these charges must annihilate all physical states. For a single excitation this leads to the conditions
\begin{equation}
  a_p c_p = 0 , \qquad
  b_p d_p = 0.
\end{equation}
Solving this by setting $c_p = d_p = 0$ would give back the left-moving representation from section~\ref{sec:su112-alg}. This means that to get a non-trivial physical state we need to consider several excitations. Let us therefore start by considering a state with two excitations with momenta $p$ and $q$. We then need to know the action on the states
\begin{equation*}
  \ket{\phi_p \phi_q} , \qquad
  \ket{\phi_p \psi_q} , \qquad
  \ket{\psi_p \phi_q} , \qquad
  \ket{\psi_p \psi_q} ,
\end{equation*}
where the subscript labels the momentum of the excitations. We will also allow the two excitations to carry different charges under $\genB$ and $\genM$.

It will now be important that the various coefficients in the representation are momentum dependent, so that 
\begin{equation}
  \genQ_{\smallL} \ket{\phi_p \phi_q} = a_p \ket{\psi_p \phi_q} + a_q \ket{\phi_p \psi_q} ,
\end{equation}
with $a_p$ and $a_q$ functions of $p$ and $q$, respectively. When a vacuum site is inserted or removed we always do so to the immediate right of the excitation we act on. Hence we have
\begin{equation}
  \genQ_{\smallR} \ket{\psi_p \psi_q} = c_p \ket{\phi_p Z^+ \psi_q} - c_q \ket{\psi_p \phi_q Z^+} ,
\end{equation}
where we in the second term picked up a minus sign when letting the supercharge $\genQ_{\smallR}$ pass through the fermionic excitation $\psi_p$. As we noted in the last section, we can now move the $Z^+$ insertion in the first term to the right. In doing this we pick up an additional phase factor so that the above can be written as
\begin{equation}
  \genQ_{\smallR} \ket{\psi_p \psi_q} = e^{-iq} c_p \ket{\phi_p \psi_q Z^+}  - c_q \ket{\psi_p \phi_q Z^+} .
\end{equation}
In the same way we can calculate the action of $\genP = \acomm{\genQ_{\smallL}}{\genQ_{\smallR}}$ on the state $\ket{\phi_p \phi_q}$,
\begin{align*}
  \acomm{\genQ_{\smallL}}{\genQ_{\smallR}} \ket{\phi_p \phi_q} 
  &= \genQ_{\smallR} \left(
    a_p \ket{\psi_p \phi_q} + a_q \ket{\phi_p \psi_q}
  \right) \\
  &= a_p c_p \ket{\phi_p Z^+ \phi_q} + a_q c_q \ket{\phi_p \phi_q Z^+} \\
  &= \left( e^{-iq} a_p c_p + a_q c_q \right) \ket{\phi_p \phi_q Z^+} .
\end{align*}
The generator $\genP$ should annihilate a physical state. If we let
\begin{equation}\label{eq:physical-state-ac}
  a_p c_p = \frac{h}{2} e^{+i\gamma} ( e^{-ip} - 1 ) ,
\end{equation}
where $h$ and $\gamma$ are two momentum independent real constants, we get
\begin{equation}
  e^{-iq} a_p c_p + a_q c_q = \frac{h}{2} e^{+i\gamma} ( e^{-i(p+q)} -1 ).
\end{equation}
Hence the above choice makes the action of $\genP$ vanish on the above two excitation state provided $e^{i(p+q)} = 1$.

Performing the same analysis for the central charge $\genK$, we find the same condition for a physical state, provided that\footnote{%
  Here we have also imposed unitarity which means that $(\genP)^\dag = \genK$.%
} %
\begin{equation}\label{eq:physical-state-bd}
    b_p d_p = \frac{h}{2} e^{-i\gamma} ( e^{+ip} - 1 ) .
\end{equation}
It is straightforward to generalize these calculations to a state with $n$ excitations carrying momentum $p_1,\dotsc,p_n$. We then find that a physical state satisfies
\begin{equation}
  \exp \Big(i\sum_{k=1}^n p_k\Big) = 1.
\end{equation}

Finally we want to calculate the action of the central charges $\genH_{\smallL}$ and $\genH_{\smallR}$. In doing so we will require that the combination $\genM= \genH_{\smallL} - \genH_{\smallR}$ takes the same value as in the non-deformed algebra,\footnote{%
  In particular, the eigenvalue of $\genM$ does not depend on the momentum $p$ of the excitation. In principle it could depend on the coupling constant $h$.%
} %
while $\genH = \genH_{\smallL} + \genH_{\smallR}$ is allowed to receive corrections. We will denote the eigenvalue of $\genM$ by $s$. In terms of the coefficients of the representation it is given by
\begin{equation}\label{eq:no-spin-correction}
  a_p b_p - c_p d_p = s .
\end{equation}
As we already mentioned, we will assume that this charge does not depend on the value of the momentum $p$. We also note that we can use~\eqref{eq:no-spin-correction}, to rewrite the shortening condition~\eqref{eq:shortening-condition} as
\begin{equation}
  (\genH_{\smallL} + \genH_{\smallR})^2 - 4\genP\genK = (\genH_{\smallL} - \genH_{\smallR})^2 = s^2.
\end{equation}
Using the condition~\eqref{eq:no-spin-correction} together with the form of the representation coefficients found in~\eqref{eq:physical-state-ac} and~\eqref{eq:physical-state-bd} we can now write the dispersion relations, \ie, the eigenvalue of $\genH$ for a single excitation, as
\begin{equation}
  E(p) = a_p b_p + c_p d_p = \sqrt{s^2 + 4h^2\sin^2\frac{p}{2}} .
\end{equation}

The above representation can be conveniently parametrized using the spectral parameters $x^\pm_p$ satisfying
\begin{equation}
  \frac{x_p^+}{x_p^-} = e^{ip} , \qquad
  \left(x_p^+ + \frac{1}{x_p^+}\right) - \left(x_p^- + \frac{1}{x_p^-}\right) = \frac{2is}{h}.
\end{equation}
We can then write the coefficients of the representation constructed above as
\begin{align}
  a &= \sqrt{\frac{h}{2}} \, \eta_p \, e^{+i\frac{\gamma+\delta}{2}} , \qquad &
  c &= -\sqrt{\frac{h}{2}} \, \frac{i \eta_p}{x_p^+} \, e^{+i\frac{\gamma-\delta}{2}} , \\
  b &= \sqrt{\frac{h}{2}} \, \eta_p \, e^{-i\frac{\gamma+\delta}{2}} , \qquad &
  d &= +\sqrt{\frac{h}{2}} \, \frac{i \eta_p}{x_p^-} \, e^{-i\frac{\gamma-\delta}{2}},
\end{align}
where we have introduced an additional arbitrary phase $\delta$ and
\begin{equation}\label{eq:eta-p-def}
  \eta_p = \sqrt{i(x_p^- - x_p^+)}.
\end{equation}

\subsection{The right-moving representation}

For the right-movers $\bar{\phi}_p$ and $\bar{\psi}_p$, the role of $\genQ_{\smallL}$, $\genQ_{\smallR}$ and $\genS_{\smallL}$, $\genS_{\smallR}$ is exchanged. Similar to~\eqref{eq:chiral-rep} we can then write the representation
\begin{equation}
  \begin{aligned}
    \genQ_{\smallR} \ket{\bar{\phi}_p} &= \bar{a}_p \ket{\bar{\psi}_p}, \qquad &
    \genQ_{\smallR} \ket{\bar{\psi}_p} &= 0 , \\
    \genS_{\smallR} \ket{\bar{\phi}_p} &= 0 , \qquad &
    \genS_{\smallR} \ket{\bar{\psi}_p} &= \bar{b}_p \ket{\bar{\phi}_p} , \\
    \genQ_{\smallL} \ket{\bar{\phi}_p} &= 0 , \qquad &
    \genQ_{\smallL} \ket{\bar{\psi}_p} &= \bar{c}_p \ket{\bar{\phi}_p Z^+} , \\
    \genS_{\smallL} \ket{\bar{\phi}_p} &= \bar{d}_p \ket{\bar{\psi}_p Z^-} , \qquad &
    \genS_{\smallL} \ket{\bar{\phi}_p} &= 0 ,
  \end{aligned}
\end{equation}
with the automorphisms acting as
\begin{equation}
  \begin{aligned}
    \genB_{\smallL} \ket{\fixedspaceL{\psi_p}{\bar{\phi}_p}} &= -\tfrac{L}{2} \ket{\bar{\phi}_p} , &
    \genB_{\smallR} \ket{\fixedspaceL{\psi_p}{\bar{\phi}_p}} &= \left(-\tfrac{L}{2} + B\right) \ket{\bar{\phi}_p} , \\
    \genB_{\smallL} \ket{\bar{\psi}_p} &= -\tfrac{L}{2} \ket{\bar{\psi}_p} , &
    \genB_{\smallR} \ket{\bar{\psi}_p} &= \left(-\tfrac{L}{2} - \tfrac{1}{2} + B\right) \ket{\bar{\psi}_p} .
  \end{aligned}
\end{equation}
The action of the central charges is 
\begin{equation}
  \begin{aligned}
    \genH_{\smallL} \ket{\bar{\phi}_p} &= \bar{c}_p \bar{d}_p \ket{\bar{\phi}_p} , \quad &
    \genP \ket{\bar{\phi}_p} &= \bar{a}_p \bar{c}_p \ket{\bar{\phi}_p Z^+} , \\
    \genH_{\smallR} \ket{\bar{\phi}_p} &= \bar{a}_p \bar{b}_p \ket{\bar{\phi}_p} , \quad &
    \genK \ket{\bar{\phi}_p} &= \bar{b}_p \bar{d}_p \ket{\bar{\phi}_p Z^-} .
  \end{aligned}
\end{equation}
Note in particular that
\begin{equation}
  \genM \ket{\bar{\phi}_p} = (\genH_{\smallL} - \genH_{\smallR}) \ket{\bar{\phi}_p} = -(\bar{a}_p \bar{b}_p - \bar{c}_p \bar{d}_p) \ket{\bar{\phi}_p} .
\end{equation}
To match the notation for the left-movers, we will denote this eigenvalue by $-s$.

As in the left-moving case, a physical state should be annihilated by $\genP$ and $\genK$. This leads us again to the condition that the total momentum should vanish (modulo $2\pi$). We then find that the coefficients of the representation take the same solution as previously, namely\footnote{%
  In principle some phases may be chosen differently in the two
  representations. For convenience we choose them to be identical in the two
  cases.%
}%
\begin{align}
  \bar{a} &= \sqrt{\frac{h}{2}} \, \eta_p \, e^{+i\frac{\gamma+\delta}{2}} , \qquad &
  \bar{c} &= -\sqrt{\frac{h}{2}} \, \frac{i \eta_p}{x_p^+} \, e^{+i\frac{\gamma-\delta}{2}} , \\
  \bar{b} &= \sqrt{\frac{h}{2}} \, \eta_p \, e^{-i\frac{\gamma+\delta}{2}} , \qquad &
  \bar{d} &= +\sqrt{\frac{h}{2}} \, \frac{i \eta_p}{x_p^-} \, e^{-i\frac{\gamma-\delta}{2}}.
\end{align}
Hence, we will from now on drop the bars on these coefficients.

\paragraph{Shortening condition.}
The right-moving representation again satisfies the condition\footnote{%
  Note that we consider the generators $\genQ_{\smallL}$ and $\genS_{\smallR}$ to be the
  positive simple roots. Hence the state $\ket{\bar{\psi}_p}$, and not
  $\ket{\bar{\phi}_p}$, is of highest weight.%
}%
\begin{equation}
  (\genH_{\smallR} \genQ_{\smallL} - \genP \genS_{\smallR}) \ket{\bar{\psi}_p} = 0.
\end{equation}
This is the same condition as for the left-moving representation. However, the small momentum limit is different. In the current case when we set the $\genP$ to zero, we obtain the anti-chirality condition $\genQ_{\smallL} \ket{\bar{\psi}_p} = 0$ from section~\ref{sec:su112-alg}.

\subsection{Two-particle states}
\label{sec:2-part-states}
Let us finally comment on the action of the central generators on states with one excitation in each sector. For the state $\ket{\phi_p \bar{\phi}_q}$, we find
\begin{equation}
  \genP \ket{\phi_p \bar{\phi}_q} = \frac{h}{2} e^{-i(p+q)} ( 1 - e^{i(p+q)} ) \ket{\phi_p \bar{\phi}_q Z^+} .
\end{equation}
Hence a physical state, for which the central extension vanishes, has zero total momentum,
\begin{equation}
  e^{i(P_L + P_R)} = 1,
\end{equation}
where $P_L$ and $P_R$ are the momentum in the left- and right-moving sector, respectively. Note that even if this is an apparent contradiction with the momentum constraint used in~\cite{Babichenko:2009dk}, the difference is only due to different conventions for the Cartan generators in the two sectors~\cite{Borsato:2012ss}.

\paragraph{Bound states.}

In general, two excitations together form a four-dimensional long multiplet of $\algSU(1|1)^2$. However, if the momenta $p$ and $q$ are related by $x_p^+ = x_q^-$, then the two-particle state $\ket{\phi_p \phi_q}$ satisfies
\begin{equation}
  (\genH_{\smallR}\genQ_{\smallL} - \genP\genS_{\smallR}) \ket{\phi_p \phi_q} = 0 , \qquad x_p^+ = x_q^-.
\end{equation}
Hence the tensor product of the two representations becomes short at this point. This corresponds to the appearance of a bound state with energy
\begin{equation}
  E_{s_1}(p) + E_{s_2}(q) = \sqrt{(s_1+s_2)^2 + 4h^2\sin^2\frac{p+q}{2}},
\end{equation}
where $s_1$ and $s_2$ are the masses of the two excitations. Note that the representation theory allows for bound states between two left-moving or two right-moving excitations with different masses. Whether such state are actually present in the $\algD{\alpha}^2$ spin-chain hence depends on the pole structure of the S-matrix. Note that these bound state do not appear in the Bethe equations of~\cite{Babichenko:2009dk,OhlssonSax:2011ms}.

\paragraph{Singlet states.}

It is interesting to note that there are two-particle states that transform as singlets under the extended $\algSU(1|1)^2$ algebra. Such a state is constructed from one left- and one-right mover with the same mass. In particular we consider the states
\begin{equation}
    \ket{\mathsf{1}^{LR}_{p\bar{p}}} = \ket{\phi_p \bar{\phi}_{\bar{p}} Z^+} + \Xi_{p\bar{p}} \ket{\psi_p \bar{\psi}_{\bar{p}}} , \qquad
    \ket{\mathsf{1}^{RL}_{p\bar{p}}} = \ket{\bar{\phi}_p \phi_{\bar{p}} Z^+} + \Xi_{p\bar{p}} \ket{\bar{\psi}_p \psi_{\bar{p}}} ,
\end{equation}
where
\begin{equation}
  \Xi_{p\bar{p}} = i x_{\bar{p}}^+ \frac{\eta_p}{\eta_{\bar{p}}} .
\end{equation}
It is straightforward to check that both $\ket{\mathsf{1}^{LR}_{p\bar{p}}}$ and $\ket{\mathsf{1}^{RL}_{p\bar{p}}}$ are annihilated by all the generators of the centrally extended $\algSU(1|1)^2$ algebra, provided we identify the Zhukovsky variables of the two excitations by
\begin{equation}
  x_{\bar{p}}^\pm = \frac{1}{x_p^\pm}.
\end{equation}
In order to emphasize that the two momenta are related, we have denoted them by $p$ and $\bar{p}$. It is clear that this identifications means that the singlet states are non-physical, since one of the constituent excitations has a negative energy. Hence, we can view the singlet states as vacuum fluctuations. These states will be useful when we consider crossing symmetry in section~\ref{sec:crossing-equations}.

The above states are closely related to the singlet $\algPSU(2|2)$ state observed in~\cite{Beisert:2005tm}. In that case, the additional $\algSU(2) \times \algSU(2)$ symmetry means that we have to combine that above state into a single singlet state $\ket{\mathsf{1}^{LR}_{pq}} - \ket{\mathsf{1}^{RL}_{pq}}$. For further comments on the relation between the symmetry considered here and the centrally extended $\algPSU(2|2)$, see section~\ref{sec:psu22-s-matrix} and appendix~\ref{sec:psu22-algebra}.

\section{The S-matrix}
\label{sec:S-matrix}

In the previous two sections we constructed the algebra and corresponding representations in which the excitations of the spin-chain transform. In this section we will consider scattering between two such excitations. 

In discussing scattering processes it will be important whether the two excitations transform in a left- or right-moving representation, and whether they have the same or different eigenvalues under the charge $\genM$. We will therefore consider excitations transforming in four different representations, which we will label by L, $\text{L}'$, R and $\text{R}'$. The representations L and $\text{L} '$ are left-moving, while R and $\text{R}'$ are right-moving. The mass of the excitations, \ie, the eigenvalues under the charge $\genM$, is $\pm s$ for the L and R representations, and $\pm s'$ for $\text{L}'$ and $\text{R}'$. In particular, in the $\algD{\alpha}^2$ spin-chain we have $s=\alpha$ and $s'=1-\alpha$. Later we will also denote the four representations L, $\text{L}'$, R and $\text{R}'$ by $1$, $3$, $\bar{1}$ and $\bar{3}$, respectively.

\paragraph{Energy and momentum conservation.}

Under a scattering process, the total energy and momentum of a two-particle state are preserved. Such a general process can be written as
\begin{equation}
  \ket{\mathcal{X}^{\text{(in)}}_{p_{\text{in}}}\, \mathcal{Y}^{\text{(in)}}_{q_{\text{in}}}} \to \ket{\mathcal{X}^{\text{(out)}}_{q_{\text{out}}}\,\mathcal{Y}^{\text{(out)}}_{p_{\text{out}}}} ,
\end{equation}
where the fields $\mathcal{X},\mathcal{Y}$ indicate generic excitations. We will always use a notation where the excitations in the in- and out-states are ordered such that $p_{\text{in}} > q_{\text{in}}$ and $p_{\text{out}} > q_{\text{out}}$. Energy and momentum conservation then take the form
\begin{equation}
  \label{eq:p-E-conservation}
  p_{\text{in}} + q_{\text{in}} = p_{\text{out}} + q_{\text{out}} , \qquad
  E_{s_{\text{in}}^{}}(p_{\text{in}}) + E_{s_{\text{in}}'}(q_{\text{in}}) = E_{s_{\text{out}}^{}}(q_{\text{out}}) + E_{s_{\text{out}}'}(p_{\text{out}}) ,
\end{equation}
where we have explicitly indicated the mass of the excitation in the dispersion relation
\begin{equation}
  E_s(p) = \sqrt{s^2 + 4h^2\sin^2\frac{p}{2}} .
\end{equation}
In all the scattering processes the individual masses are conserved. Hence we have
\begin{equation}\label{eq:mass-conservation}
   s_{\text{out}}^{} =  s_{\text{in}}', \ s_{\text{out}}' = s_{\text{in}}^{},
  \qquad \text{or} \qquad
  s_{\text{out}}^{} = s_{\text{in}}^{} ,\ s_{\text{out}}' = s_{\text{in}}'.
\end{equation}
If the two incoming particles have the same mass, so that $s_{\text{in}}=s_{\text{in}}'=s_{\text{out}}=s_{\text{out}}'$, the momenta of the two particles are preserved, and the solution of~\eqref{eq:p-E-conservation} is given by $p_{\text{out}} = p_{\text{in}}$ and $q_{\text{out}} = q_{\text{in}}$.\footnote{%
  Clearly $p_{\text{out}} = q_{\text{in}}$ and $q_{\text{out}} = p_{\text{in}}$ also solves~\eqref{eq:p-E-conservation}. However, this solution has the excitations in the out-state in the wrong order, since $p_{\text{out}} < q_{\text{out}}$.%
} %
Note that there still can be a non-trivial permutation in flavor space.

If, on the other hand, the masses of the two incoming particles are different, $s_{\text{in}} \neq s_{\text{in}}'$, there are two solutions to~\eqref{eq:p-E-conservation}, depending on the relation of masses in~\eqref{eq:mass-conservation}. For $s^{}_{\text{in}}=s_{\text{out}}'$ and $s_{\text{in}}'=s_{\text{out}}$ the solution is again given by $p_{\text{out}} = p_{\text{in}}\equiv p$ and $q_{\text{out}} = q_{\text{in}}\equiv q$. We then have a process 
\begin{equation}
    \ket{\mathcal{X}_p^{\text{(in)}} \, \mathcal{Y}_q^{\text{(in)}}} \to \ket{\mathcal{Y}_q^{\text{(out)}} \, \mathcal{X}_p^{\text{(out)}}} ,
\end{equation}
where the excitations $\mathcal{X}^{\text{(in)}}$ and $\mathcal{X}^{\text{(out)}}$ have mass $s\equiv s_{\text{in}}^{}$ while $\mathcal{Y}^{\text{(in)}}$ and $\mathcal{Y}^{\text{(out)}}$ have mass $s'\equiv s'_{\text{in}}$. With respect to the mass of the excitations, this process acts as a \emph{transmission}.

The second case in~\eqref{eq:mass-conservation}, with  $s^{}_{\text{in}}=s_{\text{out}}$ and $s_{\text{in}}'=s_{\text{out}}'$ gives a solution to~\eqref{eq:p-E-conservation} in which the momenta of the outgoing excitations are not a simple permutation of $p_{\text{in}}$ and $q_{\text{in}}$. The general solution in this case is quite complicated, and we will not write it down explicitly. The corresponding process takes the form of a \emph{reflection}
\begin{equation}
  \ket{\mathcal{X}_{p_{\text{in}}}^{\text{(in)}} \, \mathcal{Y}_{q_{\text{in}}}^{\text{(in)}}} \to \ket{\mathcal{X}_{q_{\text{out}}}^{\text{(out)}} \, \mathcal{Y}_{p_{\text{out}}}^{\text{(out)}} },
\end{equation}
where again we have used the same letter to indicate excitation of the same mass (but possibly different flavors). Notice that now $p_{\text{out}}(p_{\text{in}},q_{\text{in}})\neq p_{\text{in}}$, and similarly for $q$, as we are picking the non-trivial solution of~\eqref{eq:p-E-conservation}.

The general S-matrix for the case of excitations of different masses includes processes of both a transmission and a reflection term. However, as we will see below, the S-matrix discussed in this paper will turn out to be reflectionless.

\paragraph{Invariance under the symmetry algebra.}

The two-particle S-matrix $\Smat$ should be compatible with the symmetries of the theory, and should hence satisfy
\begin{equation}\label{eq:S-comm-alg}
  \comm{\gen{J}_1 + \gen{J}_2}{\Smat_{12}} = 0 ,
\end{equation}
where $\gen{J}$ is any generator of the centrally extended $\algSU(1|1)^2$ algebra, and the subscript on $\gen{J}$ indicates that the generator acts on the first or second component of a two-particle state.

Equation \eqref{eq:S-comm-alg} does not change if we multiply the S-matrix by a scalar factor, that is thus undetermined at this point. In the case of the $\algPSU(2|2)$ spin-chain there is one scalar factor that is constrained by unitarity, physical unitarity and crossing symmetry (see later).  It is important to note that in our case, at least in principle, we get a different scalar factor for each one of the independent sectors. We will explain later how it is possible to relate them.

\paragraph{Unitarity and physical unitarity.}

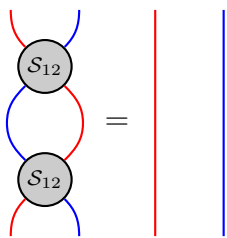
\begin{figure}
  \centering
\begin{tikzpicture}
  \begin{scope}[xshift=-0.95cm]
    \coordinate (i1) at (-0.45cm,0);
    \coordinate (i2) at (+0.45cm,0);

    \node (v1) at (0,0.75cm) [S-mat] {$\scriptstyle \Smat_{12}$};

    \coordinate (m1) at (-0.5cm,1.5cm);
    \coordinate (m2) at (+0.5cm,1.5cm);

    \node (v2) at (0,2.25cm) [S-mat] {$\scriptstyle \Smat_{12}$};

    \coordinate (o1) at (-0.45cm,3cm);
    \coordinate (o2) at (+0.45cm,3cm);

    \draw [thick,red]  [out=90,in=270-45] (i1) to (v1);
    \draw [thick,blue] [out=90,in=270+45] (i2) to (v1);

    \draw [thick,blue] [out=90+45,in=270] (v1) to (m1);
    \draw [thick,red]  [out=90-45,in=270] (v1) to (m2);

    \draw [thick,blue] [out=90,in=270-45] (m1) to (v2);
    \draw [thick,red]  [out=90,in=270+45] (m2) to (v2);

    \draw [thick,red]  [out=90+45,in=270] (v2) to (o1);
    \draw [thick,blue] [out=90-45,in=270] (v2) to (o2);
  \end{scope}
  \node at (0,1.5cm) {$=$};
  \begin{scope}[xshift=+0.95cm]
    \coordinate (i1) at (-0.45cm,0);
    \coordinate (i2) at (+0.45cm,0);

    \coordinate (o1) at (-0.45cm,3cm);
    \coordinate (o2) at (+0.45cm,3cm);

    \draw [thick,red]  (i1) to (o1);
    \draw [thick,blue] (i2) to (o2);
  \end{scope}
\end{tikzpicture}

  \caption{Unitarity. Acting twice with the S-matrix $\Smat$ on a two-particle state gives back the original state.}
  \label{fig:unitarity}
\end{figure}

We will impose two additional constraints on the S-matrix, related to its analytical structure. The first constraint is that of \emph{unitarity}, which means that acting twice with the S-matrix on a two-particle state gives back the original state, as illustrated in figure~\ref{fig:unitarity}.%
In terms of the operator $\Smat$, this constraint is expressed as
\begin{equation}
  \Smat_{12} \, \Smat_{12}= 1.
\end{equation}
The above relation takes a more familiar form when written in a matrix basis, which is discussed in more detail in appendix~\ref{sec:string-frame}. Here we just note that unitarity of the S-matrix $\mat{S}_{pq}$, where $p$ and $q$ are the momenta of the excitations in a two-particle state, is expressed as
\begin{equation}
  \mat{S}_{qp} \, \mat{S}_{pq} = \matId .
\end{equation}

The second constraint we impose is a reality constraint on the S-matrix, which we will refer to as \emph{physical unitarity}. While this constraint can also be expressed in terms of the operator $\Smat$ by introducing its conjugate $\Smat^{\dag}$, it is more useful to again go to a matrix basis, where it takes the form
\begin{equation}
    \mat{S}^{\dag}_{pq} \, \mat{S}_{pq} = \matId .
\end{equation}
Hence, the condition of physical unitarity tells us that $\mat{S}$ is unitary as a matrix.

\paragraph{Discrete LR-symmetry.}

The two copies of $\algSU(1|1)$ that compose the symmetry algebra are on equal footing and it is arbitrary what we call left- and what we call right-moving. This allows us to require a discrete $\Integers_2$ symmetry at the level of the S-matrix: scattering processes that differ only by the exchange of the flavors left and right should give the same result. We will refer to this as LR-symmetry. This discrete symmetry will be very useful in constraining the solution for the S-matrix.

This symmetry is a special feature of the $\algD{\alpha}^2$ spin-chain, due to the direct product form of the symmetry group. In the $\algSU(2|3)$ spin-chain of~\cite{Beisert:2003ys,Beisert:2005tm} the discrete LR-symmetry is replaced by a continuous $\algSU(2) \times \algSU(2)$ symmetry. As we will see in section~\ref{sec:psu22-s-matrix}, this difference in symmetry explains the difference between the S-matrices of the two models.

\paragraph{Yang-Baxter equation.}

\begin{figure}
  \centering
\begin{tikzpicture}
  \begin{scope}[xshift=-1.4cm]
    \coordinate (i1) at (-0.9cm,0);
    \coordinate (i2) at (-0,    0);
    \coordinate (i3) at (+0.9cm,0);

    \coordinate (o1) at (-0.9cm,3cm);
    \coordinate (o2) at (-0,    3cm);
    \coordinate (o3) at (+0.9cm,3cm);

    \node (v1) at (-0.45cm,0.75cm) [S-mat] {$\scriptstyle \Smat_{12}$};
    \node (v2) at ( 0.3cm, 1.5cm)    [S-mat] {$\scriptstyle \Smat_{23}$};
    \node (v3) at (-0.45cm,2.25cm) [S-mat] {$\scriptstyle \Smat_{12}$};

    \draw [thick,red]   [out=90,in=270-45] (i1) to (v1);
    \draw [thick,blue]  [out=90,in=270+45] (i2) to (v1);
    \draw [thick,violet] [out=90,in=270+45] (i3) to (v2);

    \draw [thick,red]   [out=90-45,in=270-45] (v1) to (v2);
    \draw [thick,blue]  [out=90+45,in=270-45] (v1) to (v3);
    \draw [thick,violet] [out=90+45,in=270+45] (v2) to (v3);

    \draw [thick,violet] [out=90+45,in=270] (v3) to (o1);
    \draw [thick,blue]  [out=90-45,in=270] (v3) to (o2);
    \draw [thick,red]   [out=90-45,in=270] (v2) to (o3);
  \end{scope}
  \node at (0,1.5cm) {$=$};
  \begin{scope}[xshift=+1.4cm]
    \coordinate (i1) at (-0.9cm,0);
    \coordinate (i2) at (-0,    0);
    \coordinate (i3) at (+0.9cm,0);

    \coordinate (o1) at (-0.9cm,3cm);
    \coordinate (o2) at (-0,    3cm);
    \coordinate (o3) at (+0.9cm,3cm);

    \node (v1) at (+0.45cm,0.75cm) [S-mat] {$\scriptstyle \Smat_{23}$};
    \node (v2) at (-0.3cm, 1.5cm)  [S-mat] {$\scriptstyle \Smat_{12}$};
    \node (v3) at (+0.45cm,2.25cm) [S-mat] {$\scriptstyle \Smat_{23}$};

    \draw [thick,red]    [out=90,in=270-45] (i1) to (v2);
    \draw [thick,blue]   [out=90,in=270-45] (i2) to (v1);
    \draw [thick,violet] [out=90,in=270+45] (i3) to (v1);

    \draw [thick,violet] [out=90+45,in=270+45] (v1) to (v2);
    \draw [thick,blue]   [out=90-45,in=270+45] (v1) to (v3);
    \draw [thick,red]    [out=90-45,in=270-45] (v2) to (v3);

    \draw [thick,violet] [out=90+45,in=270] (v2) to (o1);
    \draw [thick,blue]  [out=90+45,in=270] (v3) to (o2);
    \draw [thick,red]   [out=90-45,in=270] (v3) to (o3);
  \end{scope}
\end{tikzpicture}%

  \caption{The Yang-Baxter equation relates the two different ways of scattering three particles to each other.}
  \label{fig:Yang-Baxter}
\end{figure}
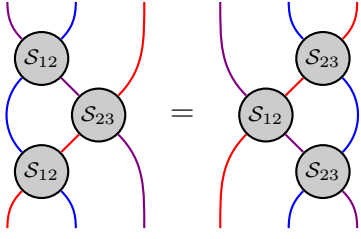

The final interesting property of our S-matrix is that is satisfies the Yang-Baxter (YB) equation
\begin{equation}
\Smat_{12} \, \Smat_{23} \, \Smat_{12} = \Smat_{23} \, \Smat_{12} \, \Smat_{23} .
\end{equation}
This equation is illustrated in figure~\ref{fig:Yang-Baxter}.%
We have written the YB equation as an operator equation, relating the two different orderings in which three two-particle S-matrices can act on a three-particle state. Since the spin-chain S-matrix involves length-changing processes, some care is needed when acting on a three particle state. The addition or removal of vacuum sites gives rise to extra phase factors in the YB equation. In section~\ref{sec:Yang-Baxter} we will discuss in more detail the exact form of the above equation in a matrix basis.

\subsection{Excitations with the same mass}
\label{sec:S-matrix-same-mass}

We start by discussing the case in which the excitations that scatter have the same mass. As explained before we will consider three different sectors separately, namely the left-left (LL), the right-right (RR) and the left-right (LR) sectors.

\paragraph{The LL sector.}

As an ansatz for the S-matrix between two particles transforming in the left-moving multiplet we use
\begin{equation}
  \begin{aligned}
    \Smat \ket{\fixedspaceL{\psi_p \psi_q}{\phi_p \phi_q}} 
    &= \fixedspaceR{D^{\smallLL}_{pq}}{A^{\smallLL}_{pq}} \ket{\fixedspaceL{\psi_p \psi_q}{\phi_q \phi_p}} , \qquad &
    \Smat \ket{\fixedspaceL{\psi_p \psi_q}{\phi_p \psi_q}} 
    &= \fixedspaceR{D^{\smallLL}_{pq}}{B^{\smallLL}_{pq}} \ket{\fixedspaceL{\psi_p \psi_q}{\psi_q \phi_p}} + C^{\smallLL}_{pq} \ket{\fixedspaceL{\psi_p \psi_q}{\phi_q \psi_p}}, \\
    \Smat \ket{\fixedspaceL{\psi_p \psi_q}{\psi_p \psi_q}} &= \fixedspaceR{D^{\smallLL}_{pq}}{F^{\smallLL}_{pq}} \ket{\fixedspaceL{\psi_p \psi_q}{\psi_q \psi_p}} , \qquad &
    \Smat \ket{\fixedspaceL{\psi_p \psi_q}{\psi_p \phi_q}} 
    &= \fixedspaceR{D^{\smallLL}_{pq}}{D^{\smallLL}_{pq}} \ket{\fixedspaceL{\psi_p \psi_q}{\phi_q \psi_p}} + E^{\smallLL}_{pq} \ket{\fixedspaceL{\psi_p \psi_q}{\psi_q \phi_p}}, \\
  \end{aligned}
\end{equation}
This ansatz is constructed so that the bosonic charges $\genH_{\smallL}$, $\genH_{\smallR}$, $\genB_{\smallL}$, $\genB_{\smallR}$, $\genP$ and $\genK$ are preserved. By requiring that $\Smat_{12}$ also commutes with the supercharges $\genQ_{\smallL}$, $\genS_{\smallL}$, $\genQ_{\smallR}$ and $\genS_{\smallR}$ we find that the coefficients take the form
\begin{equation}
  \begin{aligned}
    A^{\smallLL}_{pq} &= S^{\smallLL}_{pq} \frac{x_q^+ - x_p^-}{x_q^- - x_p^+} , \qquad &
    B^{\smallLL}_{pq} &= S^{\smallLL}_{pq} \frac{x_q^+ - x_p^+}{x_q^- - x_p^+} , \qquad &
    C^{\smallLL}_{pq} &= S^{\smallLL}_{pq} \frac{x_q^+ - x_q^-}{x_q^- - x_p^+} \frac{\eta_p}{\eta_q} , \\
    F^{\smallLL}_{pq} &= - S^{\smallLL}_{pq} , \qquad &
    D^{\smallLL}_{pq} &= S^{\smallLL}_{pq} \frac{x_q^- - x_p^-}{x_q^- - x_p^+} , \qquad &
    E^{\smallLL}_{pq} &= S^{\smallLL}_{pq} \frac{x_p^+ - x_p^-}{x_q^- - x_p^+} \frac{\eta_q}{\eta_p} ,
  \end{aligned}
\end{equation}
where $S^{\smallLL}_{pq}$ is an arbitrary scalar factor. Unitarity and physical unitarity are satisfied if the scalar factor satisfies, respectively,
\begin{equation}
  S^{\smallLL}_{pq} S^{\smallLL}_{qp} = 1, \qquad
  (S^{\smallLL}_{pq})^* S^{\smallLL}_{qp}=1.
\end{equation}
This amounts to $S^{\smallLL}_{12}$ being a phase that is anti-symmetric in the exchange
of the two momenta.

In~\cite{Beisert:2005wm} the two-particle S-matrix in a spin-chain where the excitations transform under a single copy of $\algSU(1|1)$ was written down. In doing this, the Yang-Baxter equation was imposed. The S-matrix we have found here perfectly matches the result of that paper. However, here we have only required that the S-matrix commutes with the full, centrally extended, symmetry algebra.

\paragraph{The RR sector.}

The calculation for the S-matrix in the RR sector goes exactly in the same way as in LL. The ansatz that preserves the bosonic charges takes the same form as before, except that now the excitations $\phi$ and $\psi$ are replaced by $\bar{\phi}, \bar{\psi}$. Hence, we have
\begin{equation}
  \begin{aligned}
    \Smat \ket{\fixedspaceL{\bar{\psi}_p \bar{\psi}_q}{\bar{\phi}_p \bar{\phi}_q}} 
    &= \fixedspaceR{D^{\smallRR}_{pq}}{A^{\smallRR}_{pq}} \ket{\fixedspaceL{\bar{\psi}_p \bar{\psi}_q}{\bar{\phi}_q \bar{\phi}_p}} , \qquad &
    \Smat \ket{\fixedspaceL{\bar{\psi}_p \bar{\psi}_q}{\bar{\phi}_p \bar{\psi}_q}} 
    &= \fixedspaceR{D^{\smallRR}_{pq}}{B^{\smallRR}_{pq}} \ket{\fixedspaceL{\bar{\psi}_p \bar{\psi}_q}{\bar{\psi}_q \bar{\phi}_p}} 
    + \fixedspaceR{D^{\smallRR}_{pq}}{C^{\smallRR}_{pq}} \ket{\fixedspaceL{\bar{\psi}_p \bar{\psi}_q}{\bar{\phi}_q \bar{\psi}_p}}, \\
    \Smat \ket{\fixedspaceL{\bar{\psi}_p \bar{\psi}_q}{\bar{\psi}_p \bar{\psi}_q}} &= 
    \fixedspaceR{D^{\smallRR}_{pq}}{F^{\smallRR}_{pq}} \ket{\fixedspaceL{\bar{\psi}_p \bar{\psi}_q}{\bar{\psi}_q \bar{\psi}_p}} , \qquad &
    \Smat \ket{\fixedspaceL{\bar{\psi}_p \bar{\psi}_q}{\bar{\psi}_p \bar{\phi}_q}} 
    &= \fixedspaceR{D^{\smallRR}_{pq}}{D^{\smallRR}_{pq}} \ket{\fixedspaceL{\bar{\psi}_p \bar{\psi}_q}{\bar{\phi}_q \bar{\psi}_p}} 
    +  \fixedspaceR{D^{\smallRR}_{pq}}{E^{\smallRR}_{pq}} \ket{\fixedspaceL{\bar{\psi}_p \bar{\psi}_q}{\bar{\psi}_q \bar{\phi}_p}}, \\
  \end{aligned}
\end{equation}

Requiring that $\Smat$ also commutes with the supercharges it is not surprising to find
\begin{equation}
  \begin{aligned}
    A^{\smallRR}_{pq} &= S^{\smallRR}_{pq} \frac{x_q^+ - x_p^-}{x_q^- - x_p^+} , \qquad &
    B^{\smallRR}_{pq} &= S^{\smallRR}_{pq} \frac{x_q^+ - x_p^+}{x_q^- - x_p^+} , \qquad &
    C^{\smallRR}_{pq} &= S^{\smallRR}_{pq} \frac{x_q^+ - x_q^-}{x_q^- - x_p^+} \frac{\eta_p}{\eta_q} , \\
    F^{\smallRR}_{pq} &= - S^{\smallRR}_{pq} , &
    D^{\smallRR}_{pq} &= S^{\smallRR}_{pq} \frac{x_q^- - x_p^-}{x_q^- - x_p^+} , \qquad &
    E^{\smallRR}_{pq} &= S^{\smallRR}_{pq} \frac{x_p^+ - x_p^-}{x_q^- - x_p^+} \frac{\eta_q}{\eta_p} . \\
  \end{aligned}
\end{equation}
This is the same solution as the one for LL up to the fact that now $S^{\smallRR}_{pq}$ is in principle a different scalar factor.
We can identify this factor with the previous one by requiring LR-symmetry. Imposing for example $A^{\smallRR}_{pq}=A^{\smallLL}_{pq}$, it follows that
\begin{equation}
  S^{\smallRR}_{pq}=S^{\smallLL}_{pq}.
\end{equation}
We recall that we only required that the masses of the two excitations are equal, but we never imposed their actual value.

\paragraph{The LR and RL sectors.}

In the LR and RL sectors the most general ansatz for the S-matrix that preserves the bosonic charges contains both a \emph{transmission} part $\Tmat$ and a \emph{reflection} part $\Rmat$,
\begin{equation}
  \Smat = \Tmat + \Rmat,
\end{equation}
where ``transmission'' and ``reflection'' refers to the chirality of the two excitations. Hence, the transmission $\Tmat$ takes an in-state consisting of a left- and a right-mover, such as $\ket{\phi_p \bar{\psi}_q}$, into an out-state with a right- and a left-mover, $\ket{\bar{\psi}_q \phi_p}$. Taking into account the conservation of the bosonic charges, we can write an ansatz for such an operator as
\begin{equation}\label{eq:T-ansatz-LR}
  \begin{aligned}
    \Tmat \ket{\fixedspaceL{\psi_p\bar{\psi}_q}{\phi_p \bar{\phi}_q}} 
    &= \fixedspaceR{A^{\smallLR}_{pq}}{A^{\smallLR}_{pq}} \ket{\fixedspaceL{\bar{\psi}_q\psi_p}{\bar{\phi}_q\phi_p}} + \fixedspaceR{A^{\smallLR}_{pq}}{B^{\smallLR}_{pq}} \ket{\bar{\psi}_q \psi_p Z^-}, \qquad &
    \Tmat \ket{\fixedspaceL{\psi_p\bar{\psi}_q}{\phi_p\bar{\psi}_q}} 
    &= \fixedspaceR{A^{\smallLR}_{pq}}{C^{\smallLR}_{pq}} \ket{\fixedspaceL{\bar{\psi}_q\psi_p}{\bar{\psi}_q\phi_p}}, \\
    \Tmat \ket{\fixedspaceL{\psi_p\bar{\psi}_q}{\psi_p\bar{\psi}_q}} 
    &= \fixedspaceR{A^{\smallLR}_{pq}}{E^{\smallLR}_{pq}} \ket{\fixedspaceL{\bar{\psi}_q\psi_p}{\bar{\psi}_q\psi_p}} + \fixedspaceR{A^{\smallLR}_{pq}}{F^{\smallLR}_{pq}} \ket{\bar{\phi}_q\phi_p Z^+}, \qquad &
    \Tmat \ket{\fixedspaceL{\psi_p\bar{\psi}_q}{\psi_p\bar{\phi}_q}} 
    &= \fixedspaceR{A^{\smallLR}_{pq}}{D^{\smallLR}_{pq}} \ket{\fixedspaceL{\bar{\psi}_q\psi_p}{\bar{\phi}_q\psi_p}}.
  \end{aligned}
\end{equation}
Similarly, the reflection takes a left- and a right-mover, $\ket{\phi_p \bar{\psi}_q}$, back into a left- and a right-mover, $\ket{\phi_q \bar{\psi}_p}$. The corresponding ansatz reads
\begin{equation}\label{eq:R-ansatz-LR}
  \begin{aligned}
    \Rmat \ket{\fixedspaceL{\psi_p\bar{\psi}_q}{\phi_p\bar{\phi}_q}} 
    &= \fixedspaceR{\tilde{A}^{\smallLR}_{pq}}{\tilde{A}^{\smallLR}_{pq}} \ket{\fixedspaceL{\psi_q\bar{\psi}_p}{\phi_q\bar{\phi}_p}}  + \tilde{B}^{\smallLR}_{pq} \ket{\psi_q\bar{\psi}_p Z^-}, \qquad &
    \Rmat \ket{\fixedspaceL{\psi_p\bar{\psi}_q}{\phi_p\bar{\psi}_q}} 
    &= \fixedspaceR{\tilde{A}^{\smallLR}_{pq}}{\tilde{C}^{\smallLR}_{pq}} \ket{\fixedspaceL{\psi_q\bar{\psi}_p}{\phi_q\bar{\psi}_p}}, \\
    \Rmat \ket{\fixedspaceL{\psi_p\bar{\psi}_q}{\psi_p\bar{\psi}_q}} 
    &= \fixedspaceR{\tilde{A}^{\smallLR}_{pq}}{\tilde{E}^{\smallLR}_{pq}} \ket{\fixedspaceL{\psi_q\bar{\psi}_p}{\psi_q\bar{\psi}_p}} + \tilde{F}^{\smallLR}_{pq} \ket{\phi_q\bar{\phi}_p Z^+}, \qquad &
    \Rmat \ket{\fixedspaceL{\psi_p\bar{\psi}_q}{\psi_p\bar{\phi}_q}} 
    &= \fixedspaceR{\tilde{A}^{\smallLR}_{pq}}{\tilde{D}^{\smallLR}_{pq}} \ket{\fixedspaceL{\psi_q\bar{\psi}_p}{\psi_q\bar{\phi}_p}}.
  \end{aligned}
\end{equation}
Commuting $\Tmat$ with the supercharges $\genQ_i$ and $\genS_i$, we find
\begin{equation}\label{eq:T-solution-LR}
  \begin{aligned}
    A^{\smallLR}_{pq} &= +\tau^{\smallLR}_{pq} \frac{1-\frac{1}{x_p^+ x_q^-}}{1-\frac{1}{x_p^- x_q^-}}, \quad &
    B^{\smallLR}_{pq} &= -\tau^{\smallLR}_{pq} \frac{\eta_p \eta_q}{x_p^- x_q^-} \frac{1}{1-\frac{1}{x_p^- x_q^-}}, \quad &
    C^{\smallLR}_{pq} &= \tau^{\smallLR}_{pq} , \\
    E^{\smallLR}_{pq} &= -\tau^{\smallLR}_{pq} \frac{1-\frac{1}{x_p^- x_q^+}}{1-\frac{1}{x_p^- x_q^-}}, \quad &
    F^{\smallLR}_{pq} &= -\tau^{\smallLR}_{pq} \frac{\eta_p \eta_q}{x_p^+ x_q^+} \frac{1}{1-\frac{1}{x_p^- x_q^-}}, \quad &
    D^{\smallLR}_{pq} &= \tau^{\smallLR}_{pq} \frac{1-\frac{1}{x_p^+ x_q^+}}{1-\frac{1}{x_p^- x_q^-}}.
  \end{aligned}
\end{equation}
Similarly we find that the coefficients of $\Rmat$ are given by
\begin{equation}\label{eq:R-solution-LR}
  \begin{aligned}
    \tilde{A}^{\smallLR}_{pq} &= \rho^{\smallLR}_{pq} \frac{\eta_p}{\eta_q} \frac{1-\frac{x_q^+}{x_p^+}\frac{1}{x_p^- x_q^-}}{1-\frac{1}{x_p^- x_q^-}} , \quad &
    \tilde{B}^{\smallLR}_{pq} &= \rho^{\smallLR}_{pq} \frac{i}{x_p^- x_q^-} \frac{x_p^+ - x_q^+ }{1-\frac{1}{x_p^- x_q^-}}, \quad &
    \tilde{C}^{\smallLR}_{pq} &= \rho^{\smallLR}_{pq} , \\
    \tilde{E}^{\smallLR}_{pq} &= \rho^{\smallLR}_{pq} \frac{\eta_p}{\eta_q} \frac{1-\frac{x_q^-}{x_p^-}\frac{1}{x_p^+ x_q^+}}{1-\frac{1}{x_p^+ x_q^+}} , \quad &
    \tilde{F}^{\smallLR}_{pq} &= \rho^{\smallLR}_{pq} \frac{i}{x_p^+ x_q^+} \frac{x_p^- - x_q^- }{1-\frac{1}{x_p^+ x_q^+}} , \quad &
    \tilde{D}^{\smallLR}_{pq} &= \rho^{\smallLR}_{pq} .
  \end{aligned}
\end{equation}
The coefficients $\tau^{\smallLR}_{pq}$ and $\rho^{\smallLR}_{pq}$ are two scalar factors that for now are left undetermined.

If we instead consider an in-state with the excitations in the opposite order, \ie, with the first excitation right-moving and the second left-moving, the ansatz again takes a form similar to~\eqref{eq:T-ansatz-LR} and~\eqref{eq:R-ansatz-LR}, but with the $\phi$ and $\bar{\phi}$, and $\psi$ and $\bar{\psi}$ exchanged. We hence find
\begin{equation}\label{eq:T-ansatz-RL}
  \begin{aligned}
    \Tmat \ket{\fixedspaceL{\bar{\psi}_p\psi_q}{\bar{\phi}_p\phi_q}} 
    &= \fixedspaceR{A^{\smallRL}_{pq}}{A^{\smallRL}_{pq}} \ket{\fixedspaceL{\psi_q\bar{\psi}_p}{\phi_q\bar{\phi}_p}} + \fixedspaceR{A^{\smallRL}_{pq}}{B^{\smallRL}_{pq}} \ket{\psi_q\bar{ \psi}_p Z^-}, \qquad &
    \Tmat \ket{\fixedspaceL{\bar{\psi}_p\psi_q}{\bar{\phi}_p\psi_q}} 
    &= \fixedspaceR{A^{\smallRL}_{pq}}{C^{\smallRL}_{pq}} \ket{\fixedspaceL{\psi_q\bar{\psi}_p}{\psi_q\bar{\phi}_p}}, \\
    \Tmat \ket{\fixedspaceL{\bar{\psi}_p\psi_q}{\bar{\psi}_p\psi_q}} 
    &= \fixedspaceR{A^{\smallRL}_{pq}}{E^{\smallRL}_{pq}} \ket{\fixedspaceL{\psi_q\bar{\psi}_p}{\psi_q\bar{\psi}_p}} + \fixedspaceR{A^{\smallRL}_{pq}}{F^{\smallRL}_{pq}} \ket{\phi_q\bar{\phi}_p Z^+}, \qquad &
    \Tmat \ket{\fixedspaceL{\bar{\psi}_p\psi_q}{\bar{\psi}_p\phi_q}} 
    &= \fixedspaceR{A^{\smallRL}_{pq}}{D^{\smallRL}_{pq}} \ket{\fixedspaceL{\psi_q\bar{\psi}_p}{\phi_q\bar{\psi}_p}}.
  \end{aligned}
\end{equation}
and
\begin{equation}\label{eq:R-ansatz-RL}
  \begin{aligned}
    \Rmat \ket{\fixedspaceL{\bar{\psi}_p\psi_q}{\bar{\phi}_p\phi_q}} 
    &= \fixedspaceR{\tilde{A}^{\smallRL}_{pq}}{\tilde{A}^{\smallRL}_{pq}} \ket{\fixedspaceL{\bar{\psi}_q\psi_p}{\bar{\phi}_q\phi_p}}  + \tilde{B}^{\smallRL}_{pq} \ket{\bar{\psi}_q\psi_p Z^-}, \qquad &
    \Rmat \ket{\fixedspaceL{\bar{\psi}_p\psi_q}{\bar{\phi}_p\psi_q}} 
    &= \fixedspaceR{\tilde{A}^{\smallRL}_{pq}}{\tilde{C}^{\smallRL}_{pq}} \ket{\fixedspaceL{\bar{\psi}_q\psi_p}{\bar{\phi}_q\psi_p}}, \\
    \Rmat \ket{\fixedspaceL{\bar{\psi}_p\psi_q}{\bar{\psi}_p\psi_q}} 
    &= \fixedspaceR{\tilde{A}^{\smallRL}_{pq}}{\tilde{E}^{\smallRL}_{pq}} \ket{\fixedspaceL{\bar{\psi}_q\psi_p}{\bar{\psi}_q\psi_p}} + \tilde{F}^{\smallRL}_{pq} \ket{\bar{\phi}_q\phi_p Z^+}, \qquad &
    \Rmat \ket{\fixedspaceL{\bar{\psi}_p\psi_q}{\bar{\psi}_p\phi_q}} 
    &= \fixedspaceR{\tilde{A}^{\smallRL}_{pq}}{\tilde{D}^{\smallRL}_{pq}} \ket{\fixedspaceL{\bar{\psi}_q\psi_p}{\bar{\psi}_q\phi_p}}.
  \end{aligned}
\end{equation}
In this case the coefficients are given by
\begin{equation}\label{eq:T-solution-RL}
  \begin{aligned}
    A^{\smallRL}_{pq} &= +\tau^{\smallRL}_{pq} \frac{1-\frac{1}{x_p^+ x_q^-}}{1-\frac{1}{x_p^+ x_q^+}} , \quad &
    B^{\smallRL}_{pq} &= -\tau^{\smallRL}_{pq} \frac{\eta_p \eta_q}{x_p^- x_q^-} \frac{1}{1-\frac{1}{x_p^+ x_q^+}} , \quad &
    C^{\smallRL}_{pq} &= \tau^{\smallRL}_{pq} \frac{1-\frac{1}{x_p^- x_q^-}}{1-\frac{1}{x_p^+ x_q^+}}, \\
    E^{\smallRL}_{pq} &= -\tau^{\smallRL}_{pq} \frac{1-\frac{1}{x_p^- x_q^+}}{1-\frac{1}{x_p^+ x_q^+}} , \quad &
    F^{\smallRL}_{pq} &= -\tau^{\smallRL}_{pq} \frac{\eta_p \eta_q}{x_p^+ x_q^+} \frac{1}{1-\frac{1}{x_p^+ x_q^+}} , \quad &
    D^{\smallRL}_{pq} &= \tau^{\smallRL}_{pq}  ,
  \end{aligned}
\end{equation}
and
\begin{equation}\label{eq:R-solution-RL}
  \begin{aligned}
    \tilde{A}^{\smallRL}_{pq} &= \rho^{\smallRL}_{pq} \frac{\eta_p}{\eta_q} \frac{1-\frac{x_q^+}{x_p^+}\frac{1}{x_p^- x_q^-}}{1-\frac{1}{x_p^- x_q^-}} , \quad &
    \tilde{B}^{\smallRL}_{pq} &= \rho^{\smallRL}_{pq} \frac{i}{x_p^- x_q^-} \frac{x_p^+ - x_q^+ }{1-\frac{1}{x_p^- x_q^-}} , \quad &
    \tilde{C}^{\smallRL}_{pq} &= \rho^{\smallRL}_{pq} , \\
    \tilde{E}^{\smallRL}_{pq} &= \rho^{\smallRL}_{pq} \frac{\eta_p}{\eta_q} \frac{1-\frac{x_q^-}{x_p^-}\frac{1}{x_p^+ x_q^+}}{1-\frac{1}{x_p^+ x_q^+}} , \quad &
    \tilde{F}^{\smallRL}_{pq} &= \rho^{\smallRL}_{pq} \frac{i}{x_p^+ x_q^+} \frac{x_p^- - x_q^- }{1-\frac{1}{x_p^+ x_q^+}} , \quad &
    \tilde{D}^{\smallRL}_{pq} &= \rho^{\smallRL}_{pq} .
  \end{aligned}
\end{equation}
In order to further restrict the form of the S-matrix, we impose unitarity, physical unitarity and LR-symmetry. As shown in more detail in appendix~\ref{sec:LR-S-matrix}, this leads to the S-matrix taking the form of either a pure transmission or a pure reflection. In other words, we have either $\rho^{\smallLR}_{pq} = \rho^{\smallRL}_{pq} = 0$ or $\tau^{\smallLR}_{pq} = \tau^{\smallRL}_{pq} = 0$. 
In order to match our result with the near-BMN limit analysis performed in~\cite{Rughoonauth:2012qd}, we choose the pure transmission S-matrix.
In this case the LR-symmetry also imposes
\begin{equation}\label{eq:tau-LR-RL}
  \tau^{\smallLR}_{pq} = \zeta_{pq} S^{\smallLR}_{pq} , \qquad
  \tau^{\smallRL}_{pq} = \frac{1}{\zeta_{pq}} S^{\smallLR}_{pq} , \qquad
  \zeta_{pq} = \sqrt{\frac{1-\frac{1}{x_p^- x_q^-}}{1-\frac{1}{x_p^+ x_q^+}}} ,
\end{equation}
where $S^{\smallLR}_{pq}$ is a scalar factor that is common for the two sectors with one left-moving and one right-moving excitation. Note that this gives, $A^{\smallLR}_{pq} = A^{\smallRL}_{pq}$, $B^{\smallLR}_{pq} = B^{\smallRL}_{pq}$, \textit{etc}.

\subsection{The \texorpdfstring{$\algPSU(2|2)$}{psu(2|2)} S-matrix}
\label{sec:psu22-s-matrix}

In the calculation above we imposed invariance under the LR-symmetry. In other words, the obtained S-matrix is invariant under the simultaneous exchange of the excitations $\phi_p$, $\bar{\phi}_p$ and $\psi_p$, $\bar{\psi}_p$. Alternatively, we can require that the S-matrix takes a symmetric combination of a left- and a right-moving scalar to a state containing only scalars, and an anti-symmetric combination of a left- and a right-moving scalar into anti-symmetric combinations of scalars and fermions. Hence we require the structure of the S-matrix to be
\begin{equation}
  \begin{aligned}
    \Smat \left(\ket{\phi_p \bar{\phi}_q} + \ket{\bar{\phi}_p \phi_q}\right) &= \# \left(\ket{\phi_p \bar{\phi}_q} + \ket{\bar{\phi}_p \phi_q}\right) , \\
    \Smat \left(\ket{\phi_p \bar{\phi}_q} - \ket{\bar{\phi}_p \phi_q}\right) 
    &=  \# \left(\ket{\phi_p \bar{\phi}_q} - \ket{\bar{\phi}_p \phi_q}\right) + \# \left(\ket{\psi_p \bar{\psi}_q} - \ket{\bar{\phi}_p \phi_q}\right) .
  \end{aligned}
\end{equation}
This corresponds to introducing an additional $\algSU(2) \times \algSU(2)$ symmetry, with the two scalars transforming as a doublet under one of the $\algSU(2)$ factors and the fermions as a double under the other $\algSU(2)$ factor. The full symmetry is then enlarged to a centrally extended $\algPSU(2|2)$ algebra studied in~\cite{Beisert:2005tm,Beisert:2006qh}, and the S-matrix coincides with the one found in these references. The explicit embedding of the $\algSU(1|1)^2$ algebra and its central extension into the extended $\algPSU(2|2)$ is given in appendix~\ref{sec:psu22-algebra}.

\subsection{Scattering of excitations with different mass}
\label{sec:S-matrix-diff-mass}

So far we have discussed the case in which we scatter two excitations with the same mass. As noted in the beginning of this section, when the two excitations have different masses there are two possibilities for the momenta of the outgoing excitations. The full S-matrix can then be written as a sum of a transmission part and a reflection part. However, by imposing unitarity, physical unitarity and LR-symmetry, we find again that the S-matrix is either a pure transmission or a pure reflection. For the reasons explained before, we will consider the case of pure transmission. Since the reflection part of the S-matrix is fairly complicated, we will only write down the part of the S-matrix involving transmission processes.

In the following, to distinguish the excitations with different $s$, we will use $\phi$, $\phi'$ to denote the bosons and $\psi$, $\psi'$ for the fermions. We also choose to use the variables $x^\pm$ for $(\phi, \psi)$ and $z^\pm$ for $(\phi',\psi')$,
\begin{equation}
  x^+ + \frac{1}{x^+} - x^- - \frac{1}{x^-} = \frac{2 i  s}{h}, \qquad
  z^+ + \frac{1}{z^+} - z^- - \frac{1}{z^-} = \frac{2 i s'}{h}.
\end{equation}
As before, the corresponding right excitations will be denoted with a bar $(\bar{\phi}, \bar{\psi})$, $(\bar{\phi}', \bar{\psi}')$. The results in this section do not depend on the actual values of $s$ and $s'$. When we later discuss the $\algD{\alpha}^2$ spin-chain we will have $s=\alpha$ and $s'=1-\alpha$.

\paragraph{The \texorpdfstring{$\text{LL}'$}{LL'} and \texorpdfstring{$\text{RR}'$}{RR'} sectors.}

In a scattering process the two excitations with different mass will always be exchanged, in order to have conservation for the bosonic charges, in particular the energy. From this it follows that we can really think of the momentum as a label attached to an excitation with a certain mass.
If we also recall that the S-matrix in the LL sector for the case of same mass was found without imposing the quadratic constraint on $x^\pm$, it is easy to understand that the S-matrix in the sector that we are currently considering is just a generalization of the previous one, where we use the variables $z^\pm$ when the momentum considered is attached to excitations of the kind $(\phi', \psi')$ or $(\bar{\phi}', \bar{\psi}')$. For later convenience regarding the Bethe equations that appear in~\cite{Borsato:2012ss}, to write the S-matrix elements we choose nevertheless a different normalization as before: we want the S-matrix element corresponding to the boson-boson interaction to be equal to the unspecified scalar factor.
Explicitly, in the $\text{LL}'$ sector we have
\begin{equation}
  \begin{aligned}
    \Smat \ket{\fixedspaceL{\psi^{}_p\psi'_q}{\phi^{}_p\phi'_q}} 
    &= \fixedspaceR{A^{\smallLLp}_{pq}}{A^{\smallLLp}_{pq}} \ket{\fixedspaceR{\psi'_q\psi^{}_p}{\phi'_q\phi^{}_p}} , \qquad &
    \Smat \ket{\fixedspaceL{\phi^{}_p\psi'_q}{\phi^{}_p\psi'_q}} 
    &= \fixedspaceR{A^{\smallLLp}_{pq}}{B^{\smallLLp}_{pq}} \ket{\fixedspaceR{\psi'_q\phi^{}_p}{\psi'_q\phi^{}_p}} 
    +  \fixedspaceR{A^{\smallLLp}_{pq}}{C^{\smallLLp}_{pq}} \ket{\fixedspaceR{\psi'_q\phi^{}_p}{\phi'_q\psi^{}_p}} , \\
    \Smat \ket{\fixedspaceL{\psi^{}_p\psi'_q}{\psi^{}_p\psi'_q}} 
    &= \fixedspaceR{A^{\smallLLp}_{pq}}{F^{\smallLLp}_{pq}} \ket{\fixedspaceR{\psi'_q\psi^{}_p}{\psi'_q\psi^{}_p}} , \qquad &
    \Smat \ket{\fixedspaceL{\phi^{}_p\psi'_q}{\psi^{}_p\phi'_q}} 
    &= \fixedspaceR{A^{\smallLLp}_{pq}}{D^{\smallLLp}_{pq}} \ket{\fixedspaceR{\psi'_q\phi^{}_p}{\phi'_q\psi^{}_p}}
    +  \fixedspaceR{A^{\smallLLp}_{pq}}{E^{\smallLLp}_{pq}} \ket{\fixedspaceR{\psi'_q\phi^{}_p}{\psi'_q\phi^{}_p}} ,
  \end{aligned}
\end{equation}
where the S-matrix elements are given by
\begin{equation}
  \begin{aligned}
    A^{\smallLLp}_{pq} &= +S^{\smallLLp}_{pq}  , \quad &
    B^{\smallLLp}_{pq} &= S^{\smallLLp}_{pq} \frac{z_q^+ - x_p^+}{z_q^+ - x_p^-} , \quad &
    C^{\smallLLp}_{pq} &= S^{\smallLLp}_{pq} \frac{z_q^+ - z_q^-}{z_q^+ - x_p^-} \frac{\eta_{p}}{\eta_{q}} , \\
    F^{\smallLLp}_{pq} &= - S^{\smallLLp}_{pq} \frac{z_q^- - x_p^+}{z_q^+ - x_p^-}, &
    D^{\smallLLp}_{pq} &= S^{\smallLLp}_{pq} \frac{z_q^- - x_p^-}{z_q^+ - x_p^-} , &
    E^{\smallLLp}_{pq} &= S^{\smallLLp}_{pq} \frac{x_p^+ - x_p^-}{z_q^+ - x_p^-} \frac{\eta_{q}}{\eta_{p}} ,
  \end{aligned}
\end{equation}
where $S^{\smallLLp}_{pq}$ is again a scalar factor. To obtain the S-matrix in the L${}'$L we simply exchange $x^\pm$ and $z^\pm$ in the above relations. We denote the corresponding scalar by $S^{\smallLpL}_{pq}$. Unitarity is satisfied if the two scalar factors satisfy
\begin{equation}
  S^{\smallLpL}_{qp}  S^{\smallLLp}_{pq}=1,
\end{equation}
which means that there is only one scalar factor in the $\text{LL}'$ and L${}'$L sectors.

The computations for the RR${}'$ sector are exactly the same and we choose the same normalization. In principle one gets two new factors $S^{\smallRRp}_{pq}$ and $S^{\smallRpR}_{pq}$, which are again related by unitarity, but LR-symmetry imposes $S^{\smallRRp}_{pq}=S^{\smallLLp}_{pq}$ and $S^{\smallRpR}_{pq} = S^{\smallLpL}_{pq}$.

\paragraph{The \texorpdfstring{$\text{LR}'$}{LR'} and \texorpdfstring{$\text{RL}'$}{RL'} sectors.}

The final sectors that we consider consist of one left- and one right-moving excitation with different masses. Let us write explicitly the result for the elements of the pure transmission S-matrix
\begin{equation}
  \begin{aligned}
    \Smat \ket{\fixedspaceR{\psi_p\bar{\psi}'_q}{\phi_p\bar{\phi}'_q}} 
    &= \fixedspaceR{A^{\smallLRp}_{pq}}{A^{\smallLRp}_{pq}} \ket{\fixedspaceR{\bar{\psi}'_q\psi_p}{\bar{\phi}'_q\phi_p}}  
    +  \fixedspaceR{A^{\smallLRp}_{pq}}{B^{\smallLRp}_{pq}} \ket{\fixedspaceR{\bar{\psi}'_q\psi_p Z^-}{\bar{\psi}'_q\psi_p Z^-}} , \quad &
    \Smat \ket{\fixedspaceR{\phi_p\bar{\psi}'_q}{\phi_p\bar{\psi}'_q}} 
    &= \fixedspaceR{A^{\smallLRp}_{pq}}{C^{\smallLRp}_{pq}} \ket{\fixedspaceR{\bar{\phi}'_q\psi_p}{\bar{\psi}'_q\phi_p}} , \\
    \Smat \ket{\fixedspaceR{\psi_p\bar{\psi}'_q}{\psi_p\bar{\psi}'_q}} 
    &= \fixedspaceR{A^{\smallLRp}_{pq}}{E^{\smallLRp}_{pq}} \ket{\fixedspaceR{\bar{\psi}'_q\psi_p}{\bar{\psi}'_q\psi_p}} 
    +  \fixedspaceR{A^{\smallLRp}_{pq}}{F^{\smallLRp}_{pq}} \ket{\fixedspaceR{\bar{\psi}'_q\psi_p Z^-}{\bar{\phi}'_q\phi_p Z^+}} , \quad &
    \Smat \ket{\fixedspaceR{\phi_p\bar{\psi}'_q}{\psi_p\bar{\phi}'_q}} 
    &= \fixedspaceR{A^{\smallLRp}_{pq}}{D^{\smallLRp}_{pq}} \ket{\fixedspaceR{\bar{\phi}'_q\psi_p}{\bar{\phi}'_q\psi_p}}, 
  \end{aligned}
\end{equation}
and
\begin{equation}
  \begin{aligned}
    \Smat \ket{\fixedspaceR{\bar{\psi}'_p\psi_q}{\bar{\phi}'_p\phi_q}} 
    &= \fixedspaceR{A^{\smallRpL}_{pq}}{A^{\smallRpL}_{pq}} \ket{\fixedspaceR{\psi_q\bar{\psi}'_p}{\phi_q\bar{\phi}'_p}}  
    +  \fixedspaceR{A^{\smallRpL}_{pq}}{B^{\smallRpL}_{pq}} \ket{\fixedspaceR{\psi_q\bar{\psi}'_p Z^-}{\psi_q\bar{\psi}'_p Z^-}} , \quad &
    \Smat \ket{\fixedspaceR{\bar{\phi}'_p\psi_q}{\bar{\phi}'_p\psi_q}} 
    &= \fixedspaceR{A^{\smallRpL}_{pq}}{C^{\smallRpL}_{pq}} \ket{\fixedspaceR{\phi_q\bar{\psi}'_p}{\psi_q\bar{\phi}'_p}} , \\
    \Smat \ket{\fixedspaceR{\bar{\psi}'_p\psi_q}{\bar{\psi}'_p\psi_q}} 
    &= \fixedspaceR{A^{\smallRpL}_{pq}}{E^{\smallRpL}_{pq}} \ket{\fixedspaceR{\psi_q\bar{\psi}'_p}{\psi_q\bar{\psi}'_p}} 
    +  \fixedspaceR{A^{\smallRpL}_{pq}}{F^{\smallRpL}_{pq}} \ket{\fixedspaceR{\psi_q\bar{\psi}'_p Z^-}{\phi_q\bar{\phi}'_p Z^+}} , \quad &
    \Smat \ket{\fixedspaceR{\bar{\phi}'_p\psi_q}{\bar{\psi}'_p\phi_q}} 
    &= \fixedspaceR{A^{\smallRpL}_{pq}}{D^{\smallRpL}_{pq}} \ket{\fixedspaceR{\phi_q\bar{\psi}'_p}{\phi_q\bar{\psi}'_p}}, 
  \end{aligned}
\end{equation}
where the coefficients take the form
\begin{equation}
  \begin{aligned}
    A^{\smallLRp}_{pq} &=  +\tau^{\smallLRp}_{pq} \frac{1-\frac{1}{x_p^+ z_q^-}}{1-\frac{1}{x_p^- z_q^-}} , \  &
    B^{\smallLRp}_{pq} &= -\tau^{\smallLRp}_{pq} \frac{\eta_{p} \eta_{q}}{x_p^- z_q^-}  \frac{1}{1-\frac{1}{x_p^- z_q^-}} , \  &
    C^{\smallLRp}_{pq} &=  \tau^{\smallLRp}_{pq} , \\
    E^{\smallLRp}_{pq} &= -\tau^{\smallLRp}_{pq} \frac{1-\frac{1}{x_p^- z_q^+}}{1-\frac{1}{x_p^- z_q^-}}, \  &
    F^{\smallLRp}_{pq} &= -\tau^{\smallLRp}_{pq} \frac{\eta_{p} \eta_{q}}{x_p^+ z_q^+}  \frac{1}{1-\frac{1}{x_p^- z_q^-}} , \ &
    D^{\smallLRp}_{pq} &=  \tau^{\smallLRp}_{pq} \frac{1-\frac{1}{x_p^+ z_q^+}}{1-\frac{1}{x_p^- z_q^-}},
  \end{aligned}
\end{equation}
and
\begin{equation}
  \begin{aligned}
    A^{\smallRpL}_{pq} &= +\tau^{\smallRpL}_{pq} \frac{1-\frac{1}{z_p^+ x_q^-}}{1-\frac{1}{z_p^+ x_q^+}} , \  &
    B^{\smallRpL}_{pq} &= -\tau^{\smallRpL}_{pq} \frac{\eta_{p} \eta_{q}}{z_p^- x_q^-}  \frac{1}{1-\frac{1}{z_p^+ x_q^+}} , \  &
    C^{\smallRpL}_{pq} &=  \tau^{\smallRpL}_{pq} \frac{1-\frac{1}{z_p^- x_q^-}}{1-\frac{1}{z_p^+ x_q^+}} , \\
    E^{\smallRpL}_{pq} &= -\tau^{\smallRpL}_{pq} \frac{1-\frac{1}{z_p^- x_q^+}}{1-\frac{1}{z_p^+ x_q^+}}  , \  &
    F^{\smallRpL}_{pq} &= -\tau^{\smallRpL}_{pq} \frac{\eta_{p} \eta_{q}}{z_p^+ x_q^+}  \frac{1}{1-\frac{1}{z_p^+ x_q^+}} , \ &
    D^{\smallRpL}_{pq} &=  \tau^{\smallRpL}_{pq},
  \end{aligned}
\end{equation}
Imposing unitarity and LR-symmetry we find that the scalar factors $\tau^{\smallRLp}_{pq}$ and $\tau^{\smallRpL}_{pq}$ are given by
\begin{equation}
  \tau^{\smallLRp}_{pq} = \zeta^{\smallLRp}_{pq} S^{\smallLRp}_{pq} , \qquad
  \tau^{\smallRpL}_{pq} = \frac{1}{\zeta^{\smallRpL}_{pq}} S^{\smallLRp}_{pq} ,
\end{equation}
where
\begin{equation}
\zeta^{\smallLRp}_{pq}=\sqrt{\frac{1-\frac{1}{x_p^- z_q^-}}{1-\frac{1}{x_p^+ z_q^+}}} , \qquad
\zeta^{\smallRpL}_{pq}=\sqrt{\frac{1-\frac{1}{z_p^- x_q^-}}{1-\frac{1}{z_p^+ x_q^+}}} ,
\end{equation}
and $S^{\smallLRp}_{pq}$ is the common anti-symmetric phase factor of the sector.
Also in this case the S-matrix elements are a generalization of the ones found in the case of same masses.

\subsection{The S-matrix of the \texorpdfstring{$\algD{\alpha}^2$}{d(2,1;a) x d(2,1;a)} spin-chain}
\label{sec:d21a-S-matrix}

Now that we have derived all the S-matrix coefficients, we will slightly rewrite the S-matrix in a notation suitable for discussing the $\algD{\alpha}^2$ symmetric spin-chain discussed in section~\ref{sec:d21a-spin-chain}, and the corresponding Bethe ansatz equations. As noted in that section, the excitations of the spin-chain transform in four multiplets of $\algSU(1|1)^2$. There are two left-moving representations, which we will denote by $1$ and $3$. The masses of the excitations in these representations are $\alpha$ and $1-\alpha$, respectively. Similarly we denote the two right-moving representations $\bar{1}$ and $\bar{3}$, with the excitations again carrying mass $\alpha$ and $1-\alpha$. The excitations in these representations will then be denoted
\begin{equation*}
  (\phi^1_p,\psi^1_p) , \qquad
  (\phi^3_p,\psi^3_p) , \qquad
  (\bar{\phi}^{\bar{1}}_p,\bar{\psi}^{\bar{1}}_p) , \qquad
  (\bar{\phi}^{\bar{3}}_p,\bar{\psi}^{\bar{3}}_p) .
\end{equation*}

The S-matrix elements for scattering between excitations in the various representations are all given in sections~\ref{sec:S-matrix-same-mass} and~\ref{sec:S-matrix-diff-mass}. We only need to specify the scalar factors appearing in these matrix elements. The scalar factors that appear in the Bethe ansatz construction~\cite{Borsato:2012ss} are the inverse of the ones introduced at the level of the S-matrix. It will therefore be convenient to introduce, \eg, $S^{1\bar{3}}_{pq} = (S^{\smallLRp}_{pq})^{-1}$. In principle we then end up with 16 different phases $S^{ij}$, $i,j \in \{1,3,\bar{1},\bar{3}\}$. However, in the previous sections we have already imposed unitarity, which relates the phases $S^{ij}$ and $S^{ji}$, as well as LR-symmetry, which relates $S^{ij}$ and $S^{\bar{\imath}\bar{\jmath}}$ (where we identify $\bar{\bar{\imath}}=i$). We thus end up with six remaining independent phases
\begin{equation}
    S^{11}, \qquad S^{33}, \qquad S^{13}, \qquad S^{1\bar{1}}, \qquad S^{1\bar{3}}, \qquad S^{3\bar{3}},
\end{equation}
which amounts to four different functions, as $S^{11}$, $S^{1\bar{1}}$ and $S^{33}$, $S^{3\bar{3}}$ respectively should have the same functional form up to specifing the value of the mass.
In the next subsection we will further relate these phases by using crossing symmetry.

\subsection{Crossing equations}
\label{sec:crossing-equations}

\begin{figure}
  \centering
  \begin{tikzpicture}
    \begin{scope}[xshift=-1.4cm]
      \coordinate (i1) at (-0.9cm,0);

      \coordinate (o1) at (-0.9cm,3cm);
      \coordinate (o2) at (-0,    3cm);
      \coordinate (o3) at (+0.9cm,3cm);

      \node (v1) at (-0.375cm,1.125cm) [S-mat] {$\scriptstyle \Smat_{12}$};
      \node (v2) at (+0.375cm,1.875cm) [S-mat] {$\scriptstyle \Smat_{23}$};

      \node (v3) at ($(0,1.5cm)+(0.75cm,-0.75cm)$)  [circle,draw=black,fill=black,minimum size=3pt,inner sep=0] {};

      \draw [thick,blue] [out=90,in=270-45] (i1) to (v1);
      \draw [thick,blue] (v1) to (v2);
      \draw [thick,blue] [out=90-45,in=270] (v2) to (o3);

      \draw [thick,red] [out=270,in=135] (o1) to (v1);
      \draw [thick,red] [out=135,in=270] (v2) to (o2);

      \draw [thick,red] [out=270+45,in=180+45-30] (v1) to (v3);
      \draw [thick,red] [out=0+45+30,in=315] (v3) to (v2);

      \node at (v3) [anchor=north] {$\scriptstyle \mathsf{1}$};
    \end{scope}
    \node at (0,1.5cm) {$=$};
    \begin{scope}[xshift=+1.4cm]
      \coordinate (i1) at (-0.9cm,0);

      \coordinate (o1) at (-0.9cm,3cm);
      \coordinate (o2) at (-0,    3cm);
      \coordinate (o3) at (+0.9cm,3cm);

      \node (v) at (-0.45cm,1.95cm) [circle,draw=black,fill=black,minimum size=3pt,inner sep=0] {};

      \draw [thick,blue] [out=90,in=270] (i1) to (o3);
      \draw [thick,red] [out=270,in=90+60] (o1) to (v);
      \draw [thick,red] [out=90-60,in=270] (v) to (o2);

      \node at (v) [anchor=north] {$\scriptstyle \mathsf{1}$};
    \end{scope}
  \end{tikzpicture}

  \caption{The scattering of a fundamental excitation with a singlet is trivial.}
  \label{fig:singlet}
\end{figure}
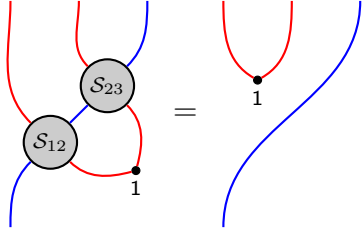

In order to derive the crossing equations, we will follow an argument first given in~\cite{Beisert:2005tm}. The idea is to scatter an excitation with the one of the singlet states constructed in section~\ref{sec:2-part-states}. By $\algSU(1|1)^2$ symmetry, the singlet should scatter trivially with an excitation carrying charges under this algebra. This gives us an equation for the scalar factors involved. As explained, the singlet is composed of one left- and one right-moving excitation with the same mass, 
\begin{equation}
  \begin{aligned}
    \ket{\textsf{1}^{1\bar{1}}_{q\bar{q}}} &= \ket{\phi^1_q \bar{\phi}^{\bar{1}}_{\bar{q}} Z^+} + \Xi_{q\bar{q}} \ket{\psi^1_q \bar{\psi}^{\bar{1}}_{\bar{q}}}, \qquad &
    \ket{\textsf{1}^{3\bar{3}}_{q\bar{q}}} &= \ket{\phi^3_q \bar{\phi}^{\bar{3}}_{\bar{q}} Z^+} + \Xi_{q\bar{q}} \ket{\psi^3_q \bar{\psi}^{\bar{3}}_{\bar{q}}}, \\
    \ket{\textsf{1}^{\bar{1}1}_{q\bar{q}}} &= \ket{\bar{\phi}^{\bar{1}}_q \phi^1_{\bar{q}} Z^+} + \Xi_{q\bar{q}} \ket{\bar{\psi}^{\bar{1}}_q \psi^1_{\bar{q}}}, \qquad &
    \ket{\textsf{1}^{\bar{3}3}_{q\bar{q}}} &= \ket{\bar{\phi}^{\bar{3}}_q \phi^3_{\bar{q}} Z^+} + \Xi_{q\bar{q}} \ket{\bar{\psi}^{\bar{3}}_q \psi^3_{\bar{q}}}, 
  \end{aligned}
\end{equation}
where $\Xi_{q\bar{q}}$ is given by
\begin{equation}
  \Xi_{q\bar{q}}= i x_{\bar{q}}^+ \frac{\eta_q}{\eta_{\bar{q}}} ,
\end{equation}
and the Zhukovsky variables identified by
\begin{equation}
  x_{\bar{q}}^\pm= \frac{1}{x_q^\pm}.
\end{equation}
 The crossing equations can be found by scattering an arbitrary excitation $\mathcal{X}_r$ with one of the four singlets
\begin{equation}\label{eq:scatt-singl}
  \Smat_{23} \Smat_{12} \ket{\mathcal{X}_p^{} \textsf{1}^{ij}_{q\bar{q}}} = X^{ij}_{p,q\bar{q}} \ket{\textsf{1}^{ij}_{q\bar{q}} \mathcal{X}_p^{} },
\end{equation}
and requiring that $X^{ij}_{p,q\bar{q}} = 1$ after the identification $x_{\bar{q}}^\pm= 1/x_q^\pm$. This is illustrated in figure~\ref{fig:singlet}.

If we scatter an excitation of type 1 with the singlets $\ket{\textsf{1}^{1\bar{1}}_{q\bar{q}}}$ and $\ket{\textsf{1}^{\bar{1}1}_{q\bar{q}}}$ we get the two equations\footnote{With respects to the first versions of this paper, we have rewritten the crossing equations in such a way that $p,q$ are on the real line and $\bar{q}$ is shifted upward by half of the imaginary period of the rapidity torus, following a standard convention.}
\begin{equation}\label{eq:cross-11}
  S^{11}_{pq} \, S^{1\bar{1}}_{p\bar{q}} = \frac{x^-_p-x^+_q}{x^-_p-x^-_q}\sqrt{\frac{x^+_p}{x^-_p}\frac{x^-_p - x^-_q}{x^+_p - x^+_q}} , \qquad
  S^{11}_{p\bar{q}} \, S^{1\bar{1}}_{pq} = \frac{1-\frac{1}{x^+_px^+_q}}{1-\frac{1}{x^+_px^-_q}}\sqrt{\frac{1-\frac{1}{x^-_p x^-_q}}{1-\frac{1}{x^+_p x^+_q}}} .
\end{equation}
If we scatter an excitation of type $\bar{1}$ with the same singlets we get
\begin{equation}
  S^{\bar{1}\bar{1}}_{pq} \, S^{\bar{1}1}_{p\bar{q}} = \frac{x^-_p-x^+_q}{x^-_p-x^-_q}\sqrt{\frac{x^+_p}{x^-_p}\frac{x^-_p - x^-_q}{x^+_p - x^+_q}} , \qquad
  S^{\bar{1}\bar{1}}_{p\bar{q}} \, S^{\bar{1}1}_{pq} = \frac{1-\frac{1}{x^+_px^+_q}}{1-\frac{1}{x^+_px^-_q}}\sqrt{\frac{1-\frac{1}{x^-_p x^-_q}}{1-\frac{1}{x^+_p x^+_q}}} .
\end{equation}
We could also have found these equations from \eqref{eq:cross-11} by imposing LR-symmetry. Similarly, we can scatter excitations of type 3 or $\bar{3}$ with the singlets made out of 3 and $\bar{3}$ and obtain equations taking the same form as the ones presented above. We can also study the case in which the excitation scatters with a singlet with a different mass. We get for example
\begin{equation}
  \label{eq:cross-31}
  S^{31}_{pq} \, S^{3\bar{1}}_{p\bar{q}} = \frac{z^+_p-x^-_q}{z^-_p-x^-_q}\sqrt{\frac{z^+_p}{z^-_p}\frac{z^-_p - x^-_q}{z^+_p - x^+_q}} , \qquad
  S^{31}_{p\bar{q}} \, S^{3\bar{1}}_{pq} =\frac{z^+_p}{z^-_p}\frac{1-\frac{1}{z^+_px^+_q}}{1-\frac{1}{z^-_px^+_q}}\sqrt{\frac{1-\frac{1}{z^-_px^-_q}}{1-\frac{1}{z^+_px^+_q}}} .
\end{equation}
The apparent discrepancy between~\eqref{eq:cross-11} and \eqref{eq:cross-31} is due to a different choice of normalization for the S-matrix in the sectors 11 and 13. Note that the crossing equations found here differ from those of~\cite{David:2010yg}.

\subsection{The Yang-Baxter equation}
\label{sec:Yang-Baxter}

In an integrable model, any $N$-body scattering event can be broken down in a sequence of two body scattering events. In general, such a factorization can be performed in several different orderings. For consistency, the result should be independent of the ordering of the two-particle S-matrix. It is sufficient to check this for the case of three-particle scattering. This leads to the Yang-Baxter (YB) equation. In terms of the operator $\Smat$ acting on three spin-chain excitations, this equation takes the form
\begin{equation}\label{eq:YB-op}
  \Smat_{12} \, \Smat_{23} \, \Smat_{12} = \Smat_{23} \, \Smat_{12} \, \Smat_{23} ,
\end{equation}
as illustrated is figure~\ref{fig:Yang-Baxter}. 
Motivated by some information on the physics, namely the near-BMN limit studied in~\cite{Rughoonauth:2012qd}, we chose a reflectionless S-matrix. A check shows that this choice satisfies YB, while we note that YB would not be satisfied if we allowed for a reflection S-matrix in some sectors.
 
It is instructive to write equation~\eqref{eq:YB-op} in a matrix basis. This is straightforward in the case of three particles of the same chirality. For definiteness, let us consider three left-moving excitations with momenta $p$, $q$ and $r$. Equation~\eqref{eq:YB-op} can then be written as
\begin{equation}\label{eq:YB-mat-LL}
  \mat{S}^{\smallLL}_{qr}\otimes \matId \, \cdot \, \matId\otimes\mat{S}^{\smallLL}_{pr} \, \cdot \, \mat{S}^{\smallLL}_{pq}\otimes\matId
   =
   \matId\otimes\mat{S}^{\smallLL}_{pq} \, \cdot \, \mat{S}^{\smallLL}_{pr}\otimes\matId \, \cdot \, \matId\otimes\mat{S}^{\smallLL}_{qr} .
\end{equation}
However, when acting with~\eqref{eq:YB-op} on a three-particle state containing particles of differing chiralities, we have to account for the presence of length-changing effects. For instance, scattering magnons of momentum $p$ and $q$ may result in two magnon with momenta $q,p$ plus the addition (or removal) of one vacuum site, \eg, 
\begin{equation}
 \ket{\phi_p \bar{\phi}_q} \mapsto \#\ket{\bar{\psi}_q \psi_p Z^-} ,
 \qquad \text{or} \qquad
 \ket{\psi_p\bar{\psi}_q} \mapsto \#\ket{\bar{\phi}_q \phi_p Z^+} .
\end{equation}
We can take care of the extra (missing) vacuum sites by moving them to the far right of the chain, but since this requires commuting them past the third magnon, this results in an additional factor of $e^{-ir}$ ($e^{+ir}$). As a result, we must modify YB equation (\ref{eq:YB-mat-LL}) to account for such phases:
\begin{equation}\label{eq:YB-mat-twist}
  \matId\otimes\mat{S}_{pq} \, \cdot \,
  \left(\mat{F}_q^{\phantom{1}}\mat{S}_{pr}\mat{F}_q^{-1}\right) \otimes \matId \, \cdot \,
  \matId\otimes\mat{S}_{qr}
  =
  \left(\mat{F}_p^{\phantom{1}}\mat{S}_{qr}\mat{F}_p^{{-1}}\right) \otimes \matId \, \cdot \,
  \matId\otimes\mat{S}_{pr} \, \cdot \,
  \left(\mat{F}_r^{\phantom{1}} \mat{S}_{pq}\mat{F}_r^{{-1}}\right) \otimes \matId ,
\end{equation}
where the transformation $\mat{F}$ implements a twist depending on the momentum of the third magnon.

The fact that the all-loop S-matrix for the dynamical spin-chain satisfies a modified (``twisted'') form of the Yang-Baxter equation, due to the length-changing effects is familiar from the case of the $\mathcal{N}=4$ SYM spin-chain.
As discussed in detail in~\cite{Arutyunov:2006yd}, the modifications to YB equation can be seen as coming from a twist in the Zamolodchikov-Faddeev (ZF) algebra that encodes the factorization of scattering. By redefining the ZF creation and annihilation operators one finds a twisted ZF algebra, with a two-body S-matrix $\mat{S}$ that satisfies a twisted YB equation\footnote{%
Our notations differ slightly from the ones of \cite{Arutyunov:2006yd} as our S-matrix permutes the momenta of the magnons.%
}. %
If $\mat{U}(p)$ is a non-local unitary operator that implements the twist on the one-particle basis, we have  that the twisted YB equation (\ref{eq:YB-mat-twist}) can be rewritten as the ordinary YB equation for a different S-matrix $\widehat{\mat{S}}$ using
\begin{equation}\label{eq:twistZF}
  \widehat{\mat{S}}_{pq} = \mat{U}^\dagger(p) \otimes \matId \, \cdot \, \mat{S}_{pq} \, \cdot \, \mat{U}(q) \otimes \matId .
\end{equation}
The factor $\mat{F}_p$ in~\eqref{eq:YB-mat-twist} is then related to $\mat{U}(p)$ by
\begin{equation}
    \mat{F}_p = \mat{U}(p) \otimes \mat{U}(p) .
\end{equation}
The twisted and untwisted ZF algebras are isomorphic, and their S-matrices must be physically, but not unitarily, equivalent. In particular, since $\widehat{\mat{S}}$ satisfies the canonical YB equation, in certain cases it may be more convenient to work directly with it. Such a choice of basis is sometimes referred to as ``string theory basis'', as opposed to the ``spin-chain basis'' that we have used here. Its explicit form can be found from the one given in section~\ref{sec:S-matrix} by using~\eqref{eq:twistZF} and specifying $\mat{U}(p)$ to act in the basis $\big(\phi,\psi,\bar{\phi},\bar{\psi}\big)$ by
\begin{equation}
  \mat{U}(p)=\diag \bigl(e^{-\frac{ip}{2}},1,e^{-\frac{ip}{2}},1\bigr) .
\end{equation}
The explicit form of the matrix elements can be found in appendix~\ref{sec:string-frame}.

\section{Discussion and outlook}
We have derived the all-loop S-matrix of an alternating spin-chain having extended $\algSU(1|1)^2$ symmetry and a discrete $\Integers_2$ left-right symmetry. The S-matrix is essentially unique and satisfies Yang-Baxter equation, which points to the quantum integrability of the resulting theory.

To completely fix the S-matrix one would have to fix the scalar factors $S^{ij}$ by means of the crossing equations of section \ref{sec:crossing-equations}, which boil down to four functional equations for $S^{ii}$, $S^{i\bar{\imath}}$, $S^{ij}$ and $S^{i\bar{\jmath}}$. Postponing a more careful analysis of the such equations to a separate work \cite{Borsato:2012ss}, let us mention that they take a form that is quite different from the one of their $\AdS_5/\CFT_4$ and $\AdS_4/\CFT_3$ counterparts. In particular, it appears that an ansatz involving simple functions and the BES dressing phase does not solve our equations. Furthermore, in light of the possible presence of bound states of particles of different masses described in section \ref{sec:2-part-states}, one might have to pick a non-minimal solution. Finally it is worth pointing out that such phases should also appear when scattering modes of different masses, which is a novel feature that was not accounted for in the Bethe ansatz equations conjectured earlier.

The S-matrix we derived should describe the massive modes of strings on $\AdS_3\times \Sphere^3\times \Sphere^3\times \Sphere^1$ through its Bethe equations \cite{Borsato:2012ss}, which give the asymptotic spectrum once the abelian phases are determined. It remains to see how the two missing massless modes can be incorporated back in this picture. Following the strategy of~\cite{Sax:2012jv}, it would be interesting to analyse the representations of the centrally extended algebra and the S-matrix in the $\alpha \to 0$ limit, and in that way trace the origin of the massless modes that appear in this limit. Understanding the massless modes would open the door to a plethora of possible developments, including the computation of the exact (\ie, non-asymptotic) spectrum by means of the thermodynamical Bethe ansatz and a better grasp of the dual CFT. An important step in that direction would be the construction of the bound state S-matrix. As noted in the end of section~\ref{sec:2-part-states}, bound states transform in a two-dimensional short multiplet similar to that of the fundamental excitations. Hence, the form of the S-matrix of these states should be the same that of the S-matrix discussed here, up to the scalar factors.

Finally, given the existence of a family of classically integrable backgrounds interpolating between the pure RR and pure NSNS one \cite{Cagnazzo:2012se}, understanding the integrable structure of these theories might provide insights on the relation between integrability and representation theory techniques.

\section*{Acknowledgments}

We would like to thank G.~Arutyunov, B.~Stefa\'nski, S.~van Tongeren, A.~Torrielli and K.~Zarembo for valuable discussions, and G.~Arutyunov for his useful comments on the manuscript. The authors acknowledge support by the Netherlands Organization for Scientific Research (NWO) under the VICI grant 680-47-602.
The work by the authors is also part of the ERC Advanced grant research programme No. 246974, ``Supersymmetry: a window to non-perturbative physics''.

\appendix

\section{The \texorpdfstring{$\algD{\alpha}$}{d(2,1;a)} algebra}
\label{sec:d21a-algebra}

The generators of $\algD{\alpha}$ satisfy the commutation relations
\begin{equation*}
  \begin{aligned}
    \comm{\gen{J}_0}{\gen{J}_\pm} &= \pm \gen{J}_\pm , &
    \comm{\gen{J}_+}{\gen{J}_-} &= 2 \gen{J}_0 , &
    \comm{\gen{J}_0}{\gen{Q}_{\pm\beta\dot{\beta}}} &= \pm\frac{1}{2} \gen{Q}_{\pm\beta\dot{\beta}} , &
    \comm{\gen{J}_\pm}{\gen{Q}_{\mp\beta\dot{\beta}}} &= \gen{Q}_{\pm\beta\dot{\beta}} , \\
    \comm{\gen{L}_5}{\gen{L}_\pm} &= \pm \gen{L}_\pm , &
    \comm{\gen{L}_+}{\gen{L}_-} &= 2 \gen{L}_5 , &
    \comm{\gen{L}_5}{\gen{Q}_{a\pm\dot{\beta}}} &= \pm\frac{1}{2} \gen{Q}_{a\pm\dot{\beta}} , &
    \comm{\gen{L}_\pm}{\gen{Q}_{a\mp\dot{\beta}}} &= \gen{Q}_{a\pm\dot{\beta}} , \\
    \comm{\gen{R}_8}{\gen{R}_\pm} &= \pm \gen{R}_\pm , &
    \comm{\gen{R}_+}{\gen{R}_-} &= 2 \gen{R}_8 , &
    \comm{\gen{R}_8}{\gen{Q}_{a\beta\pm}} &= \pm\frac{1}{2} \gen{Q}_{a\beta\pm} , &
    \comm{\gen{R}_\pm}{\gen{Q}_{a\beta\mp}} &= \gen{Q}_{a\beta\pm} ,
  \end{aligned}
\end{equation*}
\begin{equation}
  \begin{aligned}
    \acomm{\gen{Q}_{\pm++}}{\gen{Q}_{\pm--}} &= \pm \gen{J}_{\pm} , &
    \acomm{\gen{Q}_{\pm+-}}{\gen{Q}_{\pm-+}} &= \mp \gen{J}_{\pm} , \\
    \acomm{\gen{Q}_{+\pm+}}{\gen{Q}_{-\pm-}} &= \mp \alpha \gen{L}_{\pm} , &
    \acomm{\gen{Q}_{+\pm-}}{\gen{Q}_{-\pm+}} &= \pm \alpha \gen{L}_{\pm} , \\
    \acomm{\gen{Q}_{++\pm}}{\gen{Q}_{--\pm}} &= \mp (1-\alpha) \gen{R}_{\pm} , &
    \acomm{\gen{Q}_{+-\pm}}{\gen{Q}_{-+\pm}} &= \pm (1-\alpha) \gen{R}_{\pm} , \\
  \end{aligned}
\end{equation}
\begin{equation*}
  \begin{aligned}
    \acomm{\gen{Q}_{+\pm\pm}}{\gen{Q}_{-\mp\mp}} &= - \gen{J}_0 \pm \alpha\gen{L}_5 \pm (1-\alpha)\gen{R}_8 , \\
    \acomm{\gen{Q}_{+\pm\mp}}{\gen{Q}_{-\mp\pm}} &= + \gen{J}_0 \mp \alpha\gen{L}_5 \pm (1-\alpha)\gen{R}_8 .
  \end{aligned}
\end{equation*}
The non-vanishing action of the $\algD{\alpha}$ generators on these states of the $(-\tfrac{\alpha}{2};\tfrac{1}{2};0)$ representation is given by
\begin{equation}
  \begin{gathered}
    \begin{aligned}
      \gen{L}_5 \ket{\phi_{\pm}^{(n)}} &= \pm \frac{1}{2} \ket{\phi_{\pm}^{(n)}} , &
      \gen{L}_+ \ket{\phi_{-}^{(n)}} &= \ket{\phi_{+}^{(n)}} , &
      \gen{L}_- \ket{\phi_{+}^{(n)}} &= \ket{\phi_-^{(n)}} , \\
      \gen{R}_8 \ket{\psi_{\pm}^{(n)}} &= \pm \frac{1}{2} \ket{\psi_{\pm}^{(n)}} , &
      \gen{R}_+ \ket{\psi_{-}^{(n)}} &= \ket{\psi_+^{(n)}} , &
      \gen{R}_- \ket{\psi_{+}^{(n)}} &= \ket{\psi_-^{(n)}} ,
    \end{aligned} \\
    \begin{aligned}
      \gen{J}_0 \ket{\phi_{\beta}^{(n)}} &= - \left( \tfrac{\alpha}{2} + n \right) \ket{\phi_{\beta}^{(n)}} , &
      \gen{J}_0 \ket{\psi_{\dot\beta}^{(n)}} &= - \left( \tfrac{\alpha}{2} + \tfrac{1}{2} + n \right) \ket{\psi_{\dot\beta}^{(n)}} , \\
      \gen{J}_+ \ket{\phi_{\beta}^{(n)}} &= +\sqrt{(n - 1 + \alpha)n} \ket{\phi_{\beta}^{(n-1)}} , &
      \gen{J}_+ \ket{\psi_{\dot\beta}^{(n)}} &= +\sqrt{(n + \alpha)n} \ket{\psi_{\dot\beta}^{(n-1)}} , \\
      \gen{J}_- \ket{\phi_{\beta}^{(n-1)}} &= -\sqrt{(n - 1 + \alpha) n} \ket{\phi_{\beta}^{(n)}} , &
      \gen{J}_- \ket{\psi_{\dot\beta}^{(n-1)}} &= -\sqrt{(n + \alpha) n} \ket{\psi_{\dot\beta}^{(n)}} ,
    \end{aligned} \\
    \begin{aligned}
      \gen{Q}_{-\pm\dot\beta} \ket{\phi_{\mp}^{(n)}} &= \pm \sqrt{n+\alpha} \ket{\psi_{\dot\beta}^{(n)}} , &
      \gen{Q}_{+\pm\dot\beta} \ket{\phi_{\mp}^{(n)}} &= \pm \sqrt{n} \ket{\psi_{\dot\beta}^{(n-1)}} , \\
      \gen{Q}_{-\beta\pm} \ket{\psi_{\mp}^{(n)}} &= \mp \sqrt{n+1} \ket{\phi_{\beta}^{(n+1)}} , &
      \gen{Q}_{+\beta\pm} \ket{\psi_{\mp}^{(n)}} &= \mp \sqrt{n+\alpha} \ket{\phi_{\beta}^{(n)}} .
    \end{aligned}
  \end{gathered}
\end{equation}
On the $(-\tfrac{1-\alpha}{2};0;\tfrac{1}{2})$ representation the generators act as
\begin{equation}
  \begin{gathered}
    \begin{aligned}
      \gen{L}_5 \ket{\tilde{\psi}_{\pm}^{(n)}} &= \pm \frac{1}{2} \ket{\tilde{\psi}_{\pm}^{(n)}} , &
      \gen{L}_+ \ket{\tilde{\psi}_{-}^{(n)}} &= \ket{\tilde{\psi}_+^{(n)}} , &
      \gen{L}_- \ket{\tilde{\psi}_{+}^{(n)}} &= \ket{\tilde{\psi}_-^{(n)}} , \\
      \gen{R}_8 \ket{\tilde{\phi}_{\pm}^{(n)}} &= \pm \frac{1}{2} \ket{\tilde{\phi}_{\pm}^{(n)}} , &
      \gen{R}_+ \ket{\tilde{\phi}_{-}^{(n)}} &= \ket{\tilde{\phi}_+^{(n)}} , &
      \gen{R}_- \ket{\tilde{\phi}_{+}^{(n)}} &= \ket{\tilde{\phi}_-^{(n)}} ,
    \end{aligned} \\
    \begin{aligned}
      \gen{J}_0 \ket{\tilde{\phi}_{\dot\gamma}^{(n)}} &= - \left( \tfrac{1-\alpha}{2} + n \right) \ket{\tilde{\phi}_{\dot\gamma}^{(n)}} , &
      \gen{J}_0 \ket{\tilde{\psi}_{\gamma}^{(n)}} &= - \left( \tfrac{1 - \alpha}{2} + \tfrac{1}{2} + n \right) \ket{\tilde{\psi}_{\gamma}^{(n)}} , \\
      \gen{J}_+ \ket{\tilde{\phi}_{\dot\gamma}^{(n)}} &= +\sqrt{(n - \alpha) n} \ket{\tilde{\phi}_{\dot\gamma}^{(n-1)}} , & 
      \gen{J}_+ \ket{\tilde{\psi}_{\gamma}^{(n)}} &= +\sqrt{(n + 1 - \alpha) n} \ket{\tilde{\psi}_{\gamma}^{(n-1)}} , \\
      \gen{J}_- \ket{\tilde{\phi}_{\dot\gamma}^{(n-1)}} &= -\sqrt{(n - \alpha) n} \ket{\tilde{\phi}_{\dot\gamma}^{(n)}} , &
      \gen{J}_- \ket{\tilde{\psi}_{\gamma}^{(n-1)}} &= -\sqrt{(n + 1 - \alpha) n} \ket{\tilde{\psi}_{\gamma}^{(n)}} ,
    \end{aligned} \\
    \begin{aligned}
      \gen{Q}_{-\gamma\pm} \ket{\tilde{\phi}_{\mp}^{(n)}} &= \pm \sqrt{n+1-\alpha} \ket{\tilde{\psi}_{\gamma}^{(n)}} , &
      \gen{Q}_{+\gamma\pm} \ket{\tilde{\phi}_{\mp}^{(n)}} &= \pm \sqrt{n} \ket{\tilde{\psi}_{\gamma}^{(n-1)}} , \\
      \gen{Q}_{-\pm\dot\gamma} \ket{\tilde{\psi}_{\mp}^{(n)}} &= \mp \sqrt{n+1} \ket{\tilde{\phi}_{\dot\gamma}^{(n+1)}} , &
      \gen{Q}_{+\pm\dot\gamma} \ket{\tilde{\psi}_{\mp}^{(n)}} &= \mp \sqrt{n+1-\alpha} \ket{\tilde{\phi}_{\dot\gamma}^{(n)}} .
    \end{aligned}
  \end{gathered}
\end{equation}

\section{Embedding into \texorpdfstring{$\algPSU(2|2)$}{psu(2|2)}}
\label{sec:psu22-algebra}

The centrally extended $\algSU(1|1)^2$ algebra can be embedded into the centrally extended $\algPSU(2|2)$ algebra discussed in~\cite{Beisert:2005tm,Beisert:2006qh}. There are several equivalent ways to do this. We can for example consider only the diagonal generators, which can be identified as
\begin{equation}
  \begin{gathered}
    \hat{\genQ}^1{}_1 = \genQ_{\smallL} , \qquad
    \hat{\genQ}^2{}_2 = \genQ_{\smallR} , \qquad
    \hat{\genS}^1{}_1 = \genS_{\smallL} , \qquad
    \hat{\genS}^2{}_2 = \genS_{\smallR} , \\
    \hat{\gen{L}}^1{}_1 = -\hat{\gen{L}}^2{}_2 = \tfrac{1}{2} \genM - \genB , \qquad
    \hat{\gen{R}}^1{}_1 = -\hat{\gen{R}}^2{}_2 = \tfrac{1}{2} \genM + \genB , \\
    \hat{\gen{C}} = \tfrac{1}{2} \genH , \qquad
    \hat{\genP} = -\genP , \qquad
    \hat{\gen{K}} = -\genK .
  \end{gathered}
\end{equation}
Here the generators of $\algPSU(2|2)$ have been indicated by a hat. For the fundamental representations we can identify $\phi^1$ and $\psi^1$ as left-moving excitations, while $\phi^2$ and $\psi^2$ are right moving. For the charges $B=1/4$, $s=1/2$, this reproduces the fundamental $(\rep{2}|\rep{2})$ representation of the centrally extended $\algPSU(2|2)$ algebra.

\section{The S-matrix in the LR sector}
\label{sec:LR-S-matrix}

As noted in section~\ref{sec:S-matrix-same-mass}, the ansatz in~\eqref{eq:T-ansatz-LR}, \eqref{eq:R-ansatz-LR}, \eqref{eq:T-ansatz-RL} and~\eqref{eq:R-ansatz-RL}, for the S-matrix for two particles of the same mass but of different chirality involves both a transmission and a reflection part. Requiring that the S-matrix commutes with the supercharges we are left with the solution for the coefficients of the S-matrix given in equations~\eqref{eq:T-solution-LR}, \eqref{eq:R-solution-LR}, \eqref{eq:T-solution-RL} and~\eqref{eq:R-solution-RL}. This solution depends on four functions $\tau^{\smallLR}_{pq}$, $\tau^{\smallRL}_{pq}$, $\rho^{\smallLR}_{pq}$ and $\rho^{\smallRL}_{pq}$, which we can further constrain by requiring that unitarity and physical unitarity are satisfied. For definiteness we parameterize each function as a product of a positive real function of the two momenta and two phases, respectively symmetric and anti-symmetric in the exchange of $p$ and $q$. We then have
\begin{equation}
  \begin{aligned}
    \tau^{\smallLR}_{pq} &= r^{\smallLR}_{pq} \exp\bigl[ i\bigl(\theta^{\smallLR}_{pq} + \vartheta^{\smallLR}_{pq} \bigr) \bigr] , \qquad &
    \rho^{\smallLR}_{pq} &= \tilde{r}^{\smallLR}_{pq} \exp\bigl[ i\bigl (i \tilde{\theta}^{\smallLR}_{pq} + \tilde{\vartheta}^{\smallLR}_{pq} \bigr) \bigr] , \\
    \tau^{\smallRL}_{pq} &= r^{\smallRL}_{pq} \exp\bigl[ i\bigl (i \theta^{\smallRL}_{pq} + \vartheta^{\smallRL}_{pq} \bigr) \bigr] , \qquad &
    \rho^{\smallRL}_{pq} &= \tilde{r}^{\smallRL}_{pq} \exp\bigl[ i\bigl (i \tilde{\theta}^{\smallRL}_{pq} + \tilde{\vartheta}^{\smallRL}_{pq} \bigr) \bigr] ,
  \end{aligned}
\end{equation}
where, for example,
\begin{equation}
  \abs{\tau^{\smallLR}_{pq}} = r^{\smallLR}_{pq} , \qquad
  \theta^{\smallLR}_{qp} = -\theta^{\smallLR}_{qp} , \qquad
  \vartheta^{\smallLR}_{qp} = +\vartheta^{\smallLR}_{qp} .
\end{equation}
We first solve the linear constraint
\begin{equation}
  \mat{S}^{\dag}_{pq} = \mat{S}_{qp} ,
\end{equation}
which gives\footnote{%
  After imposing the other quadratic equations, it turns out that for all of the functions $r^{ij}_{qp}=r^{ij}_{pq}$. Here we already use this fact to write the solutions.%
} %
\begin{equation}\label{eq:LR-S-sol-linear-constr}
  \begin{gathered}
    \bigl(e^{i\tilde{\vartheta}^{\smallLR}_{pq}}\bigr)^2  = 1, \qquad  
    e^{i\tilde{\vartheta}^{\smallRL}_{pq}} = \pm e^{i\tilde{\vartheta}^{\smallLR}_{pq}} , \qquad
    r^{\smallLR}_{pq} = r^{\smallRL}_{pq} \equiv r_{pq} , \\
    e^{i\theta^{\smallRL}_{pq}} = e^{i\theta^{\smallLR}_{pq}} , \qquad 
    e^{i\vartheta^{\smallRL}_{pq}} = e^{-i\vartheta^{\smallLR}_{pq}} , 
  \end{gathered}
\end{equation}
At this point it is possible to impose either unitarity or physical unitarity and solve the corresponding quadratic equations: 
\begin{equation}
  \mat{S}_{qp} \mat{S}_{pq} =  \matId , \qquad
  \mat{S}^{\dag}_{pq} \mat{S}_{pq} = \matId.
\end{equation}
There are three different solutions of these equations,
\begin{enumerate}
\item a pure transmission ($\tilde{r}^{\smallLR}_{pq} = \tilde{r}^{\smallRL}_{pq} = 0$), 
\item a pure reflection ($r^{\smallLR}_{pq} = r^{\smallRL}_{pq} = 0$),
\item a family of solutions that interpolates between transmission and reflection:

\begin{equation}
  \tilde{r}^{\smallLR}_{pq} = \tilde{r}^{\smallRL}_{pq} \equiv \tilde{r}_{pq}, \qquad 
  (r_{pq})^2 + (\tilde{r}_{pq})^2  =1, \qquad
  \tilde{\theta}^{\smallLR}_{pq} + \tilde{\theta}^{\smallRL}_{pq} = \theta^{\smallLR}_{pq} + \theta^{\smallRL}_{pq} ,
\end{equation}
where the last relation is possible only if one chooses $e^{i\tilde{\vartheta}^{\smallLR}_{pq}} = -e^{i\tilde{\vartheta}^{\smallRL}_{pq}}$ in~\eqref{eq:LR-S-sol-linear-constr}. This is the reason why we wrote the solutions 1 and 2 above as separate from 3: solution 2 is different from what one gets by imposing $r_{pq} = 0$ in solution 3.\footnote{%
  This distinction is important when imposing the discrete LR-symmetry. Requiring this symmetry in solution 3 would leave only transmission, while one can find a non trivial reflection S-matrix from solution 2 that satisfies LR-symmetry by choosing $\tilde{\vartheta}^{\smallLR}_{pq}  = \tilde{\vartheta}^{\smallRL}_{pq}$.%
} %

As noted in section~\ref{sec:psu22-s-matrix}, we can recover the $\algPSU(2|2)$ found in~\cite{Beisert:2005tm} from the S-matrix considered here. In the above notation it corresponds to
\begin{equation}
  \begin{aligned}
    \tilde{r}_{pq} & = \sqrt{\frac{(x_p^- - x_p^+)(x_q^- - x_q^+)}{(x_q^- -x_p^+)(x_p^- - x_q^+)}}, \qquad &
    e^{i\tilde{\theta}^{\smallLR}_{pq}} & = S^{\text{B}}_{pq}  \sqrt{\frac{x_p^- - x_q^+}{x_q^- -x_p^+}}, \\
    e^{i\vartheta^{\smallLR}_{pq}} & = \sqrt{\frac{1-\frac{1}{x_p^- x_q^-}}{1-\frac{1}{x_p^+ x_q^+}}}, &
    e^{i\theta^{\smallLR}_{pq}} & = S^{\text{B}}_{pq}  \sqrt{\frac{x_p^- - x_q^+}{x_p^+ -x_q^-}},
  \end{aligned}
\end{equation}
where $S^{\text{B}}_{pq}$ is the dressing factor for the $\algPSU(2|2)$ S-matrix.
\end{enumerate}
Here we want also the discrete LR-symmetry to be satisfied. To do this we consider for example the processes
\begin{equation}
  \begin{aligned}
    \Smat \ket{\phi_p\bar{\psi}_q} 
    &= \fixedspaceR{C^{\smallLR}_{pq}}{C^{\smallLR}_{pq}} \ket{\fixedspaceL{\bar{\psi}_q\psi_p}{\bar{\psi}_q\phi_p}}
    + \fixedspaceR{\tilde{C}^{\smallLR}_{pq}}{\tilde{C}^{\smallLR}_{pq}} \ket{\fixedspaceL{\psi_q\bar{\psi}_p}{\phi_q\bar{\psi}_p}}, \\
    \Smat \ket{\bar{\phi}_p\psi_q} 
    &= \fixedspaceR{C^{\smallRL}_{pq}}{C^{\smallRL}_{pq}} \ket{\fixedspaceL{\bar{\psi}_q\psi_p}{\bar{\psi}_q\phi_p}}
    + \fixedspaceR{\tilde{C}^{\smallRL}_{pq}}{\tilde{C}^{\smallRL}_{pq}} \ket{\fixedspaceL{\psi_q\bar{\psi}_p}{\phi_q\bar{\psi}_p}},
  \end{aligned}
\end{equation}
and impose $C^{\smallLR}_{pq} = C^{\smallRL}_{pq}$ and $\tilde{C}^{\smallLR}_{pq} = \tilde{C}^{\smallRL}_{pq}$. This leaves only two possible solutions, a pure transmission and a pure reflection. The same result can be found by first imposing the LR-symmetry and then requiring that unitarity is satisfied. 

In order to choose between the two cases of pure transmission and pure reflection we compare our results with the perturbative calculation in~\cite{Rughoonauth:2012qd}, where the tree-level $\AdS_3$ string theory S-matrix for the scalars was found to be reflectionless. 
We also check the Yang-Baxter (YB) equation
\begin{equation}
  \Smat_{12} \Smat_{23} \Smat_{12} =\Smat_{23} \Smat_{12} \Smat_{23} 
\end{equation}
and we find that an S-matrix with a pure transmission in the LR sector does satisfy the YB equation, while one with a pure reflection does not.

\section{A second central extension}
\label{sec:central-extension-II}

In this section we will consider an alternative central extension of  the $\algSU(1|1)^2$ algebra, and give its representation and the resulting S-matrix.
In this second central extension it is more difficult to think of only one spin-chain, but it is rather natural to think of one spin-chain for the left movers and another one for the right movers. The reason is that scattering processes with length-changing effects will now add one site in a spin-chain and subtract one site in the other.
Nevertheless it is still possible to uniformize the notation by interpreting $\ZpIIext$ ($\ZmIIext$) as the addition (removal) of one site in the chain for the left movers and the removal (addition) of one site in the chain for the right movers. We will thus have
\begin{equation}
  \begin{aligned}
    \ket{\ZpIIext \phi_p} &= e^{- ip}\ket{\phi_p \ZpIIext} , & \qquad   
    \ket{\ZmIIext \phi_p} &= e^{+ ip}\ket{\phi_p \ZmIIext} , \\
    \ket{\ZpIIext \bar{\phi}_p} &= e^{+ ip}\ket{\bar{\phi}_p \ZpIIext} , &  
    \ket{\ZmIIext \bar{\phi}_p} &= e^{- ip}\ket{\bar{\phi}_p \ZmIIext} . \\
  \end{aligned}
\end{equation}

\subsection{Algebra}

In the second central extension of $\algSU(1|1)^2$ we allow for different anti-commutators to be non-vanishing, namely
\begin{equation}
  \begin{aligned}
    \acomm{\genQ_{\smallL}}{\genS_{\smallL}} &= \genH_{\smallL} , &
    \acomm{\genQ_{\smallR}}{\genS_{\smallR}} &= \genH_{\smallR} , \\
    \acomm{\genQ_{\smallL}}{\genQ_{\smallR}} &= 0 , &
    \acomm{\genS_{\smallL}}{\genS_{\smallR}} &= 0 , \\
    \acomm{\genQ_{\smallL}}{\genS_{\smallR}} &= \genP , &
    \acomm{\genS_{\smallL}}{\genQ_{\smallR}} &= \genK .
  \end{aligned}
\end{equation}
To find representations of this algebra we use a similar ansatz as before. For the case of two $\phi$ representations we write
\begin{align}
  \genQ_{\smallL} \ket{\phi} &= a \ket{\psi} , &
  \genQ_{\smallL} \ket{\psi} &= 0 , \\
  \genS_{\smallL} \ket{\phi} &= 0 , &
  \genS_{\smallL} \ket{\psi} &= b \ket{\phi} , \\
  \genQ_{\smallR} \ket{\phi} &= c \ket{\psi \ZmIIext} , &
  \genQ_{\smallR} \ket{\psi} &= 0 , \\
  \genS_{\smallR} \ket{\phi} &= 0 , &
  \genS_{\smallR} \ket{\psi} &= d \ket{\phi \ZpIIext} ,
\end{align}
This leads to a representation very similar to that of the first central extension. Indeed we only need to exchange the role of $c$ and $d$ and the coefficients now read
\begin{align}
  a_p &= e^{+i\gamma_L} \sqrt{\frac{ih}{2}(x^-_p - x^+_p)} , &
  c_p &= \frac{i e^{-i\delta_L}}{x^-_p} \sqrt{\frac{ih}{2}(x^-_p - x^+_p)} , \\
  b_p &= e^{-i\gamma_L} \sqrt{\frac{ih}{2}(x^-_p - x^+_p)} , &
  d_p &= -\frac{i e^{+i\delta_L}}{x^+_p} \sqrt{\frac{ih}{2}(x^-_p - x^+_p)} 
\end{align}
On the state $\ket{\phi_p \phi_q}$ we then get
\begin{equation}
  \genP \ket{\phi_p \phi_q} = \left( e^{-iq} a_p d_p + a_q d_q \right) \ket{\phi_p \phi_q}.
\end{equation}
As before this vanishes provided $e^{i(p+q)} = 1$.

For the right-movers we use the ans\"{a}tze 
\begin{align}
  \genQ_{\smallR} \ket{\bar{\phi}} &= \bar{a} \ket{\bar{\psi}}, &
  \genQ_{\smallR} \ket{\bar{\psi}} &= 0 , \\
  \genS_{\smallR} \ket{\bar{\phi}} &= 0 , &
  \genS_{\smallR} \ket{\bar{\psi}} &= \bar{b} \ket{\bar{\phi}} , \\
  \genQ_{\smallL} \ket{\bar{\phi}} &= \bar{c} \ket{\bar{\psi} \ZpIIext}, &
  \genQ_{\smallL} \ket{\bar{\psi}} &= 0 , \\
  \genS_{\smallL} \ket{\bar{\phi}} &= 0 , &
  \genS_{\smallL} \ket{\bar{\psi}} &= \bar{d} \ket{\bar{\phi} \ZmIIext}.
\end{align}
The coefficients are given by
\begin{align}
  \bar{a}_p &= e^{+i\gamma_R} \sqrt{\frac{ih}{2}(x^-_p - x^+_p)} , &
  \bar{c}_p &= \frac{i e^{-i\delta_R}}{x^-_p} \sqrt{\frac{ih}{2}(x^-_p - x^+_p)} , \\
  \bar{b}_p &= e^{-i\gamma_R} \sqrt{\frac{ih}{2}(x^-_p - x^+_p)}, &
  \bar{d}_p &= -\frac{i e^{+i\delta_R}}{x^+_p} \sqrt{\frac{ih}{2}(x^-_p - x^+_p)} ,
\end{align}
We now consider one left-moving and one right-moving excitation. We get
\begin{align}
  \genP \ket{\phi_p \bar{\phi}_q}
  &= 
  a_p d_p \ket{\phi_p \ZpIIext \bar{\phi}_q}
  +
  \bar{b}_q \bar{c}_q \ket{\phi_p  \bar{\phi}_q \ZpIIext} \\
  &= (e^{iq} a_p d_p + \bar{b}_q \bar{c}_q) \ket{\phi_p \bar{\phi}_q \ZpIIext}.
\end{align}
Inserting the expressions for the coefficients we then get
\begin{equation}
  e^{iq} a_p d_p + \bar{b}_q \bar{c}_q = \frac{h}{2} e^{i(\gamma_L + \delta_L)} (e^{-i(p-q)} - e^{+iq}) + \frac{h}{2} e^{i(\gamma_R + \delta_R)} ( e^{+iq} - 1 ) .
\end{equation}
In order for this to vanish we need 
\begin{equation}
  \gamma_L + \delta_L = \gamma_R + \delta_R,
\end{equation}
but now the condition on the momenta reads
\begin{equation}
  e^{i(p-q)} = 1.
\end{equation}
Hence the left-movers and right-movers in these representations contribute to the total momentum with different signs.

\subsection{The S-matrix}

As in the first central extension, also in this case we can identify independent sectors for the S-matrix. It is easy to understand that the results for the S-matrix in the LL and RR sectors will be the same as in the first central extension. Hence, we refer to section~\ref{sec:S-matrix-same-mass} for these S-matrix elements.

The results will be different in the LR sector. In particular, after imposing that it commutes with the supercharges, we find a \emph{unique} S-matrix. 
It is interesting to note that when we require that the S-matrix commutes with the bosonic charges, the ansatz in the LR sector is different from the one used in the first central extension: boson-boson interactions will not produce anymore fermion-fermion excitations (similarly for fermion-fermion interactions) and we have length-changing effects when scattering a boson and a fermion of different chiralities, in particular when the fermionic number is not transmitted in the scattering process.
The cases of different or same mass are formally the same.
We thus present the result for the more general case of two excitations with different mass; the S-matrix in the LR sector is
\begin{equation}\label{eq:second-ce-S-mat-LR}
  \begin{aligned}
    \Smat \ket{\fixedspaceL{\psi_p \psi_q}{\phi_p \bar{\phi}_q}} &= \fixedspaceR{D_{pq}}{A^{\smallLR}_{pq}} \ket{\fixedspaceL{\psi_p \psi_q}{\bar{\phi}_q \phi_p}} , \qquad &
    \Smat \ket{\fixedspaceL{\psi_p \psi_q}{\bar{\phi}_p \phi_q}} &= \fixedspaceR{D_{pq}}{A^{\smallRL}_{pq}} \ket{\fixedspaceL{\psi_p \psi_q}{\phi_q \bar{\phi}_p}} , \\
    \Smat \ket{\fixedspaceL{\psi_p \psi_q}{\phi_p \bar{\psi}_q}} &= \fixedspaceR{D_{pq}}{B^{\smallLR}_{pq}} \ket{\fixedspaceL{\psi_p \psi_q}{\bar{\psi}_q \phi_p}} + C^{\smallLR}_{pq} \ket{\fixedspaceL{\psi_p \psi_q}{\bar{\phi}_q \psi_p} \ZmIIext}, &
    \Smat \ket{\fixedspaceL{\psi_p \psi_q}{\bar{\phi}_p \psi_q}} &= \fixedspaceR{D_{pq}}{B^{\smallRL}_{pq}} \ket{\fixedspaceL{\psi_p \psi_q}{\psi_q \bar{\phi}_p}} + C^{\smallRL}_{pq} \ket{\fixedspaceL{\psi_p \psi_q}{\phi_q \bar{\psi}_p} \ZpIIext},\\
    \Smat \ket{\fixedspaceL{\psi_p \psi_q}{\psi_p \bar{\phi}_q}} &= \fixedspaceR{D_{pq}}{D^{\smallLR}_{pq}} \ket{\fixedspaceL{\psi_p \psi_q}{\bar{\phi}_q \psi_p}} + E^{\smallLR}_{pq} \ket{\fixedspaceL{\psi_p \psi_q}{\bar{\psi}_q \phi_p} \ZpIIext}, &
    \Smat \ket{\fixedspaceL{\psi_p \psi_q}{\bar{\psi}_p \phi_q}} &= \fixedspaceR{D_{pq}}{D^{\smallRL}_{pq}} \ket{\fixedspaceL{\psi_p \psi_q}{\phi_q \bar{\psi}_p}} + E^{\smallRL}_{pq} \ket{\fixedspaceL{\psi_p \psi_q}{\psi_q \bar{\phi}_p} \ZmIIext},\\
    \Smat \ket{\fixedspaceL{\psi_p \bar{\psi}_q}{\psi_p \bar{\psi}_q}} &= \fixedspaceR{D_{pq}}{F^{\smallLR}_{pq}} \ket{\bar{\psi}_q \psi_p}, &
    \Smat \ket{\fixedspaceL{\psi_p \psi_q}{\bar{\psi}_p \psi_q}} &= \fixedspaceR{D_{pq}}{F^{\smallRL}_{pq}} \ket{\psi_q \bar{\psi}_p}. \\
  \end{aligned}
\end{equation}
where the coefficients take the form
\begin{equation}
  \begin{aligned}
    A^{\smallLR}_{pq} &= +S^{\smallLR}_{pq} , \qquad &
    B^{\smallLR}_{pq} &= S^{\smallLR}_{pq} \frac{1+\frac{1}{x^+_p z^+_q}}{1+\frac{1}{ x^-_p z^+_q}} , &
    C^{\smallLR}_{pq} &= -S^{\smallLR}_{pq} \frac{\eta_{xp} \eta_{zq}}{x^-_p z^+_q} \frac{1}{1+\frac{1}{ x^-_p z^+_q}},\\
    F^{\smallLR}_{pq} &= -S^{\smallLR}_{pq} \frac{1+\frac{1}{x^+_p z^-_q}}{1+\frac{1}{ x^-_p z^+_q}},&
    D^{\smallLR}_{pq} &= S^{\smallLR}_{pq} \frac{1+\frac{1}{x^-_p z^-_q}}{1+\frac{1}{ x^-_p z^+_q}},&
    E^{\smallLR}_{pq} &= +S^{\smallLR}_{pq} \frac{\eta_{xp} \eta_{zq}}{x^+_p z^-_q} \frac{1}{1+\frac{1}{ x^-_p z^+_q}} . \\
  \end{aligned}
\end{equation}
The S-matrix elements with superscript $RL$ assume the same form as the ones with $LR$ after exchanging $x$ and $z$.
In principle, there are two different phases $S^{\smallLR}_{pq}$ and $S^{\smallRL}_{pq}$ that are then fixed to be equal by imposing unitarity and physical unitarity. Note that LR-symmetry is automatically satisfied in this case.
We checked that also this S-matrix satisfies the Yang-Baxter equation.

\section{The S-matrix in string and spin-chain frames}
\label{sec:string-frame}

Here we will give the explicit matrix form of the S-matrix acting on a two-magnon state. To keep our expression manageable, we will consider two excitations having the same mass. Let us pick a basis for Hilbert space of a single magnon with momentum $p$ as
\begin{equation}
 \mathcal{V}_p = \operatorname{span} \bigl(\phi_p,\psi_p,\bar{\phi}_p,\bar{\psi}_p\bigr) .
\end{equation}
The S-matrix then takes a state in $\mathcal{V}_p\otimes\mathcal{V}_q$ to a state in $\mathcal{V}_q\otimes\mathcal{V}_p$. The corresponding matrix $\mat{S}$ then takes the block form
\begin{equation}
  \mat{S}_{pq} = \!\left(\!
    \begin{array}{c|c}
      \mat{S}^{\smallLL}_{pq} & \mat{S}^{\smallRL}_{pq} \rule[-1.3ex]{0pt}{0pt} \\
      \hline
      \mat{S}^{\smallLR}_{pq} & \mat{S}^{\smallRR}_{pq}
    \end{array}\!
  \right) .
\end{equation}
We parametrize the full matrix form of $\Smat$ as
\begin{equation*}
  \newcommand{\0}{\color{black!40}0}
  \renewcommand{\arraystretch}{1.1}
  \setlength{\arraycolsep}{3pt}
  \mat{S}_{pq}=\!\left(\!
    \mbox{\footnotesize$
      \begin{array}{cccccccc|cccccccc}
        \Ael^{\smallLL}_{pq} & \0 & \0 & \0 & \0 & \0 & \0 & \0 & \0 & \0 & \0 & \0 & \0 & \0 & \0 & \0 \\
        \0 & \Cel^{\smallLL}_{pq} & \0 & \0 &\Del^{\smallLL}_{pq} & \0 & \0 & \0 & \0 & \0 & \0 & \0 & \0 & \0 & \0 & \0 \\
        \0 & \0 & \0 & \0 & \0 & \0 & \0 & \0 & \Ael^{\smallRL}_{pq}  & \0 & \0 & \0 & \0 & \Fel^{\smallRL}_{pq} & \0 & \0 \\
        \0 & \0 & \0 & \0 & \0 & \0 & \0 & \0 & \0 & \0 & \0 & \0 & \Del^{\smallRL}_{pq} & \0 & \0 & \0 \\
        \0 & \Bel^{\smallLL}_{pq} & \0 & \0 & \Eel^{\smallLL}_{pq} & \0 & \0 & \0 & \0 & \0 & \0 & \0 & \0 & \0 & \0 & \0 \\
        \0 & \0 & \0 & \0 & \0 & \Fel^{\smallLL}_{pq} & \0 & \0 & \0 & \0 & \0 & \0 & \0 & \0 & \0 & \0 \\
        \0 & \0 & \0 & \0 & \0 & \0 & \0 & \0 & \0 & \Cel^{\smallRL}_{pq} & \0 & \0 & \0 & \0 & \0 & \0 \\
        \0 & \0 & \0 & \0 & \0 & \0 & \0 & \0 & \Bel^{\smallRL}_{pq} & \0 & \0 & \0 & \0 &\Eel^{\smallRL}_{pq} & \0 & \0 \rule[-1.4ex]{0pt}{0pt} \\
        \hline
        \0 & \0 & \Ael^{\smallLR}_{pq} & \0 & \0 & \0 & \0 & \Fel^{\smallLR}_{pq}  & \0 & \0 & \0 & \0 & \0 & \0 & \0 & \0 \\
        \0 & \0 & \0 & \0 & \0 & \0 & \Del^{\smallLR}_{pq}  & \0 & \0 & \0 & \0 & \0 & \0 & \0 & \0 & \0 \\
        \0 & \0 & \0 & \0 & \0 & \0 & \0 & \0 & \0 & \0 & \Ael^{\smallRR}_{pq} & \0 & \0 & \0 & \0 & \0 \\
        \0 & \0 & \0 & \0 & \0 & \0 & \0 & \0 & \0 & \0 & \0 &\Cel^{\smallRR}_{pq} & \0 & \0 & \Del^{\smallRR}_{pq} & \0 \\
        \0 & \0 & \0 & \Cel^{\smallLR}_{pq}  & \0 & \0 & \0 & \0 & \0 & \0 & \0 & \0 & \0 & \0 & \0 & \0 \\
        \0 & \0 & \Bel^{\smallLR}_{pq}  & \0 & \0 & \0 & \0 & \Eel^{\smallLR}_{pq}  & \0 & \0 & \0 & \0 & \0 & \0 & \0 & \0 \\
        \0 & \0 & \0 & \0 & \0 & \0 & \0 & \0 & \0 & \0 & \0 & \Bel^{\smallRR}_{pq} & \0 & \0 &\Eel^{\smallRR}_{pq} & \0 \\
        \0 & \0 & \0 & \0 & \0 & \0 & \0 & \0 & \0 & \0 & \0 & \0 & \0 & \0 & \0 & \Fel^{\smallRR}_{pq}
      \end{array}$}\!
  \right).
\end{equation*}
Let us list the matrix elements in either the spin-chain or string frame. For this purpose, recalling that the two are related by~\eqref{eq:twistZF}, we introduce
\begin{equation}
  \nu_p = \begin{cases} 
    \sqrt{\frac{x_p^+}{x_p^-}} & \text{in the string frame}\,,\\
    1 & \text{in the spin-chain frame}\,. 
  \end{cases}
\end{equation}
We then have
\begin{equation}
  \begin{aligned}
    \Ael^{\smallLL}_{pq} &= +S^{\smallLL}_{pq}\frac{\nu_p}{\nu_q} \frac{x_q^+ - x_p^-}{x_q^- - x_p^+} , \quad &
    \Bel^{\smallLL}_{pq} &= S^{\smallLL}_{pq}\frac{1}{\nu_q}\frac{x_q^+ - x_p^+}{x_q^- - x_p^+} , \quad &
    \Cel^{\smallLL}_{pq} &= S^{\smallLL}_{pq} \frac{\nu_p}{\nu_q} \frac{x_q^+ - x_q^-}{x_q^- - x_p^+} \frac{\eta_p}{\eta_q} ,\\
    \Fel^{\smallLL}_{pq} &= - S^{\smallLL}_{pq} ,\quad &    
    \Del^{\smallLL}_{pq} &= S^{\smallLL}_{pq}\, \nu_p\frac{x_q^- - x_p^-}{x_q^- - x_p^+} ,\quad &
    \Eel^{\smallLL}_{pq} &= S^{\smallLL}_{pq} \frac{x_p^+ - x_p^-}{x_q^- - x_p^+} \frac{\eta_q}{\eta_p},
  \end{aligned}
\end{equation}
and a similar list of expression in the RR sector, as discussed in section \ref{sec:S-matrix-same-mass}. Furthermore,
\begin{equation}
  \begin{aligned}
    \Ael^{\smallLR}_{pq} &= \tau^{\smallLR}_{pq} \frac{\nu_p}{\nu_q} \frac{1-\frac{1}{x_p^+ x_q^-}}{1-\frac{1}{x_p^- x_q^-}}, \ &
    \Bel^{\smallLR}_{pq} &= -\tau^{\smallLR}_{pq} \frac{1}{\nu_q} \frac{\eta_p \eta_q}{x_p^- x_q^-} \frac{1}{1-\frac{1}{x_p^- x_q^-}}, \ &
    \Cel^{\smallLR}_{pq} &= \tau^{\smallLR}_{pq} \frac{1}{\nu_q} , \\
    \Eel^{\smallLR}_{pq} &= -\tau^{\smallLR}_{pq} \frac{1-\frac{1}{x_p^- x_q^+}}{1-\frac{1}{x_p^- x_q^-}}, \ &
    \Fel^{\smallLR}_{pq} &= -\tau^{\smallLR}_{pq} \,\nu_p\frac{\eta_p \eta_q}{x_p^+ x_q^+} \frac{1}{1-\frac{1}{x_p^- x_q^-}}, \ &
    \Del^{\smallLR}_{pq} &= \tau^{\smallLR}_{pq} \,\nu_p\frac{1-\frac{1}{x_p^+ x_q^+}}{1-\frac{1}{x_p^- x_q^-}},
  \end{aligned}
\end{equation}
and
\begin{equation}
  \begin{aligned}
    \Ael^{\smallRL}_{pq} &= \tau^{\smallRL}_{pq} \frac{\nu_p}{\nu_q} \frac{1-\frac{1}{x_p^+ x_q^-}}{1-\frac{1}{x_p^+ x_q^+}} , \ &
    \Bel^{\smallRL}_{pq} &= -\tau^{\smallRL}_{pq} \frac{1}{\nu_q} \frac{\eta_p \eta_q}{x_p^- x_q^-} \frac{1}{1-\frac{1}{x_p^+ x_q^+}} , \ &
    \Cel^{\smallRL}_{pq} &= \tau^{\smallRL}_{pq}  \frac{1}{\nu_q}\frac{1-\frac{1}{x_p^- x_q^-}}{1-\frac{1}{x_p^+ x_q^+}}, \\
    \Eel^{\smallRL}_{pq} &= -\tau^{\smallRL}_{pq} \frac{1-\frac{1}{x_p^- x_q^+}}{1-\frac{1}{x_p^+ x_q^+}} , \ &
    \Fel^{\smallRL}_{pq} &= -\tau^{\smallRL}_{pq} \,\nu_p\frac{\eta_p \eta_q}{x_p^+ x_q^+} \frac{1}{1-\frac{1}{x_p^+ x_q^+}} , \ &
    \Del^{\smallRL}_{pq} &= \tau^{\smallRL}_{pq} \,\nu_p,
  \end{aligned}
\end{equation}
where $\eta$ is given by~\eqref{eq:eta-p-def} and the coefficients $\tau^{\smallLR}$ and $\tau^{\smallRL}$ by~\eqref{eq:tau-LR-RL},
\begin{equation}
  \tau^{\smallLR}_{pq} = \zeta_{pq} S^{\smallLR}_{pq} , \qquad
  \tau^{\smallRL}_{pq} = \frac{1}{\zeta_{pq}} S^{\smallLR}_{pq} , \qquad
  \zeta_{pq} = \sqrt{\frac{1-\frac{1}{x_p^- x_q^-}}{1-\frac{1}{x_p^+ x_q^+}}} .
\end{equation}
Recall that $S^{\smallLL}_{pq},S^{\smallRR}_{pq},S^{\smallLR}_{pq}$ and $S^{\smallRL}_{pq}$ are undetermined antisymmetric scalar factors.

As discussed in section \ref{sec:S-matrix-diff-mass}, the S-matrix for different masses takes the same form as above when one allows for an appropriate scalar factor and introduces suitable Zhukovski variables $z^\pm$. In view of our different choice of normalization and for the reader's convenience let us nonetheless explicitly list the elements of $\Smat$ below for particles of different masses in either frame:
\begin{equation}
  \begin{aligned}
    \Ael^{\smallLLp}_{pq} &= S^{\smallLLp}_{pq}\frac{\nu_p}{\nu_q}  , \ &
    \Bel^{\smallLLp}_{pq} &= S^{\smallLLp}_{pq}\frac{1}{\nu_q} \frac{z_q^+ - x_p^+}{z_q^+ - x_p^-} , \ &
    \Cel^{\smallLLp}_{pq} &= S^{\smallLLp}_{pq} \frac{\nu_p}{\nu_q}\frac{z_q^+ - z_q^-}{z_q^+ - x_p^-} \frac{\eta_{p}}{\eta_{q}} , \\
    \Fel^{\smallLLp}_{pq} &= - S^{\smallLLp}_{pq} \frac{z_q^- - x_p^+}{z_q^+ - x_p^-}, \ &
    \Del^{\smallLLp}_{pq} &= S^{\smallLLp}_{pq} \,\nu_p\frac{z_q^- - x_p^-}{z_q^+ - x_p^-} ,\ &
    \Eel^{\smallLLp}_{pq} &= S^{\smallLLp}_{pq} \frac{x_p^+ - x_p^-}{z_q^+ - x_p^-} \frac{\eta_{q}}{\eta_{p}} ,
  \end{aligned}
\end{equation}
\begin{equation*}
  \begin{aligned}
    \Ael^{\smallLRp}_{pq} &=  \tau^{\smallLRp}_{pq}\frac{\nu_p}{\nu_q}\frac{1-\frac{1}{x_p^+ z_q^-}}{1-\frac{1}{x_p^- z_q^-}} , \  &
    \Bel^{\smallLRp}_{pq} &= -\tau^{\smallLRp}_{pq} \frac{1}{\nu_q}\frac{\eta_{p} \eta_{q}}{x_p^- z_q^-}  \frac{1}{1-\frac{1}{x_p^- z_q^-}} , \ &
    \Cel^{\smallLRp}_{pq} &=  \tau^{\smallLRp}_{pq} \frac{1}{\nu_q}, \\
    \Eel^{\smallLRp}_{pq} &= -\tau^{\smallLRp}_{pq} \frac{1-\frac{1}{x_p^- z_q^+}}{1-\frac{1}{x_p^- z_q^-}}, \ &
    \Fel^{\smallLRp}_{pq} &= -\tau^{\smallLRp}_{pq} \,\nu_p\frac{\eta_{p} \eta_{q}}{x_p^+ z_q^+}  \frac{1}{1-\frac{1}{x_p^- z_q^-}} , \ &
    \Del^{\smallLRp}_{pq} &=  \tau^{\smallLRp}_{pq} \,\nu_p\frac{1-\frac{1}{x_p^+ z_q^+}}{1-\frac{1}{x_p^- z_q^-}},
  \end{aligned}
\end{equation*}
\begin{equation*}
  \begin{aligned}
    \Ael^{\smallRpL}_{pq} &=  \tau^{\smallRpL}_{pq} \frac{\nu_p}{\nu_q}\frac{1-\frac{1}{z_p^+ x_q^-}}{1-\frac{1}{z_p^+ x_q^+}} , \  &
    \Bel^{\smallRpL}_{pq} &= -\tau^{\smallRpL}_{pq} \frac{1}{\nu_q} \frac{\eta_{p} \eta_{q}}{z_p^- x_q^-}  \frac{1}{1-\frac{1}{z_p^+ x_q^+}} , \ &
    \Cel^{\smallRpL}_{pq} &=  \tau^{\smallRpL}_{pq} \frac{1}{\nu_q}\frac{1-\frac{1}{z_p^- x_q^-}}{1-\frac{1}{z_p^+ x_q^+}} , \\
    \Eel^{\smallRpL}_{pq} &= -\tau^{\smallRpL}_{pq} \frac{1-\frac{1}{z_p^- x_q^+}}{1-\frac{1}{z_p^+ x_q^+}},\ &
    \Fel^{\smallRpL}_{pq} &= -\tau^{\smallRpL}_{pq} \,\nu_p\frac{\eta_{p} \eta_{q}}{z_p^+ x_q^+}  \frac{1}{1-\frac{1}{z_p^+ x_q^+}} , \ &
    \Del^{\smallRpL}_{pq} &=  \tau^{\smallRpL}_{pq} \,\nu_p .
  \end{aligned}
\end{equation*}

\section{Comparison with Beisert's $\algSU(1|1)$ S-matrix}
\label{sec:beisert-su11-S-matrix}

In this section we will compare the S-matrix in the main text with the $\algSU(1|1)$ symmetric S-matrix discussed in~\cite{Beisert:2005wm}. Let us consider excitations transforming under an $\algU(1|1)$ algebra of the form discussed in section~\ref{sec:u11-algebra}. The commutation relations of this algebra read
\begin{equation}\label{eq:u11-algebra}
  \acomm{\genQ}{\genS} = \genH , \qquad
  \comm{\genB}{\genQ} = - \tfrac{1}{2} \genQ , \qquad
  \comm{\genB}{\genS} = + \tfrac{1}{2} \genS .
\end{equation}
This algebra has a two-dimensional representation denoted $(\rep{1}|\rep{1})_{H,B}$, with the charges acting as
\begin{equation}\label{eq:u11-representation}
  \begin{aligned}
    \genQ \ket{\fixedspaceL{\psi}{\phi}} &= v \ket{\psi} , &
    \genS \ket{\fixedspaceL{\psi}{\phi}} &= 0 , &
    \genH \ket{\fixedspaceL{\psi}{\phi}} &= H \ket{\phi} , &
    \genB \ket{\fixedspaceL{\psi}{\phi}} &= (B - \tfrac{1}{2}) \ket{\phi} , \\
    \genQ \ket{\psi} &= 0 , &
    \genS \ket{\psi} &= H/v \ket{\phi} , &
    \genH \ket{\psi} &= H \ket{\psi} , &
    \genB \ket{\psi} &= (B - 1) \ket{\psi} .
  \end{aligned}
\end{equation}
The two-particle S-matrix acting on excitations transforming in such representations can be written in terms of projectors onto the representations appearing in the decomposition
\begin{equation}
  (\rep{1}|\rep{1})_{H,B} \otimes (\rep{1}|\rep{1})_{H',B'} = (\rep{1}|\rep{1})_{H+H',B+B'-1/2} \oplus (\rep{1}|\rep{1})^*_{H+H',B+B'-1} ,
\end{equation}
where $(\rep{1}|\rep{1})^*_{H,B}$ denotes a representation of the form~\eqref{eq:u11-representation}, but with the roles of the boson $\phi$ and the fermion $\psi$ interchanged.
Imposing that the S-matrix satisfies unitarity and the Yang-Baxter equation, the action of the R-matrix is found to be\footnote{%
  The R-matrix $\Rmat$ is closely related to the S-matrix $\Smat$, but does not exchange the scattering particles. We therefore have the relation $\Smat = \mathcal{P} \Rmat$, where $\mathcal{P}$ is a permutation acting on a two-particle state.%
}%
\begin{equation}\label{eq:beisert-su11-Rmat}
  \begin{aligned}
    \Rmat \ket{\phi_p \phi_q} &= \frac{a_q - a_p + \frac{i}{2}(H_p + H_q)}{a_q - a_p - \frac{i}{2}(H_p + H_q)} \ket{\phi_p \phi_q} , \\
    \Rmat \ket{\phi_p \psi_q} &= 
    \frac{a_q - a_p - \frac{i}{2}(H_p - H_q)}{a_q - a_p - \frac{i}{2}(H_p + H_q)} \ket{\phi_p \psi_q} + 
    \frac{iH_q}{a_q - a_p - \frac{i}{2}(H_p + H_q)} \frac{v_p}{v_q} \ket{\psi_p \phi_q} , \\
    \Rmat \ket{\psi_p \phi_q} &= 
    \frac{a_q - a_p + \frac{i}{2}(H_p - H_q)}{a_q - a_p - \frac{i}{2}(H_p + H_q)} \ket{\psi_p \phi_q} + 
    \frac{iH_p}{a_q - a_p - \frac{i}{2}(H_p + H_q)} \frac{v_q}{v_p} \ket{\phi_p \psi_q} , \\
    \Rmat \ket{\psi_p \psi_q} &= \ket{\psi_p \psi_q} ,
  \end{aligned}
\end{equation}
where $a_p$ and $a_q$ are spectral parameters corresponding to the two excitations. The exact form of these parameters will depend on the model under consideration.

In order to compare this result with the S-matrix in section~\ref{sec:S-matrix}, we consider the left-moving $\algU(1|1)$ algebra generated by the charges $\gen{Q}_L$, $\gen{S}_L$, $\gen{H}_L$ and $\gen{B}_L$. Under this algebra the left-moving excitations $(\phi_p|\psi_p)$ transform in a representation of the form~\eqref{eq:u11-representation} with the parameters $H_p$ and $v_p$ given by
\begin{equation}
  H_p = +i(x^- - x^+) , \qquad
  v_p = \eta_p .
\end{equation}
We furthermore express the spectral parameter $a_p$ in terms of $x^\pm$ as
\begin{equation}
  a_p = \frac{1}{2}(x^- + x^+)
\end{equation}
Inserting these expressions into the R-matrix~\eqref{eq:beisert-su11-Rmat}, we recover the LL sector S-matrix of section~\ref{sec:S-matrix-same-mass}.

The right-moving excitations are slightly more complicated. Firstly, the charge $\gen{Q}_L$ acts non-trivially on the fermion $\bar{\psi}_p$ and trivially on the boson $\bar{\phi}_p$. Hence the grading of the right moving $\algU(1|1)$ representation is reversed, corresponding to a representation of the type $(\rep{1}|\rep{1})^*_{H,B}$. Secondly, the left-moving supercharges act on the right-moving excitations by adding or subtracting vacuum sites. In order to have a representation of the form~\eqref{eq:u11-representation} we therefore consider the doublet $(\bar{\psi}_p|\bar{\phi}_p Z^+)$ and identify the parameters
\begin{equation}
  \bar{H}_p = -i\left(\frac{1}{\bar{x}^-} - \frac{1}{\bar{x}^+}\right) , \qquad
  \bar{v}_p = - \frac{i \eta_p}{\bar{x}^+}.
\end{equation}
Writing spectral parameter $\bar{a}_p$ as
\begin{equation}
  \bar{a}_p = \frac{1}{2} \left(\frac{1}{\bar{x}^-} + \frac{1}{\bar{x}^+}\right).
\end{equation}
the expressions in~\eqref{eq:beisert-su11-Rmat} reproduce RR sectors of the S-matrix in section~\ref{sec:S-matrix}.\footnote{%
  The inverted grading of the right-moving representation gives rise to some additional minus signs that are needed to reproduce the results of section~\ref{sec:S-matrix}.%
}%
By using a combination of the two representations above we also find the S-matrix in the LR and RL sectors.

In~\cite{Ahn:2012hw} Ahn and Bombardelli (AB) proposed another S-matrix for $\AdS_3 \times \Sphere^3 \times \Sphere^3 \times \Sphere^1$. Like above, this S-matrix can be constructed from Beisert's $\algSU(1|1)$ S-matrix. The difference between the two constructions is that the left- and right-movers in~\cite{Ahn:2012hw} transform in identical $\algSU(1|1)$ representations $(\phi|\psi)$ and $(\bar{\phi}|\bar{\psi})$, while, as discussed above, the right-movers in this paper transform in a representations where the roles of the bosons and fermions are reversed. This difference is only important when we consider the mixed LR and RL sectors. In particular, in the LR sector the S-matrix of this paper schematically acts as\footnote{%
  In order to simplify the notation we have suppressed the effects of length-changing in the spin-chain in the out states, corresponding to the string frame discussed in appendix~\ref{sec:string-frame}.%
}%
\begin{equation}\label{eq:S-LR-BOSS}
  \begin{aligned}
    \Smat^{\scriptscriptstyle\text{BOSS}} \ket{\fixedspaceL{\psi_p\bar{\psi}_q}{\phi_p \bar{\phi}_q}} 
    &= \fixedspaceR{A^{\scriptscriptstyle\text{BOSS}}_{pq}}{A^{\scriptscriptstyle\text{BOSS}}_{pq}} \ket{\fixedspaceL{\bar{\psi}_q\psi_p}{\bar{\phi}_q\phi_p}} + \fixedspaceR{A^{\scriptscriptstyle\text{BOSS}}_{pq}}{B^{\scriptscriptstyle\text{BOSS}}_{pq}} \ket{\bar{\psi}_q \psi_p}, \qquad &
    \Smat^{\scriptscriptstyle\text{BOSS}} \ket{\fixedspaceL{\psi_p\bar{\psi}_q}{\phi_p\bar{\psi}_q}} 
    &= \fixedspaceR{A^{\scriptscriptstyle\text{BOSS}}_{pq}}{C^{\scriptscriptstyle\text{BOSS}}_{pq}} \ket{\fixedspaceL{\bar{\psi}_q\psi_p}{\bar{\psi}_q\phi_p}}, \\
    \Smat^{\scriptscriptstyle\text{BOSS}} \ket{\fixedspaceL{\psi_p\bar{\psi}_q}{\psi_p\bar{\psi}_q}} 
    &= \fixedspaceR{A^{\scriptscriptstyle\text{BOSS}}_{pq}}{E^{\scriptscriptstyle\text{BOSS}}_{pq}} \ket{\fixedspaceL{\bar{\psi}_q\psi_p}{\bar{\psi}_q\psi_p}} + \fixedspaceR{A^{\scriptscriptstyle\text{BOSS}}_{pq}}{F^{\scriptscriptstyle\text{BOSS}}_{pq}} \ket{\bar{\phi}_q\phi_p}, \qquad &
    \Smat^{\scriptscriptstyle\text{BOSS}} \ket{\fixedspaceL{\psi_p\bar{\psi}_q}{\psi_p\bar{\phi}_q}} 
    &= \fixedspaceR{A^{\scriptscriptstyle\text{BOSS}}_{pq}}{D^{\scriptscriptstyle\text{BOSS}}_{pq}} \ket{\fixedspaceL{\bar{\psi}_q\psi_p}{\bar{\phi}_q\psi_p}},
  \end{aligned}
\end{equation}
while the S-matrix of~\cite{Ahn:2012hw} gives
\begin{equation}\label{eq:S-LR-AB}
  \begin{aligned}
    \Smat^{\scriptscriptstyle\text{AB}} \ket{\fixedspaceL{\psi_p\bar{\psi}_q}{\phi_p \bar{\phi}_q}} 
    &= \fixedspaceR{A^{\scriptscriptstyle\text{AB}}_{pq}}{A^{\scriptscriptstyle\text{AB}}_{pq}} \ket{\fixedspaceL{\bar{\psi}_q\psi_p}{\bar{\phi}_q\phi_p}} , \qquad &
    \Smat^{\scriptscriptstyle\text{AB}} \ket{\fixedspaceL{\psi_p\bar{\psi}_q}{\phi_p\bar{\psi}_q}} 
    &= \fixedspaceR{A^{\scriptscriptstyle\text{AB}}_{pq}}{B^{\scriptscriptstyle\text{AB}}_{pq}} \ket{\fixedspaceL{\bar{\psi}_q\psi_p}{\bar{\psi}_q\phi_p}} +
    \fixedspaceR{A^{\scriptscriptstyle\text{AB}}_{pq}}{C^{\scriptscriptstyle\text{AB}}_{pq}} \ket{\fixedspaceL{\bar{\psi}_q\psi_p}{\bar{\phi}_q\psi_p}} , \\
    \Smat^{\scriptscriptstyle\text{AB}} \ket{\fixedspaceL{\psi_p\bar{\psi}_q}{\psi_p\bar{\psi}_q}} 
    &= \fixedspaceR{A^{\scriptscriptstyle\text{AB}}_{pq}}{F^{\scriptscriptstyle\text{AB}}_{pq}} \ket{\fixedspaceL{\bar{\psi}_q\psi_p}{\bar{\psi}_q\psi_p}} , \qquad &
    \Smat^{\scriptscriptstyle\text{AB}} \ket{\fixedspaceL{\psi_p\bar{\psi}_q}{\psi_p\bar{\phi}_q}} 
    &= \fixedspaceR{A^{\scriptscriptstyle\text{AB}}_{pq}}{D^{\scriptscriptstyle\text{AB}}_{pq}} \ket{\fixedspaceL{\bar{\psi}_q\psi_p}{\bar{\phi}_q\psi_p}} +
    \fixedspaceR{A^{\scriptscriptstyle\text{AB}}_{pq}}{E^{\scriptscriptstyle\text{AB}}_{pq}} \ket{\fixedspaceL{\bar{\psi}_q\psi_p}{\bar{\psi}_q\phi_p}}.
  \end{aligned}
\end{equation}
Comparing~\eqref{eq:S-LR-AB} with the S-matrix in~\eqref{eq:second-ce-S-mat-LR}, we see that the Ahn and Bombardelli corresponds to the second central extension of the $\algSU(1|1) \times \algSU(1|1)$ algebra discussed in appendix~\ref{sec:central-extension-II}. As noted there, this algebra does not seem to have a straight forward spin-chain interpretation, since the supercharges change the lengths of the left- and right-moving parts of the spin-chain in opposite ways.

We also note that it is possible to distinguish equations~\eqref{eq:S-LR-BOSS} and~\eqref{eq:S-LR-AB} perturbatively by determining which of the two processes
\begin{equation*}
  \ket{\phi_p \bar{\phi}_q} \to \ket{\bar{\psi}_q \psi_p} , 
  \qquad \text{and} \qquad
  \ket{\phi_p \bar{\psi}_q} \to \ket{\bar{\phi}_q \psi_p}
\end{equation*}
has a non-zero amplitude.

\bibliographystyle{oos}
\bibliography{refs}

\end{document}